%% file: main.tex
\documentclass[3p]{elsarticle}
\usepackage{amsmath}
\usepackage{amssymb}
\usepackage{amsthm}
\usepackage{mathtools}
\usepackage{stmaryrd}
\usepackage[utf8]{inputenc}
\usepackage{hyperref}
\usepackage{url}
\usepackage{color}
\usepackage{multirow}
\usepackage{pgfplotstable}
\usepackage{booktabs}
\usepackage{graphicx}
\usepackage{caption}
\usepackage{subcaption}
\usepackage{float}
\usepackage{lineno}
\usepackage[utf8]{inputenc}
\usepackage{pgfplots}
\DeclareUnicodeCharacter{2212}{−}
\usepgfplotslibrary{external,groupplots,dateplot}

\usetikzlibrary{patterns,shapes.arrows}
\pgfplotsset{compat=newest}
\input{figures/test.tikzstyles}
\usepackage{todonotes}
\usepackage{appendix}
\usepackage{cleveref}
\crefname{section}{Section}{Sections}
\crefname{figure}{Figure}{Figures}
\crefname{equation}{Equation}{Equations}
\crefname{table}{Table}{Tables}
\crefname{appendix}{Appendix}{Appendices}
\crefname{algocfline}{Algorithm}{Algorithms}
\crefname{lstlisting}{Listing}{Listings}
\usepackage{flafter}
\usepackage[frozencache]{minted}
\usepackage[linesnumbered,lined,boxed,commentsnumbered]{algorithm2e}
\SetKw{KwDownto}{downto}
\SetKwComment{Comment}{$\triangleright$\ }{}
\usepackage{csquotes}
\usepackage[exponent-product=\cdot]{siunitx}
\DeclareSIUnit\Nm{Nm}  
\DeclareSIUnit\Hertz{Hz}
\DeclareSIUnit\byte{B}  
\DeclareSIUnit\flop{FLOP}  
\DeclareSIUnit\Darcy{D}  
\newcommand{\ips}[2]{\left<#1,#2\right>}
\newcommand{\ip}[2]{\left[#1,#2\right]}
\newcommand{\dd}[1]{\,\mathrm{d}#1}
\newcommand{\ck}[0]{Cauchy-Kovalewski }
\DeclareMathOperator{\diag}{diag}

\newcommand{\R}{\mathbb{R}}
\newcommand{\derivative}[2]{\mathchoice{\frac{\partial #1}{\partial #2}}{\partial_{#2} #1}{}{}}

\newtheorem{lem}{Lemma}

\title{An Efficient ADER-DG Local Time Stepping Scheme for 3D HPC Simulation of Seismic Waves in Poroelastic Media}
\author[1]{Sebastian Wolf \corref{cor1}}
\ead{wolf.sebastian@in.tum.de}
\cortext[cor1]{Corresponding author}
\affiliation[1]{organization={Department of Informatics, Technical University of Munich}, addressline={Boltzmannstr. 3}, postcode={85748}, city={Garching}, country={Germany}}
\author[2,3]{Martin Galis}
\ead{martin.galis@uniba.sk}
\affiliation[2]{organization={Faculty of Mathematics, Physics and Informatics, Comenius University in Bratislava}, addressline={Mlynska dolina F1}, postcode={842 48}, city={Bratislava}, country={Slovakia}}
\affiliation[3]{organization={Earth Science Institute, Slovak Academy of Sciences}, addressline={Dubravska cesta 9}, postalcode={840 05}, city={Bratislava}, country={Slovakia}}
\author[4]{Carsten Uphoff}
\ead{uphoff@geophysik.uni-muenchen.de}
\affiliation[4]{organization={Department of Earth and Environmental Sciences, Ludwig-Maximilians-Universität München}, addressline={Theresienstr. 41}, postcode={80333}, city={Munich}, country={Germany}}
\author[4]{Alice-Agnes Gabriel}
\author[2,3]{Peter Moczo}
\author[2]{David Gregor}
\author[1]{Michael Bader}
\date{\today}
\begin{document}
\begin{abstract}
Many applications from the fields of seismology and geoengineering require simulations of seismic waves in porous media.
Biot's theory of poroelasticity describes the coupling between solid and fluid phases and introduces a stiff reactive source term (Darcy's Law) into the elastodynamic wave equations, thereby increasing computational cost of respective numerical solvers and motivating efficient methods utilising High-Performance Computing.

We present a novel realisation of the discontinuous Galerkin scheme with Arbitrary High-Order DERivative time stepping (ADER-DG) that copes with stiff source terms.
To integrate this source term with a reasonable time step size, we utilise an element-local space-time predictor, which needs to solve medium-sized linear systems -- each with \num{1000} to \num{10000} unknowns -- in each element update (i.e., billions of times).
We present a novel block-wise back-substitution algorithm for solving these systems efficiently, thus enabling large-scale 3D simulations.
In comparison to LU decomposition, we reduce the number of floating-point operations by a factor of up to \num{25}, when using polynomials of degree \num{6}.
The block-wise back-substitution is mapped to a sequence of small matrix-matrix multiplications, for which code generators are available to generate highly optimised code. 

We verify the new solver thoroughly against analytical and semi-analytical reference solutions in problems of increasing complexity. 
We demonstrate high-order convergence of the scheme for 3D problems.
We verify the correct treatment of point sources and boundary conditions, including homogeneous and heterogeneous full space problems as well as problems with traction-free boundary conditions.
In addition, we compare against a finite difference solution for a newly defined 3D layer over half-space problem containing an internal material interface and free surface.
We find that extremely high accuracy is required to accurately resolve the slow, diffusive P-wave at a or near a free surface, while we also demonstrate that solid particle velocities are not affected by coarser resolutions.
We demonstrate that by using a clustered local time stepping scheme, time to solution is reduced by a factor of \num{6} to \num{10} compared to global time stepping.
We conclude our study with a scaling and performance analysis on the SuperMUC-NG supercomputer, demonstrating our implementation's high computational efficiency and its potential for extreme-scale simulations.
\end{abstract}

\maketitle

\section{Introduction}

Elastodynamic wave propagation in fluid-saturated porous rocks is a relevant and challenging topic in computational seismology.
Specifically in the contexts of seismic exploration, monitoring of geological reservoirs and human-induced earthquakes, it is important to study the interaction between waves, fluids and solids in the subsurface.
Applications from exploration geophysics to earthquake engineering require high-resolution 3D forward simulations of seismic wave propagation in porous media.
To better understand which information seismic waves carry about the porosity, permeability and fluid-saturation of rocks, forward simulations of seismic waves propagating in poroelastic materials are required (e.g.~\cite{carcione_computational_2010}).

Poroelastic materials consist of a solid matrix with pores that are completely fluid-filled.
Biot's theory of poroelasticity~\cite{biot_theory_1956, biot_theory_1956-1, biot_theory_1956-2, biot_mechanics_1962} describes the interaction between the fluid and the solid phase and is widely accepted and validated~\cite{plona_observation_1980, berryman_confirmation_1980, carcione_wave_2015}.
The resulting system of partial differential equations (PDEs) describes seismic wave propagation in porous media, extending the elastic model often used in computational seismology by additional quantities (e.g., fluid velocities) and, in particular, by a stiff reactive source term that is required to model viscosity effects of the fluid--solid interaction. 
For the numerical solution of the governing equations, this stiff source term is typically a key computational challenge (see \cref{sec:related-work}). 

In this study, we focus on the Discontinuous Galerkin method with Arbitrary DERivative time stepping (ADER-DG) for poroelastic materials, as introduced by~\citet{de_la_puente_discontinuous_2008}. 
The DG method combines advantages from finite volume and finite element methods~\cite{reed_triangular_1973, cockburn_tvb_1989, hesthaven_nodal_2002, dumbser_arbitrary_2006}.
DG schemes by design lead to strongly local data access patterns to advance one element in time -- only the information from this particular element and its neighbours is needed.
Therefore the DG method can be easily parallelised to be used on modern supercomputers~\cite{burstedde_extreme-scale_2010,heinecke_petascale_2014,uphoff_extreme_2017,krenz_3d_2021}.
Combined with ADER time stepping, we achieve the same high-order convergence in time as in space and can exploit local time stepping~\cite{dumbser_arbitrary_2007}.
We extend SeisSol\footnote{\url{https://www.seissol.org}}, an open-source software for modelling seismic wave propagation and earthquake source dynamics, which relies on the ADER-DG method.
SeisSol supports elastic, viscoelastic and anisotropic materials and regularised, non-associated Drucker–Prager plastic deformation~\cite{dumbser_arbitrary_2006, dumbser_arbitrary_2007, kaser_arbitrary_2007, wollherr_off-fault_2018, wolf_optimization_2020}.
It makes use of unstructured tetrahedral meshes to easily incorporate topography and complex material discontinuities.
SeisSol also allows modelling nonlinear rupture dynamics of earthquake sources~\cite{pelties_three-dimensional_2012, pelties_verification_2014, ulrich_dynamic_2019, palgunadi_dynamic_2020}.
SeisSol is optimised for the latest CPU~\cite{uphoff_extreme_2017, krenz_3d_2021} and GPU~\cite{dorozhinskii_seissol_2021} based supercomputers.

Here, we demonstrate that by using the space-time predictor variant of the ADER time stepping scheme~\cite{gassner_explicit_2011}, which makes the ADER-DG scheme locally implicit, the stiff source term can be integrated without a strict time step restriction.
The solution procedure stays overall explicit, which means that no global system has to be assembled and solved.
As a key part of the solution procedure, a medium-sized linear system with a few thousand unknowns has to be solved for each element and time step.
As extreme-scale simulations may calculate more than $10^{13}$ element updates~\cite{krenz_3d_2021}, a highly efficient solver for these systems is needed.
We show that with a standard approach, such as the LU decomposition~\cite{golub_matrix_2013}, it is not feasible to tackle large-scale poroelastic problems due to the high demands on computational power and memory.
We exploit the structure of the system to derive an efficient back-substitution algorithm.
When we compare our algorithm to a standard LU-decomposition, we see a reduction in the number of floating-point operations by a factor of up to \num{25}, when using polynomials of degree \num{6} and we expect that the speed-up factor further increases with higher degrees.
In addition, the back-substitution procedure can be expressed as a chain of small matrix-matrix multiplications (GEMMs,~\cite{blackford_updated_2001}).
For these operations, efficient implementations exist (e.g.~\cite{heinecke_libxsmm_2016}), thus high computational efficiency can be readily achieved.

In the following, we review existing approaches to solve the governing equations of poroelasticity in \cref{sec:related-work}.
Then, in \cref{sec:equations}, we summarise the governing equations of poroelasticity.
Subsequently, we outline the spatial and temporal discretisation using the space-time variant of ADER-DG in \cref{sec:discretisation}.
In \cref{sec:stp}, we present our novel solution algorithm.
The new scheme is then compared to reference solutions in a series of verification exercises, in \cref{sec:verification}.
We demonstrate the high-order convergence of our method with canonical 3D models of planar wave propagation.
We verify the accurate treatment of wave propagation in a homogeneous full-space excited by an explosive point source, free surfaces and internal material interfaces in comparison to analytical and semi-analytical reference solutions.
Finally, we present a new 3D poroelastic layer over half-space scenario, in which we verify all implementation aspects conjunctively in comparison to a finite difference method.
In \cref{sec:performance}, we examine the performance and scalability of our implementation.
We conclude with a discussion of our results in \cref{sec:discussion}.

\section{Related work}
\label{sec:related-work}
The PDEs describing wave propagation in poroelastic media are interesting from a mathematical and computational point of view because they contain a stiff source term.
Several different approaches to solve these equations have been proposed.
When simulating wave propagation in poroelastic media, the main challenge comes from the viscous coupling between the solid and the fluid, which introduces a stiff reactive source term to the equation.
Here, we shortly summarise various semi-analytical and numerical methods and approaches for the simulation of seismic wave propagation in a poroelastic medium.

For simple models, Green's function approaches are applicable to solve the equations of motion combined with Biot's constitutive law for poroelastic media.
A Green's function approach can be considered semi--analytical: 
The solution is expressed analytically as a convolution of a Green's function with a source time function.
To actually compute the solution at a given point, numerical quadrature is typically used.
\citet{diaz_analytical_2008} solve the PDEs for the solid particle velocities of a poroelastic material filled with an inviscid fluid.
They consider a homogeneous full-space, a contact of two half-spaces with distinct material or a half-space with a free surface.
\citet{karpfinger_greens_2009} consider general moment-tensor sources (monopole, dipole and double-couple), and solve for the solid particle velocities as well as the relative fluid velocities in homogeneous full-spaces.
They take inviscid and viscous fluids in the pore space into account.
While both approaches give good results for geometrically simple test cases, they are not applicable for more complicated setups.
We are not aware of analytical Green's functions that can readily account for non-planar topography or 3D heterogeneous poroelastic materials.

\citet{carcione_computational_2010} summarise the numerical methods used for wave propagation in poroelastic media.
In contrast to the Green's function approaches, the PDEs are discretised and the system of discretised equations is then solved.
\citet{carcione_aspects_1995} introduce a splitting method in time in combination with a pseudospectral element discretisation in space to overcome stability problems.
\citet{morency_spectral-element_2008} apply the spectral element method to solve the equations of poroelasticity for 2D applications.
There, the governing equations are written down in second-order form, with the solid displacements and the relative fluid displacements as principal quantities.
After the spatial discretisation, the time stepping is done using a Newmark scheme.

The finite-difference (FD) method has been applied to wave propagation in poroelastic media since the early 70s.
One of the first studies was published by~\citet{garg_compressional_1974}. 
Since then, many authors have applied a variety of FD schemes to model seismic waves and diffusion in poroelastic media. 
A detailed overview is given by~\citet{moczo_discrete_2019} and~\citet{gregor_subcell-resolution_2021, gregor_seismic_2021}. 
The three papers introduced the staggered-grid velocity--stress--pressure FD scheme with a sub-cell resolution in poroelastic media with zero, nonzero constant and frequency-dependent resistive friction.

Also, the DG framework has been used to solve the governing equations of poroelasticity~\cite{de_la_puente_discontinuous_2008, zhang_discontinuous_2019, shukla_nodal_2019, zhan_full-anisotropic_2019}.
\citet{de_la_puente_discontinuous_2008} combine ADER time stepping with the DG method using modal basis functions
(using a deprecated version of SeisSol).
They compare a splitting approach with a space-time predictor to integrate the stiff source term.
The DG method can also be combined with implicit-explicit (IMEX) Runge-Kutta time stepping --
\citet{dudley_ward_discontinuous_2017} focus mostly on the so-called high-frequency case.
\citet{shukla_nodal_2019} use operator splitting with nodal basis functions and Runge-Kutta time stepping.
\citet{zhang_discontinuous_2019} use ADER time stepping similar to SeisSol and focus on the coupling between wave propagation in elastic and poroelastic materials.
\citet{zhan_full-anisotropic_2019} also combine the DG method with Runge Kutta time stepping. 
However, they omit the stiff source term, by only considering inviscid fluids.
Most of the approaches are presented for 2D scenarios; only a few~\cite{de_la_puente_discontinuous_2008, zhan_full-anisotropic_2019} are able to solve more realistic 3D problems.


\input{Equations.tex}
\input{Discretisation.tex}
\input{STP.tex}
\input{Benchmarks.tex}
\input{Performance.tex}
\input{discussion.tex}
\section{Conclusion}
We have derived a new efficient algorithm for solving the system of equations, which arises from the discretisation of the governing equations of wave propagation in poroelastic media using the space-time variant of ADER-DG.
The algorithm relies on a block-wise back-substitution procedure, which can be efficiently implemented by chains of matrix-matrix products.
The implementation is thoroughly validated against reference solutions.
Detailed numerical tests revealed that for viscous problems with a free surface or internal material interface, a standard mesh resolution (based on an accuracy analysis of the elastic problem) is sufficient for the solid particle velocities. 
However, if the relative fluid velocities have to be calculated at or very close to the boundary, much finer spatial resolution is necessary.
Performance and scalability experiments show that our method is suited to be used on recent supercomputers.
With our focus on high-performance and scalability, we enable large-scale seismic simulations in poroelastic materials.
Since the implementation is open-source software, we envision broad applicability of our work within the geophysics and engineering communities.

\section*{Acknowledgements}
The presented work has been funded by the European Union's Horizon 2020 Research and Innovation program (ENERXICO, grant agreement No.\ 828947).
Computing resources were provided by the Leibniz Supercomputing Centre (project no.\ pr83no on SuperMUC-NG) and KAUST Supercomputing Laboratory (project no.\ k1343 on Shaheen II).
C.U.\ and A.-A.G.\ acknowledge support by the European Union’s Horizon 2020 Research and Innovation Programme under ERC StG TEAR, no. 852992 and the German Research Foundation (DFG) (grants no. GA 2465/2-1, GA 2465/3-1).
M.G.\ and A.-A.G.\ acknowledge support by KAUST-CRG (grant no.\ ORS-2017-CRG6 3389.02).
S.W.\ and M.B.\ acknowledge support by KONWIHR (project \enquote{Optimisation of SeisSol for Large Scale Simulations of Induced Earthquakes}).
M.G, P.M.\ and D.G.\ acknowledge support by the Slovak Research and Development Agency under the contract APVV-15-0560 (project ID-EFFECTS).

\section*{Data availability}
The source code of SeisSol is available as open-source software under \url{https://github.com/SeisSol/SeisSol}.
The model descriptions, simulation outputs and reference data can be found under \url{https://doi.org/10.5281/zenodo.5236133}.

\appendix
\begin{appendices}
\input{appendix.tex}
\end{appendices}

\bibliographystyle{elsarticle-num-names}
\bibliography{main}
\end{document}

%% file: figures/test.tikzstyles
\tikzstyle{large red dot}=[fill=red, draw=black, scale=0.35, shape=circle]
\tikzstyle{large green dot}=[fill={rgb,255: red,191; green,255; blue,0}, draw=black, scale=0.35, shape=circle]


%% file: Equations.tex
\section{Governing equations}
\label{sec:equations}
In the following, we outline the governing equations for wave propagation in poroelastic materials.
The final PDE system combines \num{13} unknowns -- six stress components, three solid particle velocities, the pore pressure and three relative fluid velocities -- in the vector of unknowns: 
\begin{equation*}
q = \left( \sigma_{xx}, \sigma_{yy}, \sigma_{zz}, \sigma_{xy}, \sigma_{yz}, \sigma_{xz}, u, v, w, p, u_f, v_f, w_f \right)^T.
\end{equation*}
Following~\cite{de_la_puente_discontinuous_2008, carcione_energy_2001, carcione_wave_2015}, the governing equation for wave propagation in a poroelastic medium can be written in the matrix-vector form as: \begin{equation}
    \label{eq:poroelastic-wave}
    \derivative{q}{t} + A \derivative{q}{x} + B \derivative{q}{y} + C \derivative{q}{z} = Eq,
\end{equation}
with matrices $A, B, C, E \in \R^{13 \times 13}$.
The term $Eq$ on the right-hand side of \cref{eq:poroelastic-wave} is denoted as reactive source term or viscous dissipation term.
It accounts for the dissipation of energy due to the motion of the viscous fluid relative to the solid.
The governing equations of poroelasticity form a linear hyperbolic partial differential equation with a stiff source term.

\subsection{Constitutive behaviour}
To better understand the mechanics of this PDE system and to introduce all involved quantities and parameters, we summarise its underlying physical concepts and derivation. 
Poroelastic materials are inherently heterogeneous combining a solid and a fluid phase.
They consist of a solid matrix (also called solid frame) with pores.
The pore space is then completely filled by a fluid.
The material of the solid phase is characterised by the bulk modulus $K_S$ and the density $\rho_S$.
The porosity $\phi$ describes the volume fraction occupied by the pores.
The solid matrix, including empty pores, behaves like an elastic body, i.e. its rheology can be characterised by the two Lam\'{e} parameters $\lambda_M$ and $\mu_M$. 
The fluid phase is described by its bulk modulus $K_F$, density $\rho_F$ and the viscosity $\nu$.
We also need two parameters, which describe how the solid matrix and the fluid interact: 
A fluid particle moving from one point in the solid matrix to another cannot follow the direct path, but has to follow the path dictated by the pores. The tortuosity $T$ describes how much longer this path is compared to the direct connection.
The permeability $\kappa$ is a measure of how well fluids can be transported through the pores.
All poroelastic material parameters are summarised in \cref{tab:poroelastic_parameters}.

It is now possible to define the displacement of the solid matrix $U^M_i$ and corresponding strain tensor $\epsilon^M_{ij} = \frac{1}{2}(\partial_i U^M_j + \partial_j U^M_i)$.
In addition, we take the displacement of the fluid $U^F$ and the pore pressure $p$ into account and consider the variation of the fluid content as $\zeta := - \nabla \cdot \left(\phi(U^F - U^M)\right)$.
\begin{table}
  \centering
  \caption{Material parameters used to characterise poroelastic materials.}
  \label{tab:poroelastic_parameters}
  \pgfplotstabletypeset[
    multicolumn names=l,
    string type,
    col sep=&,row sep=\\,
    header=false,
    every head row/.style={before row=\toprule,after row=\midrule},
    every last row/.style={after row=\bottomrule},
    display columns/0/.style={column name=Parameter, column type={l}},
    display columns/1/.style={column name=Symbol, column type={l}},
    display columns/2/.style={column name=Unit, column type={l}},
   ]{
    Solid Bulk modulus                & {$K_S$}       & \si{\Pa}\\
    Solid density                     & {$\rho_S$}    & \si{\kg \per \meter \tothe{3}}\\
    Matrix $1^{st}$ Lam\'e parameter  & {$\lambda_M$} & \si{\Pa}\\
    Matrix $2^{nd}$ Lam\'e parameter  & {$\mu_M$}     & \si{\Pa}\\
    Matrix permeability               & {$\kappa$}    & \si{\meter \tothe{2}}\\
    Matrix porosity                   & {$\phi$}      & \\
    Matrix tortuosity                 & {$T$}         & \\
    Fluid bulk modulus                & {$K_F$}       & \si{\Pa}\\
    Fluid density                     & {$\rho_F$}    & \si{\kg \per \meter \tothe{3}}\\
    Fluid viscosity                   & {$\nu$}       & \si{\Pa \second} \\
  }
\end{table}
To derive a constitutive law of the poroelastic material, we consider a homogenised material, i.e.\ we neglect spatial scales smaller than a pore's diameter.
For example, the effective density of the poroelastic material can be computed as $\rho = \phi \cdot \rho_F + (1-\phi) \cdot \rho_S$.
The total stress $\sigma_{ij} = \sigma^M_{ij} - \phi p \delta_{ij}$ can be observed.
We compute the bulk modulus of the solid matrix $K_M = \lambda_M+ \frac{2}{3} \mu_M$.
Finally, we consider the solid-fluid coupling modulus
\begin{equation*}
    M = \frac{K_S}{(1-K_M/K_S) - \phi (1-K_S / K_F)},
\end{equation*}
and the effective stress component
\begin{equation*}
    \alpha = 1 - K_M / K_S.
\end{equation*}
With all quantities defined, we can write down the constitutive law, relating $\sigma$ and $p$ to $\epsilon^M$ and $\zeta$~\cite[sec.~2]{carcione_energy_2001}.
\begin{equation*}
    \begin{pmatrix}
        \sigma_{xx} \\ \sigma_{yy} \\ \sigma_{zz} \\ \sigma_{yz} \\ \sigma_{xz}  \\ \sigma_{xy} \\ -p
    \end{pmatrix}
    = 
    \begin{pmatrix}
        \lambda_M + 2\mu_M + M\alpha^2 & \lambda + M\alpha^2 & \lambda + M\alpha^2 & 0 & 0 & 0 & M \alpha \\
        \lambda + M\alpha^2 & \lambda_M + 2\mu_M + M\alpha^2 & \lambda + M\alpha^2 & 0 & 0 & 0 & M \alpha \\
        \lambda + M\alpha^2 & \lambda + M\alpha^2 & \lambda_M + 2\mu_M + M\alpha^2 & 0 & 0 & 0 & M \alpha \\
        0 & 0 & 0 & \mu & 0 & 0 & 0 \\
        0 & 0 & 0 & 0 & \mu & 0 & 0 \\
        0 & 0 & 0 & 0 & 0 & \mu & 0 \\
        M \alpha & M \alpha & M \alpha & 0 & 0 & 0 & M
    \end{pmatrix}
    \begin{pmatrix}
        \epsilon^M_{xx} \\ 
        \epsilon^M_{yy} \\ 
        \epsilon^M_{zz} \\ 
        \epsilon^M_{yz} \\ 
        \epsilon^M_{xz} \\ 
        \epsilon^M_{xy} \\
         -\zeta
    \end{pmatrix}.
\end{equation*}
Note that by our sign convention, pressure is positive in compression while normal stresses are negative in compression.
\subsection{Equations of motion}
We now consider the time-dependent problem, in terms of the solid particle velocities $v_i = \derivative{U^M_i}{t}$ and the relative fluid velocities $v_i^F = \phi \derivative{(U^F_i - U^M_i)}{t}$.
The equations of motion can be combined with Darcy's law to obtain a system of PDEs~\cite[sec.~3]{carcione_energy_2001}:
\begin{equation}
    \label{eq:biot-dynamic}
    \begin{aligned}
    \sum_{j=1}^3 \derivative{\sigma_{ij}}{x_j} &= \rho \derivative{v_i}{t} + \rho_F \derivative{v_i^F}{t}\\
    -\derivative{p}{x_i} &= \rho \derivative{v_i}{t} + \frac{ \rho^F T}{\phi} \derivative{v^F_i}{t} + \frac{\nu}{\kappa} v_i^F.\\
    \end{aligned}
\end{equation}
We note that \cref{eq:biot-dynamic} is only valid for frequencies lower than Biot's characteristic frequency $f_c = \frac{1}{2 \pi} \frac{\nu \phi}{T\kappa\rho_F}$. 
In this so-called low-frequency regime, the flow of the fluid in pore space can be approximated as a laminar flow. 
For higher frequencies, the fluid flow becomes turbulent and a more complex form of Darcy's law has to be considered. 
Biot's frequency depends mainly on the porosity~$\phi$ and the permeability~$\kappa$.
Within geo-reservoirs, permeability typically ranges from\footnote{The unit Darcy (\SI{1}{\Darcy} = \SI{9.86923e-13}{\meter\tothe{2}}) is a unit for the permeability of a porous medium.} \SI{1e-4}{\milli\Darcy} to \SI{1e2}{\milli\Darcy} and porosity varies between \SI{0}{\percent} and \SI{20}{\percent} (e.g.~\cite[Fig.~14.1]{franchi_petroleum_2006}), leading to Biot's frequency ranging from tens of \si{\Hertz} to hundreds of \si{\kilo\Hertz}. 
We aim for simulations of seismic waves up to \SI{10}{\Hertz}, which means that considering the low-frequency regime is sufficient (see also, e.g.~\cite[Fig~2.1]{saxena_handbook_2018}). 
If all pores are filled with an inviscid fluid, Biot's frequency is formally zero, implying the high-frequency regime. 
However, for the inviscid case, the equations for low- and high-frequency regimes are identical~\cite{de_la_puente_discontinuous_2008}.

We follow~\cite{de_la_puente_discontinuous_2008} in combining the constitutive behaviour and the equations of motions to derive the governing equations as a hyperbolic system in first-order form in the form of \cref{eq:poroelastic-wave}.
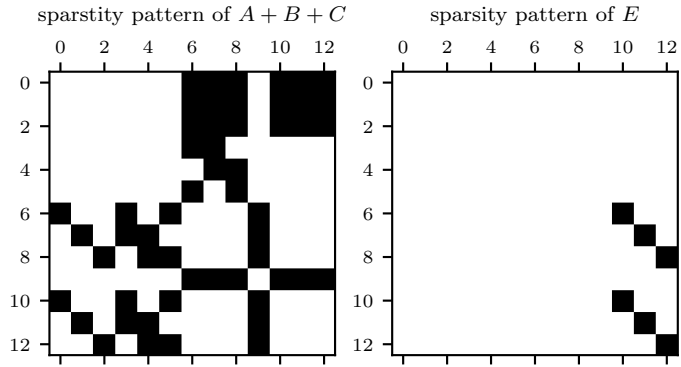
\begin{figure}
    \center{
        \input{figures/sparsity/sparsity_jacobians.pgf} 
    }
    \caption{Sparsity patterns of the matrices $A+B+C$ and $E$ of \cref{eq:poroelastic-wave}.}
    \label{fig:jacobians}
\end{figure}
We show the sparsity patterns of the matrices $A+B+C$ and $E$ in \cref{fig:jacobians}.
A detailed description of these matrices can be found in~\cite[pp.~114-115]{de_la_puente_seismic_2008}, where also the anisotropic case is discussed.
Here we restrict ourselves to the isotropic case.
At this point, we want to remark that $E$ is an upper triangular matrix i.e. $E_{ij} = 0 \quad \forall i > j$, which will be important in \cref{sec:stp}.

We can combine \cref{eq:biot-dynamic} with initial conditions.
Typically, in the initial state, the system is at rest, and seismic waves are excited by an external source term, which additionally enters \cref{eq:biot-dynamic} on the right-hand side.
See for example~\cite{karpfinger_greens_2009}, for an overview of different source types and their respective radiation patterns in a homogeneous full-space.

%% file: figures/sparsity/sparsity_jacobians.pgf
\begingroup%
\makeatletter%
\begin{pgfpicture}%
\pgfpathrectangle{\pgfpointorigin}{\pgfqpoint{3.662948in}{2.072292in}}%
\pgfusepath{use as bounding box, clip}%
\begin{pgfscope}%
\pgfsetbuttcap%
\pgfsetmiterjoin%
\pgfsetlinewidth{0.000000pt}%
\definecolor{currentstroke}{rgb}{1.000000,1.000000,1.000000}%
\pgfsetstrokecolor{currentstroke}%
\pgfsetstrokeopacity{0.000000}%
\pgfsetdash{}{0pt}%
\pgfpathmoveto{\pgfqpoint{0.000000in}{0.000000in}}%
\pgfpathlineto{\pgfqpoint{3.662948in}{0.000000in}}%
\pgfpathlineto{\pgfqpoint{3.662948in}{2.072292in}}%
\pgfpathlineto{\pgfqpoint{0.000000in}{2.072292in}}%
\pgfpathclose%
\pgfusepath{}%
\end{pgfscope}%
\begin{pgfscope}%
\pgfsetbuttcap%
\pgfsetmiterjoin%
\definecolor{currentfill}{rgb}{1.000000,1.000000,1.000000}%
\pgfsetfillcolor{currentfill}%
\pgfsetlinewidth{0.000000pt}%
\definecolor{currentstroke}{rgb}{0.000000,0.000000,0.000000}%
\pgfsetstrokecolor{currentstroke}%
\pgfsetstrokeopacity{0.000000}%
\pgfsetdash{}{0pt}%
\pgfpathmoveto{\pgfqpoint{0.307948in}{0.148611in}}%
\pgfpathlineto{\pgfqpoint{1.787494in}{0.148611in}}%
\pgfpathlineto{\pgfqpoint{1.787494in}{1.628157in}}%
\pgfpathlineto{\pgfqpoint{0.307948in}{1.628157in}}%
\pgfpathclose%
\pgfusepath{fill}%
\end{pgfscope}%
\begin{pgfscope}%
\pgfpathrectangle{\pgfqpoint{0.307948in}{0.148611in}}{\pgfqpoint{1.479545in}{1.479545in}}%
\pgfusepath{clip}%
\pgfsys@transformshift{0.307948in}{0.148611in}%
\pgftext[left,bottom]{\includegraphics[interpolate=true,width=1.480000in,height=1.480000in]{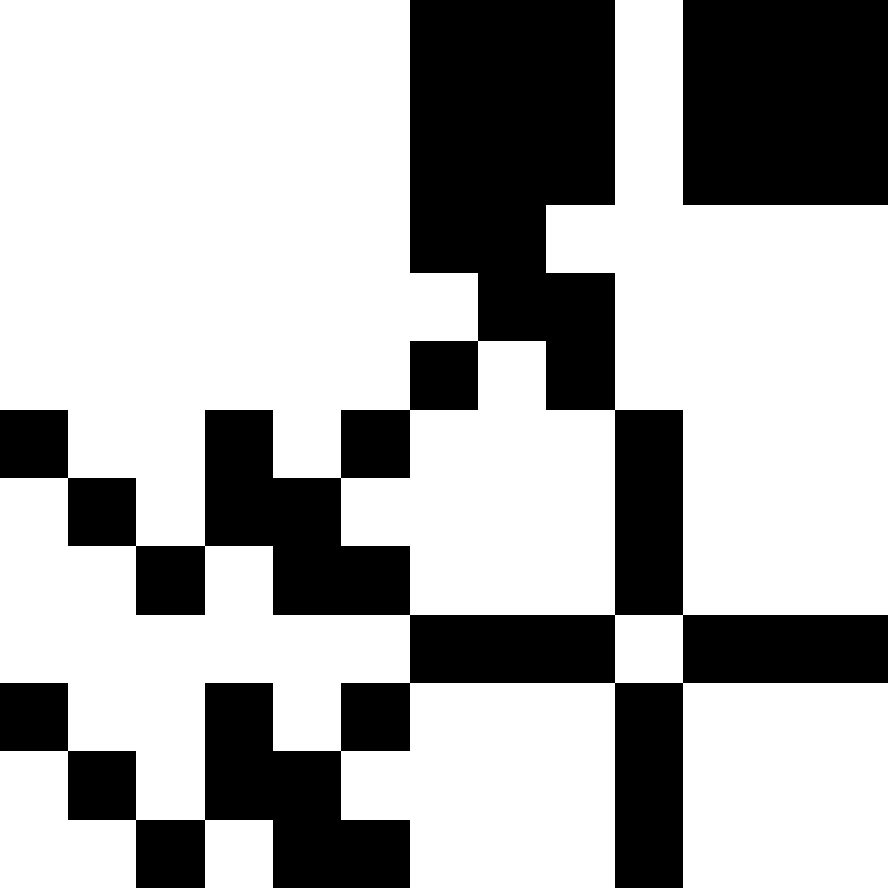}}%
\end{pgfscope}%
\begin{pgfscope}%
\pgfsetbuttcap%
\pgfsetroundjoin%
\definecolor{currentfill}{rgb}{0.000000,0.000000,0.000000}%
\pgfsetfillcolor{currentfill}%
\pgfsetlinewidth{0.803000pt}%
\definecolor{currentstroke}{rgb}{0.000000,0.000000,0.000000}%
\pgfsetstrokecolor{currentstroke}%
\pgfsetdash{}{0pt}%
\pgfsys@defobject{currentmarker}{\pgfqpoint{0.000000in}{-0.048611in}}{\pgfqpoint{0.000000in}{0.000000in}}{%
\pgfpathmoveto{\pgfqpoint{0.000000in}{0.000000in}}%
\pgfpathlineto{\pgfqpoint{0.000000in}{-0.048611in}}%
\pgfusepath{stroke,fill}%
}%
\begin{pgfscope}%
\pgfsys@transformshift{0.364854in}{0.148611in}%
\pgfsys@useobject{currentmarker}{}%
\end{pgfscope}%
\end{pgfscope}%
\begin{pgfscope}%
\pgfsetbuttcap%
\pgfsetroundjoin%
\definecolor{currentfill}{rgb}{0.000000,0.000000,0.000000}%
\pgfsetfillcolor{currentfill}%
\pgfsetlinewidth{0.803000pt}%
\definecolor{currentstroke}{rgb}{0.000000,0.000000,0.000000}%
\pgfsetstrokecolor{currentstroke}%
\pgfsetdash{}{0pt}%
\pgfsys@defobject{currentmarker}{\pgfqpoint{0.000000in}{0.000000in}}{\pgfqpoint{0.000000in}{0.048611in}}{%
\pgfpathmoveto{\pgfqpoint{0.000000in}{0.000000in}}%
\pgfpathlineto{\pgfqpoint{0.000000in}{0.048611in}}%
\pgfusepath{stroke,fill}%
}%
\begin{pgfscope}%
\pgfsys@transformshift{0.364854in}{1.628157in}%
\pgfsys@useobject{currentmarker}{}%
\end{pgfscope}%
\end{pgfscope}%
\begin{pgfscope}%
\definecolor{textcolor}{rgb}{0.000000,0.000000,0.000000}%
\pgfsetstrokecolor{textcolor}%
\pgfsetfillcolor{textcolor}%
\pgftext[x=0.364854in,y=1.725379in,,bottom]{\color{textcolor}\rmfamily\fontsize{7.000000}{8.400000}\selectfont \(\displaystyle {0}\)}%
\end{pgfscope}%
\begin{pgfscope}%
\pgfsetbuttcap%
\pgfsetroundjoin%
\definecolor{currentfill}{rgb}{0.000000,0.000000,0.000000}%
\pgfsetfillcolor{currentfill}%
\pgfsetlinewidth{0.803000pt}%
\definecolor{currentstroke}{rgb}{0.000000,0.000000,0.000000}%
\pgfsetstrokecolor{currentstroke}%
\pgfsetdash{}{0pt}%
\pgfsys@defobject{currentmarker}{\pgfqpoint{0.000000in}{-0.048611in}}{\pgfqpoint{0.000000in}{0.000000in}}{%
\pgfpathmoveto{\pgfqpoint{0.000000in}{0.000000in}}%
\pgfpathlineto{\pgfqpoint{0.000000in}{-0.048611in}}%
\pgfusepath{stroke,fill}%
}%
\begin{pgfscope}%
\pgfsys@transformshift{0.592476in}{0.148611in}%
\pgfsys@useobject{currentmarker}{}%
\end{pgfscope}%
\end{pgfscope}%
\begin{pgfscope}%
\pgfsetbuttcap%
\pgfsetroundjoin%
\definecolor{currentfill}{rgb}{0.000000,0.000000,0.000000}%
\pgfsetfillcolor{currentfill}%
\pgfsetlinewidth{0.803000pt}%
\definecolor{currentstroke}{rgb}{0.000000,0.000000,0.000000}%
\pgfsetstrokecolor{currentstroke}%
\pgfsetdash{}{0pt}%
\pgfsys@defobject{currentmarker}{\pgfqpoint{0.000000in}{0.000000in}}{\pgfqpoint{0.000000in}{0.048611in}}{%
\pgfpathmoveto{\pgfqpoint{0.000000in}{0.000000in}}%
\pgfpathlineto{\pgfqpoint{0.000000in}{0.048611in}}%
\pgfusepath{stroke,fill}%
}%
\begin{pgfscope}%
\pgfsys@transformshift{0.592476in}{1.628157in}%
\pgfsys@useobject{currentmarker}{}%
\end{pgfscope}%
\end{pgfscope}%
\begin{pgfscope}%
\definecolor{textcolor}{rgb}{0.000000,0.000000,0.000000}%
\pgfsetstrokecolor{textcolor}%
\pgfsetfillcolor{textcolor}%
\pgftext[x=0.592476in,y=1.725379in,,bottom]{\color{textcolor}\rmfamily\fontsize{7.000000}{8.400000}\selectfont \(\displaystyle {2}\)}%
\end{pgfscope}%
\begin{pgfscope}%
\pgfsetbuttcap%
\pgfsetroundjoin%
\definecolor{currentfill}{rgb}{0.000000,0.000000,0.000000}%
\pgfsetfillcolor{currentfill}%
\pgfsetlinewidth{0.803000pt}%
\definecolor{currentstroke}{rgb}{0.000000,0.000000,0.000000}%
\pgfsetstrokecolor{currentstroke}%
\pgfsetdash{}{0pt}%
\pgfsys@defobject{currentmarker}{\pgfqpoint{0.000000in}{-0.048611in}}{\pgfqpoint{0.000000in}{0.000000in}}{%
\pgfpathmoveto{\pgfqpoint{0.000000in}{0.000000in}}%
\pgfpathlineto{\pgfqpoint{0.000000in}{-0.048611in}}%
\pgfusepath{stroke,fill}%
}%
\begin{pgfscope}%
\pgfsys@transformshift{0.820098in}{0.148611in}%
\pgfsys@useobject{currentmarker}{}%
\end{pgfscope}%
\end{pgfscope}%
\begin{pgfscope}%
\pgfsetbuttcap%
\pgfsetroundjoin%
\definecolor{currentfill}{rgb}{0.000000,0.000000,0.000000}%
\pgfsetfillcolor{currentfill}%
\pgfsetlinewidth{0.803000pt}%
\definecolor{currentstroke}{rgb}{0.000000,0.000000,0.000000}%
\pgfsetstrokecolor{currentstroke}%
\pgfsetdash{}{0pt}%
\pgfsys@defobject{currentmarker}{\pgfqpoint{0.000000in}{0.000000in}}{\pgfqpoint{0.000000in}{0.048611in}}{%
\pgfpathmoveto{\pgfqpoint{0.000000in}{0.000000in}}%
\pgfpathlineto{\pgfqpoint{0.000000in}{0.048611in}}%
\pgfusepath{stroke,fill}%
}%
\begin{pgfscope}%
\pgfsys@transformshift{0.820098in}{1.628157in}%
\pgfsys@useobject{currentmarker}{}%
\end{pgfscope}%
\end{pgfscope}%
\begin{pgfscope}%
\definecolor{textcolor}{rgb}{0.000000,0.000000,0.000000}%
\pgfsetstrokecolor{textcolor}%
\pgfsetfillcolor{textcolor}%
\pgftext[x=0.820098in,y=1.725379in,,bottom]{\color{textcolor}\rmfamily\fontsize{7.000000}{8.400000}\selectfont \(\displaystyle {4}\)}%
\end{pgfscope}%
\begin{pgfscope}%
\pgfsetbuttcap%
\pgfsetroundjoin%
\definecolor{currentfill}{rgb}{0.000000,0.000000,0.000000}%
\pgfsetfillcolor{currentfill}%
\pgfsetlinewidth{0.803000pt}%
\definecolor{currentstroke}{rgb}{0.000000,0.000000,0.000000}%
\pgfsetstrokecolor{currentstroke}%
\pgfsetdash{}{0pt}%
\pgfsys@defobject{currentmarker}{\pgfqpoint{0.000000in}{-0.048611in}}{\pgfqpoint{0.000000in}{0.000000in}}{%
\pgfpathmoveto{\pgfqpoint{0.000000in}{0.000000in}}%
\pgfpathlineto{\pgfqpoint{0.000000in}{-0.048611in}}%
\pgfusepath{stroke,fill}%
}%
\begin{pgfscope}%
\pgfsys@transformshift{1.047721in}{0.148611in}%
\pgfsys@useobject{currentmarker}{}%
\end{pgfscope}%
\end{pgfscope}%
\begin{pgfscope}%
\pgfsetbuttcap%
\pgfsetroundjoin%
\definecolor{currentfill}{rgb}{0.000000,0.000000,0.000000}%
\pgfsetfillcolor{currentfill}%
\pgfsetlinewidth{0.803000pt}%
\definecolor{currentstroke}{rgb}{0.000000,0.000000,0.000000}%
\pgfsetstrokecolor{currentstroke}%
\pgfsetdash{}{0pt}%
\pgfsys@defobject{currentmarker}{\pgfqpoint{0.000000in}{0.000000in}}{\pgfqpoint{0.000000in}{0.048611in}}{%
\pgfpathmoveto{\pgfqpoint{0.000000in}{0.000000in}}%
\pgfpathlineto{\pgfqpoint{0.000000in}{0.048611in}}%
\pgfusepath{stroke,fill}%
}%
\begin{pgfscope}%
\pgfsys@transformshift{1.047721in}{1.628157in}%
\pgfsys@useobject{currentmarker}{}%
\end{pgfscope}%
\end{pgfscope}%
\begin{pgfscope}%
\definecolor{textcolor}{rgb}{0.000000,0.000000,0.000000}%
\pgfsetstrokecolor{textcolor}%
\pgfsetfillcolor{textcolor}%
\pgftext[x=1.047721in,y=1.725379in,,bottom]{\color{textcolor}\rmfamily\fontsize{7.000000}{8.400000}\selectfont \(\displaystyle {6}\)}%
\end{pgfscope}%
\begin{pgfscope}%
\pgfsetbuttcap%
\pgfsetroundjoin%
\definecolor{currentfill}{rgb}{0.000000,0.000000,0.000000}%
\pgfsetfillcolor{currentfill}%
\pgfsetlinewidth{0.803000pt}%
\definecolor{currentstroke}{rgb}{0.000000,0.000000,0.000000}%
\pgfsetstrokecolor{currentstroke}%
\pgfsetdash{}{0pt}%
\pgfsys@defobject{currentmarker}{\pgfqpoint{0.000000in}{-0.048611in}}{\pgfqpoint{0.000000in}{0.000000in}}{%
\pgfpathmoveto{\pgfqpoint{0.000000in}{0.000000in}}%
\pgfpathlineto{\pgfqpoint{0.000000in}{-0.048611in}}%
\pgfusepath{stroke,fill}%
}%
\begin{pgfscope}%
\pgfsys@transformshift{1.275343in}{0.148611in}%
\pgfsys@useobject{currentmarker}{}%
\end{pgfscope}%
\end{pgfscope}%
\begin{pgfscope}%
\pgfsetbuttcap%
\pgfsetroundjoin%
\definecolor{currentfill}{rgb}{0.000000,0.000000,0.000000}%
\pgfsetfillcolor{currentfill}%
\pgfsetlinewidth{0.803000pt}%
\definecolor{currentstroke}{rgb}{0.000000,0.000000,0.000000}%
\pgfsetstrokecolor{currentstroke}%
\pgfsetdash{}{0pt}%
\pgfsys@defobject{currentmarker}{\pgfqpoint{0.000000in}{0.000000in}}{\pgfqpoint{0.000000in}{0.048611in}}{%
\pgfpathmoveto{\pgfqpoint{0.000000in}{0.000000in}}%
\pgfpathlineto{\pgfqpoint{0.000000in}{0.048611in}}%
\pgfusepath{stroke,fill}%
}%
\begin{pgfscope}%
\pgfsys@transformshift{1.275343in}{1.628157in}%
\pgfsys@useobject{currentmarker}{}%
\end{pgfscope}%
\end{pgfscope}%
\begin{pgfscope}%
\definecolor{textcolor}{rgb}{0.000000,0.000000,0.000000}%
\pgfsetstrokecolor{textcolor}%
\pgfsetfillcolor{textcolor}%
\pgftext[x=1.275343in,y=1.725379in,,bottom]{\color{textcolor}\rmfamily\fontsize{7.000000}{8.400000}\selectfont \(\displaystyle {8}\)}%
\end{pgfscope}%
\begin{pgfscope}%
\pgfsetbuttcap%
\pgfsetroundjoin%
\definecolor{currentfill}{rgb}{0.000000,0.000000,0.000000}%
\pgfsetfillcolor{currentfill}%
\pgfsetlinewidth{0.803000pt}%
\definecolor{currentstroke}{rgb}{0.000000,0.000000,0.000000}%
\pgfsetstrokecolor{currentstroke}%
\pgfsetdash{}{0pt}%
\pgfsys@defobject{currentmarker}{\pgfqpoint{0.000000in}{-0.048611in}}{\pgfqpoint{0.000000in}{0.000000in}}{%
\pgfpathmoveto{\pgfqpoint{0.000000in}{0.000000in}}%
\pgfpathlineto{\pgfqpoint{0.000000in}{-0.048611in}}%
\pgfusepath{stroke,fill}%
}%
\begin{pgfscope}%
\pgfsys@transformshift{1.502966in}{0.148611in}%
\pgfsys@useobject{currentmarker}{}%
\end{pgfscope}%
\end{pgfscope}%
\begin{pgfscope}%
\pgfsetbuttcap%
\pgfsetroundjoin%
\definecolor{currentfill}{rgb}{0.000000,0.000000,0.000000}%
\pgfsetfillcolor{currentfill}%
\pgfsetlinewidth{0.803000pt}%
\definecolor{currentstroke}{rgb}{0.000000,0.000000,0.000000}%
\pgfsetstrokecolor{currentstroke}%
\pgfsetdash{}{0pt}%
\pgfsys@defobject{currentmarker}{\pgfqpoint{0.000000in}{0.000000in}}{\pgfqpoint{0.000000in}{0.048611in}}{%
\pgfpathmoveto{\pgfqpoint{0.000000in}{0.000000in}}%
\pgfpathlineto{\pgfqpoint{0.000000in}{0.048611in}}%
\pgfusepath{stroke,fill}%
}%
\begin{pgfscope}%
\pgfsys@transformshift{1.502966in}{1.628157in}%
\pgfsys@useobject{currentmarker}{}%
\end{pgfscope}%
\end{pgfscope}%
\begin{pgfscope}%
\definecolor{textcolor}{rgb}{0.000000,0.000000,0.000000}%
\pgfsetstrokecolor{textcolor}%
\pgfsetfillcolor{textcolor}%
\pgftext[x=1.502966in,y=1.725379in,,bottom]{\color{textcolor}\rmfamily\fontsize{7.000000}{8.400000}\selectfont \(\displaystyle {10}\)}%
\end{pgfscope}%
\begin{pgfscope}%
\pgfsetbuttcap%
\pgfsetroundjoin%
\definecolor{currentfill}{rgb}{0.000000,0.000000,0.000000}%
\pgfsetfillcolor{currentfill}%
\pgfsetlinewidth{0.803000pt}%
\definecolor{currentstroke}{rgb}{0.000000,0.000000,0.000000}%
\pgfsetstrokecolor{currentstroke}%
\pgfsetdash{}{0pt}%
\pgfsys@defobject{currentmarker}{\pgfqpoint{0.000000in}{-0.048611in}}{\pgfqpoint{0.000000in}{0.000000in}}{%
\pgfpathmoveto{\pgfqpoint{0.000000in}{0.000000in}}%
\pgfpathlineto{\pgfqpoint{0.000000in}{-0.048611in}}%
\pgfusepath{stroke,fill}%
}%
\begin{pgfscope}%
\pgfsys@transformshift{1.730588in}{0.148611in}%
\pgfsys@useobject{currentmarker}{}%
\end{pgfscope}%
\end{pgfscope}%
\begin{pgfscope}%
\pgfsetbuttcap%
\pgfsetroundjoin%
\definecolor{currentfill}{rgb}{0.000000,0.000000,0.000000}%
\pgfsetfillcolor{currentfill}%
\pgfsetlinewidth{0.803000pt}%
\definecolor{currentstroke}{rgb}{0.000000,0.000000,0.000000}%
\pgfsetstrokecolor{currentstroke}%
\pgfsetdash{}{0pt}%
\pgfsys@defobject{currentmarker}{\pgfqpoint{0.000000in}{0.000000in}}{\pgfqpoint{0.000000in}{0.048611in}}{%
\pgfpathmoveto{\pgfqpoint{0.000000in}{0.000000in}}%
\pgfpathlineto{\pgfqpoint{0.000000in}{0.048611in}}%
\pgfusepath{stroke,fill}%
}%
\begin{pgfscope}%
\pgfsys@transformshift{1.730588in}{1.628157in}%
\pgfsys@useobject{currentmarker}{}%
\end{pgfscope}%
\end{pgfscope}%
\begin{pgfscope}%
\definecolor{textcolor}{rgb}{0.000000,0.000000,0.000000}%
\pgfsetstrokecolor{textcolor}%
\pgfsetfillcolor{textcolor}%
\pgftext[x=1.730588in,y=1.725379in,,bottom]{\color{textcolor}\rmfamily\fontsize{7.000000}{8.400000}\selectfont \(\displaystyle {12}\)}%
\end{pgfscope}%
\begin{pgfscope}%
\pgfsetbuttcap%
\pgfsetroundjoin%
\definecolor{currentfill}{rgb}{0.000000,0.000000,0.000000}%
\pgfsetfillcolor{currentfill}%
\pgfsetlinewidth{0.803000pt}%
\definecolor{currentstroke}{rgb}{0.000000,0.000000,0.000000}%
\pgfsetstrokecolor{currentstroke}%
\pgfsetdash{}{0pt}%
\pgfsys@defobject{currentmarker}{\pgfqpoint{-0.048611in}{0.000000in}}{\pgfqpoint{-0.000000in}{0.000000in}}{%
\pgfpathmoveto{\pgfqpoint{-0.000000in}{0.000000in}}%
\pgfpathlineto{\pgfqpoint{-0.048611in}{0.000000in}}%
\pgfusepath{stroke,fill}%
}%
\begin{pgfscope}%
\pgfsys@transformshift{0.307948in}{1.571251in}%
\pgfsys@useobject{currentmarker}{}%
\end{pgfscope}%
\end{pgfscope}%
\begin{pgfscope}%
\definecolor{textcolor}{rgb}{0.000000,0.000000,0.000000}%
\pgfsetstrokecolor{textcolor}%
\pgfsetfillcolor{textcolor}%
\pgftext[x=0.155363in, y=1.537493in, left, base]{\color{textcolor}\rmfamily\fontsize{7.000000}{8.400000}\selectfont \(\displaystyle {0}\)}%
\end{pgfscope}%
\begin{pgfscope}%
\pgfsetbuttcap%
\pgfsetroundjoin%
\definecolor{currentfill}{rgb}{0.000000,0.000000,0.000000}%
\pgfsetfillcolor{currentfill}%
\pgfsetlinewidth{0.803000pt}%
\definecolor{currentstroke}{rgb}{0.000000,0.000000,0.000000}%
\pgfsetstrokecolor{currentstroke}%
\pgfsetdash{}{0pt}%
\pgfsys@defobject{currentmarker}{\pgfqpoint{-0.048611in}{0.000000in}}{\pgfqpoint{-0.000000in}{0.000000in}}{%
\pgfpathmoveto{\pgfqpoint{-0.000000in}{0.000000in}}%
\pgfpathlineto{\pgfqpoint{-0.048611in}{0.000000in}}%
\pgfusepath{stroke,fill}%
}%
\begin{pgfscope}%
\pgfsys@transformshift{0.307948in}{1.343629in}%
\pgfsys@useobject{currentmarker}{}%
\end{pgfscope}%
\end{pgfscope}%
\begin{pgfscope}%
\definecolor{textcolor}{rgb}{0.000000,0.000000,0.000000}%
\pgfsetstrokecolor{textcolor}%
\pgfsetfillcolor{textcolor}%
\pgftext[x=0.155363in, y=1.309871in, left, base]{\color{textcolor}\rmfamily\fontsize{7.000000}{8.400000}\selectfont \(\displaystyle {2}\)}%
\end{pgfscope}%
\begin{pgfscope}%
\pgfsetbuttcap%
\pgfsetroundjoin%
\definecolor{currentfill}{rgb}{0.000000,0.000000,0.000000}%
\pgfsetfillcolor{currentfill}%
\pgfsetlinewidth{0.803000pt}%
\definecolor{currentstroke}{rgb}{0.000000,0.000000,0.000000}%
\pgfsetstrokecolor{currentstroke}%
\pgfsetdash{}{0pt}%
\pgfsys@defobject{currentmarker}{\pgfqpoint{-0.048611in}{0.000000in}}{\pgfqpoint{-0.000000in}{0.000000in}}{%
\pgfpathmoveto{\pgfqpoint{-0.000000in}{0.000000in}}%
\pgfpathlineto{\pgfqpoint{-0.048611in}{0.000000in}}%
\pgfusepath{stroke,fill}%
}%
\begin{pgfscope}%
\pgfsys@transformshift{0.307948in}{1.116006in}%
\pgfsys@useobject{currentmarker}{}%
\end{pgfscope}%
\end{pgfscope}%
\begin{pgfscope}%
\definecolor{textcolor}{rgb}{0.000000,0.000000,0.000000}%
\pgfsetstrokecolor{textcolor}%
\pgfsetfillcolor{textcolor}%
\pgftext[x=0.155363in, y=1.082249in, left, base]{\color{textcolor}\rmfamily\fontsize{7.000000}{8.400000}\selectfont \(\displaystyle {4}\)}%
\end{pgfscope}%
\begin{pgfscope}%
\pgfsetbuttcap%
\pgfsetroundjoin%
\definecolor{currentfill}{rgb}{0.000000,0.000000,0.000000}%
\pgfsetfillcolor{currentfill}%
\pgfsetlinewidth{0.803000pt}%
\definecolor{currentstroke}{rgb}{0.000000,0.000000,0.000000}%
\pgfsetstrokecolor{currentstroke}%
\pgfsetdash{}{0pt}%
\pgfsys@defobject{currentmarker}{\pgfqpoint{-0.048611in}{0.000000in}}{\pgfqpoint{-0.000000in}{0.000000in}}{%
\pgfpathmoveto{\pgfqpoint{-0.000000in}{0.000000in}}%
\pgfpathlineto{\pgfqpoint{-0.048611in}{0.000000in}}%
\pgfusepath{stroke,fill}%
}%
\begin{pgfscope}%
\pgfsys@transformshift{0.307948in}{0.888384in}%
\pgfsys@useobject{currentmarker}{}%
\end{pgfscope}%
\end{pgfscope}%
\begin{pgfscope}%
\definecolor{textcolor}{rgb}{0.000000,0.000000,0.000000}%
\pgfsetstrokecolor{textcolor}%
\pgfsetfillcolor{textcolor}%
\pgftext[x=0.155363in, y=0.854626in, left, base]{\color{textcolor}\rmfamily\fontsize{7.000000}{8.400000}\selectfont \(\displaystyle {6}\)}%
\end{pgfscope}%
\begin{pgfscope}%
\pgfsetbuttcap%
\pgfsetroundjoin%
\definecolor{currentfill}{rgb}{0.000000,0.000000,0.000000}%
\pgfsetfillcolor{currentfill}%
\pgfsetlinewidth{0.803000pt}%
\definecolor{currentstroke}{rgb}{0.000000,0.000000,0.000000}%
\pgfsetstrokecolor{currentstroke}%
\pgfsetdash{}{0pt}%
\pgfsys@defobject{currentmarker}{\pgfqpoint{-0.048611in}{0.000000in}}{\pgfqpoint{-0.000000in}{0.000000in}}{%
\pgfpathmoveto{\pgfqpoint{-0.000000in}{0.000000in}}%
\pgfpathlineto{\pgfqpoint{-0.048611in}{0.000000in}}%
\pgfusepath{stroke,fill}%
}%
\begin{pgfscope}%
\pgfsys@transformshift{0.307948in}{0.660761in}%
\pgfsys@useobject{currentmarker}{}%
\end{pgfscope}%
\end{pgfscope}%
\begin{pgfscope}%
\definecolor{textcolor}{rgb}{0.000000,0.000000,0.000000}%
\pgfsetstrokecolor{textcolor}%
\pgfsetfillcolor{textcolor}%
\pgftext[x=0.155363in, y=0.627004in, left, base]{\color{textcolor}\rmfamily\fontsize{7.000000}{8.400000}\selectfont \(\displaystyle {8}\)}%
\end{pgfscope}%
\begin{pgfscope}%
\pgfsetbuttcap%
\pgfsetroundjoin%
\definecolor{currentfill}{rgb}{0.000000,0.000000,0.000000}%
\pgfsetfillcolor{currentfill}%
\pgfsetlinewidth{0.803000pt}%
\definecolor{currentstroke}{rgb}{0.000000,0.000000,0.000000}%
\pgfsetstrokecolor{currentstroke}%
\pgfsetdash{}{0pt}%
\pgfsys@defobject{currentmarker}{\pgfqpoint{-0.048611in}{0.000000in}}{\pgfqpoint{-0.000000in}{0.000000in}}{%
\pgfpathmoveto{\pgfqpoint{-0.000000in}{0.000000in}}%
\pgfpathlineto{\pgfqpoint{-0.048611in}{0.000000in}}%
\pgfusepath{stroke,fill}%
}%
\begin{pgfscope}%
\pgfsys@transformshift{0.307948in}{0.433139in}%
\pgfsys@useobject{currentmarker}{}%
\end{pgfscope}%
\end{pgfscope}%
\begin{pgfscope}%
\definecolor{textcolor}{rgb}{0.000000,0.000000,0.000000}%
\pgfsetstrokecolor{textcolor}%
\pgfsetfillcolor{textcolor}%
\pgftext[x=0.100000in, y=0.399381in, left, base]{\color{textcolor}\rmfamily\fontsize{7.000000}{8.400000}\selectfont \(\displaystyle {10}\)}%
\end{pgfscope}%
\begin{pgfscope}%
\pgfsetbuttcap%
\pgfsetroundjoin%
\definecolor{currentfill}{rgb}{0.000000,0.000000,0.000000}%
\pgfsetfillcolor{currentfill}%
\pgfsetlinewidth{0.803000pt}%
\definecolor{currentstroke}{rgb}{0.000000,0.000000,0.000000}%
\pgfsetstrokecolor{currentstroke}%
\pgfsetdash{}{0pt}%
\pgfsys@defobject{currentmarker}{\pgfqpoint{-0.048611in}{0.000000in}}{\pgfqpoint{-0.000000in}{0.000000in}}{%
\pgfpathmoveto{\pgfqpoint{-0.000000in}{0.000000in}}%
\pgfpathlineto{\pgfqpoint{-0.048611in}{0.000000in}}%
\pgfusepath{stroke,fill}%
}%
\begin{pgfscope}%
\pgfsys@transformshift{0.307948in}{0.205517in}%
\pgfsys@useobject{currentmarker}{}%
\end{pgfscope}%
\end{pgfscope}%
\begin{pgfscope}%
\definecolor{textcolor}{rgb}{0.000000,0.000000,0.000000}%
\pgfsetstrokecolor{textcolor}%
\pgfsetfillcolor{textcolor}%
\pgftext[x=0.100000in, y=0.171759in, left, base]{\color{textcolor}\rmfamily\fontsize{7.000000}{8.400000}\selectfont \(\displaystyle {12}\)}%
\end{pgfscope}%
\begin{pgfscope}%
\pgfsetrectcap%
\pgfsetmiterjoin%
\pgfsetlinewidth{0.803000pt}%
\definecolor{currentstroke}{rgb}{0.000000,0.000000,0.000000}%
\pgfsetstrokecolor{currentstroke}%
\pgfsetdash{}{0pt}%
\pgfpathmoveto{\pgfqpoint{0.307948in}{0.148611in}}%
\pgfpathlineto{\pgfqpoint{0.307948in}{1.628157in}}%
\pgfusepath{stroke}%
\end{pgfscope}%
\begin{pgfscope}%
\pgfsetrectcap%
\pgfsetmiterjoin%
\pgfsetlinewidth{0.803000pt}%
\definecolor{currentstroke}{rgb}{0.000000,0.000000,0.000000}%
\pgfsetstrokecolor{currentstroke}%
\pgfsetdash{}{0pt}%
\pgfpathmoveto{\pgfqpoint{1.787494in}{0.148611in}}%
\pgfpathlineto{\pgfqpoint{1.787494in}{1.628157in}}%
\pgfusepath{stroke}%
\end{pgfscope}%
\begin{pgfscope}%
\pgfsetrectcap%
\pgfsetmiterjoin%
\pgfsetlinewidth{0.803000pt}%
\definecolor{currentstroke}{rgb}{0.000000,0.000000,0.000000}%
\pgfsetstrokecolor{currentstroke}%
\pgfsetdash{}{0pt}%
\pgfpathmoveto{\pgfqpoint{0.307948in}{0.148611in}}%
\pgfpathlineto{\pgfqpoint{1.787494in}{0.148611in}}%
\pgfusepath{stroke}%
\end{pgfscope}%
\begin{pgfscope}%
\pgfsetrectcap%
\pgfsetmiterjoin%
\pgfsetlinewidth{0.803000pt}%
\definecolor{currentstroke}{rgb}{0.000000,0.000000,0.000000}%
\pgfsetstrokecolor{currentstroke}%
\pgfsetdash{}{0pt}%
\pgfpathmoveto{\pgfqpoint{0.307948in}{1.628157in}}%
\pgfpathlineto{\pgfqpoint{1.787494in}{1.628157in}}%
\pgfusepath{stroke}%
\end{pgfscope}%
\begin{pgfscope}%
\definecolor{textcolor}{rgb}{0.000000,0.000000,0.000000}%
\pgfsetstrokecolor{textcolor}%
\pgfsetfillcolor{textcolor}%
\pgftext[x=1.047721in,y=1.895132in,,base]{\color{textcolor}\rmfamily\fontsize{8.400000}{10.080000}\selectfont sparstity pattern of \(\displaystyle A+B+C\)}%
\end{pgfscope}%
\begin{pgfscope}%
\pgfsetbuttcap%
\pgfsetmiterjoin%
\definecolor{currentfill}{rgb}{1.000000,1.000000,1.000000}%
\pgfsetfillcolor{currentfill}%
\pgfsetlinewidth{0.000000pt}%
\definecolor{currentstroke}{rgb}{0.000000,0.000000,0.000000}%
\pgfsetstrokecolor{currentstroke}%
\pgfsetstrokeopacity{0.000000}%
\pgfsetdash{}{0pt}%
\pgfpathmoveto{\pgfqpoint{2.083403in}{0.148611in}}%
\pgfpathlineto{\pgfqpoint{3.562948in}{0.148611in}}%
\pgfpathlineto{\pgfqpoint{3.562948in}{1.628157in}}%
\pgfpathlineto{\pgfqpoint{2.083403in}{1.628157in}}%
\pgfpathclose%
\pgfusepath{fill}%
\end{pgfscope}%
\begin{pgfscope}%
\pgfpathrectangle{\pgfqpoint{2.083403in}{0.148611in}}{\pgfqpoint{1.479545in}{1.479545in}}%
\pgfusepath{clip}%
\pgfsys@transformshift{2.083403in}{0.148611in}%
\pgftext[left,bottom]{\includegraphics[interpolate=true,width=1.480000in,height=1.480000in]{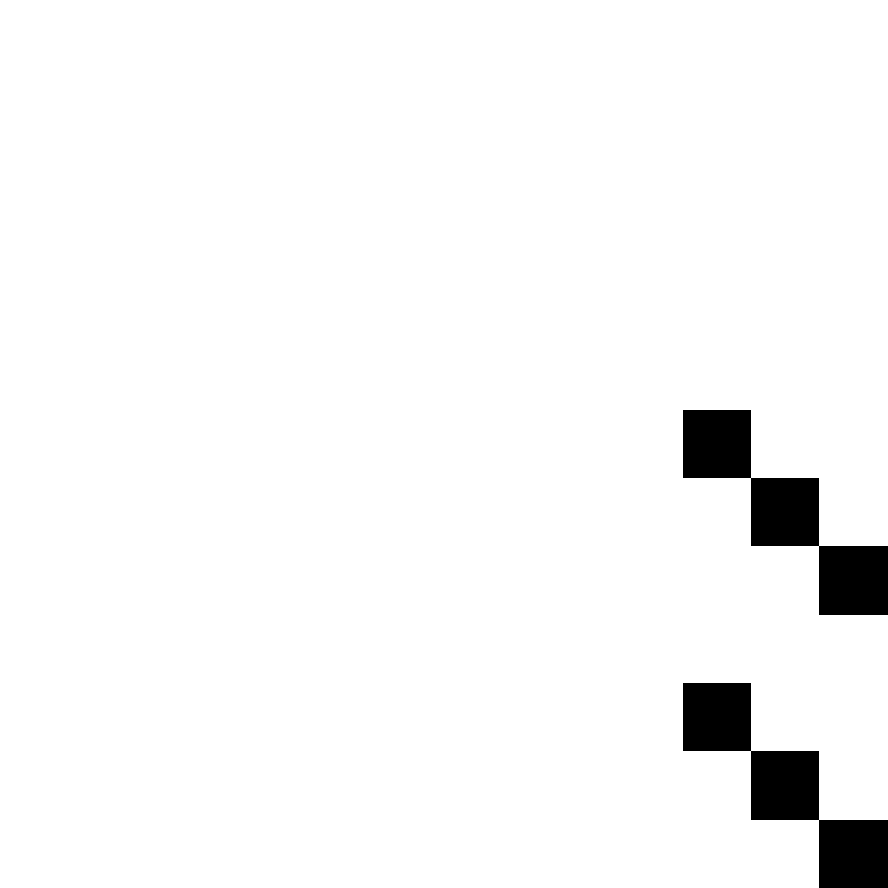}}%
\end{pgfscope}%
\begin{pgfscope}%
\pgfsetbuttcap%
\pgfsetroundjoin%
\definecolor{currentfill}{rgb}{0.000000,0.000000,0.000000}%
\pgfsetfillcolor{currentfill}%
\pgfsetlinewidth{0.803000pt}%
\definecolor{currentstroke}{rgb}{0.000000,0.000000,0.000000}%
\pgfsetstrokecolor{currentstroke}%
\pgfsetdash{}{0pt}%
\pgfsys@defobject{currentmarker}{\pgfqpoint{0.000000in}{-0.048611in}}{\pgfqpoint{0.000000in}{0.000000in}}{%
\pgfpathmoveto{\pgfqpoint{0.000000in}{0.000000in}}%
\pgfpathlineto{\pgfqpoint{0.000000in}{-0.048611in}}%
\pgfusepath{stroke,fill}%
}%
\begin{pgfscope}%
\pgfsys@transformshift{2.140308in}{0.148611in}%
\pgfsys@useobject{currentmarker}{}%
\end{pgfscope}%
\end{pgfscope}%
\begin{pgfscope}%
\pgfsetbuttcap%
\pgfsetroundjoin%
\definecolor{currentfill}{rgb}{0.000000,0.000000,0.000000}%
\pgfsetfillcolor{currentfill}%
\pgfsetlinewidth{0.803000pt}%
\definecolor{currentstroke}{rgb}{0.000000,0.000000,0.000000}%
\pgfsetstrokecolor{currentstroke}%
\pgfsetdash{}{0pt}%
\pgfsys@defobject{currentmarker}{\pgfqpoint{0.000000in}{0.000000in}}{\pgfqpoint{0.000000in}{0.048611in}}{%
\pgfpathmoveto{\pgfqpoint{0.000000in}{0.000000in}}%
\pgfpathlineto{\pgfqpoint{0.000000in}{0.048611in}}%
\pgfusepath{stroke,fill}%
}%
\begin{pgfscope}%
\pgfsys@transformshift{2.140308in}{1.628157in}%
\pgfsys@useobject{currentmarker}{}%
\end{pgfscope}%
\end{pgfscope}%
\begin{pgfscope}%
\definecolor{textcolor}{rgb}{0.000000,0.000000,0.000000}%
\pgfsetstrokecolor{textcolor}%
\pgfsetfillcolor{textcolor}%
\pgftext[x=2.140308in,y=1.725379in,,bottom]{\color{textcolor}\rmfamily\fontsize{7.000000}{8.400000}\selectfont \(\displaystyle {0}\)}%
\end{pgfscope}%
\begin{pgfscope}%
\pgfsetbuttcap%
\pgfsetroundjoin%
\definecolor{currentfill}{rgb}{0.000000,0.000000,0.000000}%
\pgfsetfillcolor{currentfill}%
\pgfsetlinewidth{0.803000pt}%
\definecolor{currentstroke}{rgb}{0.000000,0.000000,0.000000}%
\pgfsetstrokecolor{currentstroke}%
\pgfsetdash{}{0pt}%
\pgfsys@defobject{currentmarker}{\pgfqpoint{0.000000in}{-0.048611in}}{\pgfqpoint{0.000000in}{0.000000in}}{%
\pgfpathmoveto{\pgfqpoint{0.000000in}{0.000000in}}%
\pgfpathlineto{\pgfqpoint{0.000000in}{-0.048611in}}%
\pgfusepath{stroke,fill}%
}%
\begin{pgfscope}%
\pgfsys@transformshift{2.367931in}{0.148611in}%
\pgfsys@useobject{currentmarker}{}%
\end{pgfscope}%
\end{pgfscope}%
\begin{pgfscope}%
\pgfsetbuttcap%
\pgfsetroundjoin%
\definecolor{currentfill}{rgb}{0.000000,0.000000,0.000000}%
\pgfsetfillcolor{currentfill}%
\pgfsetlinewidth{0.803000pt}%
\definecolor{currentstroke}{rgb}{0.000000,0.000000,0.000000}%
\pgfsetstrokecolor{currentstroke}%
\pgfsetdash{}{0pt}%
\pgfsys@defobject{currentmarker}{\pgfqpoint{0.000000in}{0.000000in}}{\pgfqpoint{0.000000in}{0.048611in}}{%
\pgfpathmoveto{\pgfqpoint{0.000000in}{0.000000in}}%
\pgfpathlineto{\pgfqpoint{0.000000in}{0.048611in}}%
\pgfusepath{stroke,fill}%
}%
\begin{pgfscope}%
\pgfsys@transformshift{2.367931in}{1.628157in}%
\pgfsys@useobject{currentmarker}{}%
\end{pgfscope}%
\end{pgfscope}%
\begin{pgfscope}%
\definecolor{textcolor}{rgb}{0.000000,0.000000,0.000000}%
\pgfsetstrokecolor{textcolor}%
\pgfsetfillcolor{textcolor}%
\pgftext[x=2.367931in,y=1.725379in,,bottom]{\color{textcolor}\rmfamily\fontsize{7.000000}{8.400000}\selectfont \(\displaystyle {2}\)}%
\end{pgfscope}%
\begin{pgfscope}%
\pgfsetbuttcap%
\pgfsetroundjoin%
\definecolor{currentfill}{rgb}{0.000000,0.000000,0.000000}%
\pgfsetfillcolor{currentfill}%
\pgfsetlinewidth{0.803000pt}%
\definecolor{currentstroke}{rgb}{0.000000,0.000000,0.000000}%
\pgfsetstrokecolor{currentstroke}%
\pgfsetdash{}{0pt}%
\pgfsys@defobject{currentmarker}{\pgfqpoint{0.000000in}{-0.048611in}}{\pgfqpoint{0.000000in}{0.000000in}}{%
\pgfpathmoveto{\pgfqpoint{0.000000in}{0.000000in}}%
\pgfpathlineto{\pgfqpoint{0.000000in}{-0.048611in}}%
\pgfusepath{stroke,fill}%
}%
\begin{pgfscope}%
\pgfsys@transformshift{2.595553in}{0.148611in}%
\pgfsys@useobject{currentmarker}{}%
\end{pgfscope}%
\end{pgfscope}%
\begin{pgfscope}%
\pgfsetbuttcap%
\pgfsetroundjoin%
\definecolor{currentfill}{rgb}{0.000000,0.000000,0.000000}%
\pgfsetfillcolor{currentfill}%
\pgfsetlinewidth{0.803000pt}%
\definecolor{currentstroke}{rgb}{0.000000,0.000000,0.000000}%
\pgfsetstrokecolor{currentstroke}%
\pgfsetdash{}{0pt}%
\pgfsys@defobject{currentmarker}{\pgfqpoint{0.000000in}{0.000000in}}{\pgfqpoint{0.000000in}{0.048611in}}{%
\pgfpathmoveto{\pgfqpoint{0.000000in}{0.000000in}}%
\pgfpathlineto{\pgfqpoint{0.000000in}{0.048611in}}%
\pgfusepath{stroke,fill}%
}%
\begin{pgfscope}%
\pgfsys@transformshift{2.595553in}{1.628157in}%
\pgfsys@useobject{currentmarker}{}%
\end{pgfscope}%
\end{pgfscope}%
\begin{pgfscope}%
\definecolor{textcolor}{rgb}{0.000000,0.000000,0.000000}%
\pgfsetstrokecolor{textcolor}%
\pgfsetfillcolor{textcolor}%
\pgftext[x=2.595553in,y=1.725379in,,bottom]{\color{textcolor}\rmfamily\fontsize{7.000000}{8.400000}\selectfont \(\displaystyle {4}\)}%
\end{pgfscope}%
\begin{pgfscope}%
\pgfsetbuttcap%
\pgfsetroundjoin%
\definecolor{currentfill}{rgb}{0.000000,0.000000,0.000000}%
\pgfsetfillcolor{currentfill}%
\pgfsetlinewidth{0.803000pt}%
\definecolor{currentstroke}{rgb}{0.000000,0.000000,0.000000}%
\pgfsetstrokecolor{currentstroke}%
\pgfsetdash{}{0pt}%
\pgfsys@defobject{currentmarker}{\pgfqpoint{0.000000in}{-0.048611in}}{\pgfqpoint{0.000000in}{0.000000in}}{%
\pgfpathmoveto{\pgfqpoint{0.000000in}{0.000000in}}%
\pgfpathlineto{\pgfqpoint{0.000000in}{-0.048611in}}%
\pgfusepath{stroke,fill}%
}%
\begin{pgfscope}%
\pgfsys@transformshift{2.823175in}{0.148611in}%
\pgfsys@useobject{currentmarker}{}%
\end{pgfscope}%
\end{pgfscope}%
\begin{pgfscope}%
\pgfsetbuttcap%
\pgfsetroundjoin%
\definecolor{currentfill}{rgb}{0.000000,0.000000,0.000000}%
\pgfsetfillcolor{currentfill}%
\pgfsetlinewidth{0.803000pt}%
\definecolor{currentstroke}{rgb}{0.000000,0.000000,0.000000}%
\pgfsetstrokecolor{currentstroke}%
\pgfsetdash{}{0pt}%
\pgfsys@defobject{currentmarker}{\pgfqpoint{0.000000in}{0.000000in}}{\pgfqpoint{0.000000in}{0.048611in}}{%
\pgfpathmoveto{\pgfqpoint{0.000000in}{0.000000in}}%
\pgfpathlineto{\pgfqpoint{0.000000in}{0.048611in}}%
\pgfusepath{stroke,fill}%
}%
\begin{pgfscope}%
\pgfsys@transformshift{2.823175in}{1.628157in}%
\pgfsys@useobject{currentmarker}{}%
\end{pgfscope}%
\end{pgfscope}%
\begin{pgfscope}%
\definecolor{textcolor}{rgb}{0.000000,0.000000,0.000000}%
\pgfsetstrokecolor{textcolor}%
\pgfsetfillcolor{textcolor}%
\pgftext[x=2.823175in,y=1.725379in,,bottom]{\color{textcolor}\rmfamily\fontsize{7.000000}{8.400000}\selectfont \(\displaystyle {6}\)}%
\end{pgfscope}%
\begin{pgfscope}%
\pgfsetbuttcap%
\pgfsetroundjoin%
\definecolor{currentfill}{rgb}{0.000000,0.000000,0.000000}%
\pgfsetfillcolor{currentfill}%
\pgfsetlinewidth{0.803000pt}%
\definecolor{currentstroke}{rgb}{0.000000,0.000000,0.000000}%
\pgfsetstrokecolor{currentstroke}%
\pgfsetdash{}{0pt}%
\pgfsys@defobject{currentmarker}{\pgfqpoint{0.000000in}{-0.048611in}}{\pgfqpoint{0.000000in}{0.000000in}}{%
\pgfpathmoveto{\pgfqpoint{0.000000in}{0.000000in}}%
\pgfpathlineto{\pgfqpoint{0.000000in}{-0.048611in}}%
\pgfusepath{stroke,fill}%
}%
\begin{pgfscope}%
\pgfsys@transformshift{3.050798in}{0.148611in}%
\pgfsys@useobject{currentmarker}{}%
\end{pgfscope}%
\end{pgfscope}%
\begin{pgfscope}%
\pgfsetbuttcap%
\pgfsetroundjoin%
\definecolor{currentfill}{rgb}{0.000000,0.000000,0.000000}%
\pgfsetfillcolor{currentfill}%
\pgfsetlinewidth{0.803000pt}%
\definecolor{currentstroke}{rgb}{0.000000,0.000000,0.000000}%
\pgfsetstrokecolor{currentstroke}%
\pgfsetdash{}{0pt}%
\pgfsys@defobject{currentmarker}{\pgfqpoint{0.000000in}{0.000000in}}{\pgfqpoint{0.000000in}{0.048611in}}{%
\pgfpathmoveto{\pgfqpoint{0.000000in}{0.000000in}}%
\pgfpathlineto{\pgfqpoint{0.000000in}{0.048611in}}%
\pgfusepath{stroke,fill}%
}%
\begin{pgfscope}%
\pgfsys@transformshift{3.050798in}{1.628157in}%
\pgfsys@useobject{currentmarker}{}%
\end{pgfscope}%
\end{pgfscope}%
\begin{pgfscope}%
\definecolor{textcolor}{rgb}{0.000000,0.000000,0.000000}%
\pgfsetstrokecolor{textcolor}%
\pgfsetfillcolor{textcolor}%
\pgftext[x=3.050798in,y=1.725379in,,bottom]{\color{textcolor}\rmfamily\fontsize{7.000000}{8.400000}\selectfont \(\displaystyle {8}\)}%
\end{pgfscope}%
\begin{pgfscope}%
\pgfsetbuttcap%
\pgfsetroundjoin%
\definecolor{currentfill}{rgb}{0.000000,0.000000,0.000000}%
\pgfsetfillcolor{currentfill}%
\pgfsetlinewidth{0.803000pt}%
\definecolor{currentstroke}{rgb}{0.000000,0.000000,0.000000}%
\pgfsetstrokecolor{currentstroke}%
\pgfsetdash{}{0pt}%
\pgfsys@defobject{currentmarker}{\pgfqpoint{0.000000in}{-0.048611in}}{\pgfqpoint{0.000000in}{0.000000in}}{%
\pgfpathmoveto{\pgfqpoint{0.000000in}{0.000000in}}%
\pgfpathlineto{\pgfqpoint{0.000000in}{-0.048611in}}%
\pgfusepath{stroke,fill}%
}%
\begin{pgfscope}%
\pgfsys@transformshift{3.278420in}{0.148611in}%
\pgfsys@useobject{currentmarker}{}%
\end{pgfscope}%
\end{pgfscope}%
\begin{pgfscope}%
\pgfsetbuttcap%
\pgfsetroundjoin%
\definecolor{currentfill}{rgb}{0.000000,0.000000,0.000000}%
\pgfsetfillcolor{currentfill}%
\pgfsetlinewidth{0.803000pt}%
\definecolor{currentstroke}{rgb}{0.000000,0.000000,0.000000}%
\pgfsetstrokecolor{currentstroke}%
\pgfsetdash{}{0pt}%
\pgfsys@defobject{currentmarker}{\pgfqpoint{0.000000in}{0.000000in}}{\pgfqpoint{0.000000in}{0.048611in}}{%
\pgfpathmoveto{\pgfqpoint{0.000000in}{0.000000in}}%
\pgfpathlineto{\pgfqpoint{0.000000in}{0.048611in}}%
\pgfusepath{stroke,fill}%
}%
\begin{pgfscope}%
\pgfsys@transformshift{3.278420in}{1.628157in}%
\pgfsys@useobject{currentmarker}{}%
\end{pgfscope}%
\end{pgfscope}%
\begin{pgfscope}%
\definecolor{textcolor}{rgb}{0.000000,0.000000,0.000000}%
\pgfsetstrokecolor{textcolor}%
\pgfsetfillcolor{textcolor}%
\pgftext[x=3.278420in,y=1.725379in,,bottom]{\color{textcolor}\rmfamily\fontsize{7.000000}{8.400000}\selectfont \(\displaystyle {10}\)}%
\end{pgfscope}%
\begin{pgfscope}%
\pgfsetbuttcap%
\pgfsetroundjoin%
\definecolor{currentfill}{rgb}{0.000000,0.000000,0.000000}%
\pgfsetfillcolor{currentfill}%
\pgfsetlinewidth{0.803000pt}%
\definecolor{currentstroke}{rgb}{0.000000,0.000000,0.000000}%
\pgfsetstrokecolor{currentstroke}%
\pgfsetdash{}{0pt}%
\pgfsys@defobject{currentmarker}{\pgfqpoint{0.000000in}{-0.048611in}}{\pgfqpoint{0.000000in}{0.000000in}}{%
\pgfpathmoveto{\pgfqpoint{0.000000in}{0.000000in}}%
\pgfpathlineto{\pgfqpoint{0.000000in}{-0.048611in}}%
\pgfusepath{stroke,fill}%
}%
\begin{pgfscope}%
\pgfsys@transformshift{3.506042in}{0.148611in}%
\pgfsys@useobject{currentmarker}{}%
\end{pgfscope}%
\end{pgfscope}%
\begin{pgfscope}%
\pgfsetbuttcap%
\pgfsetroundjoin%
\definecolor{currentfill}{rgb}{0.000000,0.000000,0.000000}%
\pgfsetfillcolor{currentfill}%
\pgfsetlinewidth{0.803000pt}%
\definecolor{currentstroke}{rgb}{0.000000,0.000000,0.000000}%
\pgfsetstrokecolor{currentstroke}%
\pgfsetdash{}{0pt}%
\pgfsys@defobject{currentmarker}{\pgfqpoint{0.000000in}{0.000000in}}{\pgfqpoint{0.000000in}{0.048611in}}{%
\pgfpathmoveto{\pgfqpoint{0.000000in}{0.000000in}}%
\pgfpathlineto{\pgfqpoint{0.000000in}{0.048611in}}%
\pgfusepath{stroke,fill}%
}%
\begin{pgfscope}%
\pgfsys@transformshift{3.506042in}{1.628157in}%
\pgfsys@useobject{currentmarker}{}%
\end{pgfscope}%
\end{pgfscope}%
\begin{pgfscope}%
\definecolor{textcolor}{rgb}{0.000000,0.000000,0.000000}%
\pgfsetstrokecolor{textcolor}%
\pgfsetfillcolor{textcolor}%
\pgftext[x=3.506042in,y=1.725379in,,bottom]{\color{textcolor}\rmfamily\fontsize{7.000000}{8.400000}\selectfont \(\displaystyle {12}\)}%
\end{pgfscope}%
\begin{pgfscope}%
\pgfsetbuttcap%
\pgfsetroundjoin%
\definecolor{currentfill}{rgb}{0.000000,0.000000,0.000000}%
\pgfsetfillcolor{currentfill}%
\pgfsetlinewidth{0.803000pt}%
\definecolor{currentstroke}{rgb}{0.000000,0.000000,0.000000}%
\pgfsetstrokecolor{currentstroke}%
\pgfsetdash{}{0pt}%
\pgfsys@defobject{currentmarker}{\pgfqpoint{-0.048611in}{0.000000in}}{\pgfqpoint{-0.000000in}{0.000000in}}{%
\pgfpathmoveto{\pgfqpoint{-0.000000in}{0.000000in}}%
\pgfpathlineto{\pgfqpoint{-0.048611in}{0.000000in}}%
\pgfusepath{stroke,fill}%
}%
\begin{pgfscope}%
\pgfsys@transformshift{2.083403in}{1.571251in}%
\pgfsys@useobject{currentmarker}{}%
\end{pgfscope}%
\end{pgfscope}%
\begin{pgfscope}%
\definecolor{textcolor}{rgb}{0.000000,0.000000,0.000000}%
\pgfsetstrokecolor{textcolor}%
\pgfsetfillcolor{textcolor}%
\pgftext[x=1.930817in, y=1.537493in, left, base]{\color{textcolor}\rmfamily\fontsize{7.000000}{8.400000}\selectfont \(\displaystyle {0}\)}%
\end{pgfscope}%
\begin{pgfscope}%
\pgfsetbuttcap%
\pgfsetroundjoin%
\definecolor{currentfill}{rgb}{0.000000,0.000000,0.000000}%
\pgfsetfillcolor{currentfill}%
\pgfsetlinewidth{0.803000pt}%
\definecolor{currentstroke}{rgb}{0.000000,0.000000,0.000000}%
\pgfsetstrokecolor{currentstroke}%
\pgfsetdash{}{0pt}%
\pgfsys@defobject{currentmarker}{\pgfqpoint{-0.048611in}{0.000000in}}{\pgfqpoint{-0.000000in}{0.000000in}}{%
\pgfpathmoveto{\pgfqpoint{-0.000000in}{0.000000in}}%
\pgfpathlineto{\pgfqpoint{-0.048611in}{0.000000in}}%
\pgfusepath{stroke,fill}%
}%
\begin{pgfscope}%
\pgfsys@transformshift{2.083403in}{1.343629in}%
\pgfsys@useobject{currentmarker}{}%
\end{pgfscope}%
\end{pgfscope}%
\begin{pgfscope}%
\definecolor{textcolor}{rgb}{0.000000,0.000000,0.000000}%
\pgfsetstrokecolor{textcolor}%
\pgfsetfillcolor{textcolor}%
\pgftext[x=1.930817in, y=1.309871in, left, base]{\color{textcolor}\rmfamily\fontsize{7.000000}{8.400000}\selectfont \(\displaystyle {2}\)}%
\end{pgfscope}%
\begin{pgfscope}%
\pgfsetbuttcap%
\pgfsetroundjoin%
\definecolor{currentfill}{rgb}{0.000000,0.000000,0.000000}%
\pgfsetfillcolor{currentfill}%
\pgfsetlinewidth{0.803000pt}%
\definecolor{currentstroke}{rgb}{0.000000,0.000000,0.000000}%
\pgfsetstrokecolor{currentstroke}%
\pgfsetdash{}{0pt}%
\pgfsys@defobject{currentmarker}{\pgfqpoint{-0.048611in}{0.000000in}}{\pgfqpoint{-0.000000in}{0.000000in}}{%
\pgfpathmoveto{\pgfqpoint{-0.000000in}{0.000000in}}%
\pgfpathlineto{\pgfqpoint{-0.048611in}{0.000000in}}%
\pgfusepath{stroke,fill}%
}%
\begin{pgfscope}%
\pgfsys@transformshift{2.083403in}{1.116006in}%
\pgfsys@useobject{currentmarker}{}%
\end{pgfscope}%
\end{pgfscope}%
\begin{pgfscope}%
\definecolor{textcolor}{rgb}{0.000000,0.000000,0.000000}%
\pgfsetstrokecolor{textcolor}%
\pgfsetfillcolor{textcolor}%
\pgftext[x=1.930817in, y=1.082249in, left, base]{\color{textcolor}\rmfamily\fontsize{7.000000}{8.400000}\selectfont \(\displaystyle {4}\)}%
\end{pgfscope}%
\begin{pgfscope}%
\pgfsetbuttcap%
\pgfsetroundjoin%
\definecolor{currentfill}{rgb}{0.000000,0.000000,0.000000}%
\pgfsetfillcolor{currentfill}%
\pgfsetlinewidth{0.803000pt}%
\definecolor{currentstroke}{rgb}{0.000000,0.000000,0.000000}%
\pgfsetstrokecolor{currentstroke}%
\pgfsetdash{}{0pt}%
\pgfsys@defobject{currentmarker}{\pgfqpoint{-0.048611in}{0.000000in}}{\pgfqpoint{-0.000000in}{0.000000in}}{%
\pgfpathmoveto{\pgfqpoint{-0.000000in}{0.000000in}}%
\pgfpathlineto{\pgfqpoint{-0.048611in}{0.000000in}}%
\pgfusepath{stroke,fill}%
}%
\begin{pgfscope}%
\pgfsys@transformshift{2.083403in}{0.888384in}%
\pgfsys@useobject{currentmarker}{}%
\end{pgfscope}%
\end{pgfscope}%
\begin{pgfscope}%
\definecolor{textcolor}{rgb}{0.000000,0.000000,0.000000}%
\pgfsetstrokecolor{textcolor}%
\pgfsetfillcolor{textcolor}%
\pgftext[x=1.930817in, y=0.854626in, left, base]{\color{textcolor}\rmfamily\fontsize{7.000000}{8.400000}\selectfont \(\displaystyle {6}\)}%
\end{pgfscope}%
\begin{pgfscope}%
\pgfsetbuttcap%
\pgfsetroundjoin%
\definecolor{currentfill}{rgb}{0.000000,0.000000,0.000000}%
\pgfsetfillcolor{currentfill}%
\pgfsetlinewidth{0.803000pt}%
\definecolor{currentstroke}{rgb}{0.000000,0.000000,0.000000}%
\pgfsetstrokecolor{currentstroke}%
\pgfsetdash{}{0pt}%
\pgfsys@defobject{currentmarker}{\pgfqpoint{-0.048611in}{0.000000in}}{\pgfqpoint{-0.000000in}{0.000000in}}{%
\pgfpathmoveto{\pgfqpoint{-0.000000in}{0.000000in}}%
\pgfpathlineto{\pgfqpoint{-0.048611in}{0.000000in}}%
\pgfusepath{stroke,fill}%
}%
\begin{pgfscope}%
\pgfsys@transformshift{2.083403in}{0.660761in}%
\pgfsys@useobject{currentmarker}{}%
\end{pgfscope}%
\end{pgfscope}%
\begin{pgfscope}%
\definecolor{textcolor}{rgb}{0.000000,0.000000,0.000000}%
\pgfsetstrokecolor{textcolor}%
\pgfsetfillcolor{textcolor}%
\pgftext[x=1.930817in, y=0.627004in, left, base]{\color{textcolor}\rmfamily\fontsize{7.000000}{8.400000}\selectfont \(\displaystyle {8}\)}%
\end{pgfscope}%
\begin{pgfscope}%
\pgfsetbuttcap%
\pgfsetroundjoin%
\definecolor{currentfill}{rgb}{0.000000,0.000000,0.000000}%
\pgfsetfillcolor{currentfill}%
\pgfsetlinewidth{0.803000pt}%
\definecolor{currentstroke}{rgb}{0.000000,0.000000,0.000000}%
\pgfsetstrokecolor{currentstroke}%
\pgfsetdash{}{0pt}%
\pgfsys@defobject{currentmarker}{\pgfqpoint{-0.048611in}{0.000000in}}{\pgfqpoint{-0.000000in}{0.000000in}}{%
\pgfpathmoveto{\pgfqpoint{-0.000000in}{0.000000in}}%
\pgfpathlineto{\pgfqpoint{-0.048611in}{0.000000in}}%
\pgfusepath{stroke,fill}%
}%
\begin{pgfscope}%
\pgfsys@transformshift{2.083403in}{0.433139in}%
\pgfsys@useobject{currentmarker}{}%
\end{pgfscope}%
\end{pgfscope}%
\begin{pgfscope}%
\definecolor{textcolor}{rgb}{0.000000,0.000000,0.000000}%
\pgfsetstrokecolor{textcolor}%
\pgfsetfillcolor{textcolor}%
\pgftext[x=1.875455in, y=0.399381in, left, base]{\color{textcolor}\rmfamily\fontsize{7.000000}{8.400000}\selectfont \(\displaystyle {10}\)}%
\end{pgfscope}%
\begin{pgfscope}%
\pgfsetbuttcap%
\pgfsetroundjoin%
\definecolor{currentfill}{rgb}{0.000000,0.000000,0.000000}%
\pgfsetfillcolor{currentfill}%
\pgfsetlinewidth{0.803000pt}%
\definecolor{currentstroke}{rgb}{0.000000,0.000000,0.000000}%
\pgfsetstrokecolor{currentstroke}%
\pgfsetdash{}{0pt}%
\pgfsys@defobject{currentmarker}{\pgfqpoint{-0.048611in}{0.000000in}}{\pgfqpoint{-0.000000in}{0.000000in}}{%
\pgfpathmoveto{\pgfqpoint{-0.000000in}{0.000000in}}%
\pgfpathlineto{\pgfqpoint{-0.048611in}{0.000000in}}%
\pgfusepath{stroke,fill}%
}%
\begin{pgfscope}%
\pgfsys@transformshift{2.083403in}{0.205517in}%
\pgfsys@useobject{currentmarker}{}%
\end{pgfscope}%
\end{pgfscope}%
\begin{pgfscope}%
\definecolor{textcolor}{rgb}{0.000000,0.000000,0.000000}%
\pgfsetstrokecolor{textcolor}%
\pgfsetfillcolor{textcolor}%
\pgftext[x=1.875455in, y=0.171759in, left, base]{\color{textcolor}\rmfamily\fontsize{7.000000}{8.400000}\selectfont \(\displaystyle {12}\)}%
\end{pgfscope}%
\begin{pgfscope}%
\pgfsetrectcap%
\pgfsetmiterjoin%
\pgfsetlinewidth{0.803000pt}%
\definecolor{currentstroke}{rgb}{0.000000,0.000000,0.000000}%
\pgfsetstrokecolor{currentstroke}%
\pgfsetdash{}{0pt}%
\pgfpathmoveto{\pgfqpoint{2.083403in}{0.148611in}}%
\pgfpathlineto{\pgfqpoint{2.083403in}{1.628157in}}%
\pgfusepath{stroke}%
\end{pgfscope}%
\begin{pgfscope}%
\pgfsetrectcap%
\pgfsetmiterjoin%
\pgfsetlinewidth{0.803000pt}%
\definecolor{currentstroke}{rgb}{0.000000,0.000000,0.000000}%
\pgfsetstrokecolor{currentstroke}%
\pgfsetdash{}{0pt}%
\pgfpathmoveto{\pgfqpoint{3.562948in}{0.148611in}}%
\pgfpathlineto{\pgfqpoint{3.562948in}{1.628157in}}%
\pgfusepath{stroke}%
\end{pgfscope}%
\begin{pgfscope}%
\pgfsetrectcap%
\pgfsetmiterjoin%
\pgfsetlinewidth{0.803000pt}%
\definecolor{currentstroke}{rgb}{0.000000,0.000000,0.000000}%
\pgfsetstrokecolor{currentstroke}%
\pgfsetdash{}{0pt}%
\pgfpathmoveto{\pgfqpoint{2.083403in}{0.148611in}}%
\pgfpathlineto{\pgfqpoint{3.562948in}{0.148611in}}%
\pgfusepath{stroke}%
\end{pgfscope}%
\begin{pgfscope}%
\pgfsetrectcap%
\pgfsetmiterjoin%
\pgfsetlinewidth{0.803000pt}%
\definecolor{currentstroke}{rgb}{0.000000,0.000000,0.000000}%
\pgfsetstrokecolor{currentstroke}%
\pgfsetdash{}{0pt}%
\pgfpathmoveto{\pgfqpoint{2.083403in}{1.628157in}}%
\pgfpathlineto{\pgfqpoint{3.562948in}{1.628157in}}%
\pgfusepath{stroke}%
\end{pgfscope}%
\begin{pgfscope}%
\definecolor{textcolor}{rgb}{0.000000,0.000000,0.000000}%
\pgfsetstrokecolor{textcolor}%
\pgfsetfillcolor{textcolor}%
\pgftext[x=2.823175in,y=1.895132in,,base]{\color{textcolor}\rmfamily\fontsize{8.400000}{10.080000}\selectfont sparsity pattern of \(\displaystyle E\)}%
\end{pgfscope}%
\end{pgfpicture}%
\makeatother%
\endgroup%

%% file: Discretisation.tex
\section{ADER-DG discretisation}
\label{sec:discretisation}
In this section we summarise the numerical method we use to discretise \cref{eq:poroelastic-wave}.
We focus on the Discontinuous Galerkin (DG) method, which has been increasingly attractive for simulation of elastic wave propagation~\cite{chung_optimal_2006, riviere_discontinuous_2007, de_basabe_interior_2008, diaz_energy_2009, etienne_hp-adaptive_2010, wilcox_high-order_2010, antonietti_non-conforming_2012, mazzieri_speed_2013, peyrusse_high-order_2014, mercerat_nodal_2015}
We combine the DG method with Arbitrary high-order DERivative (ADER) time stepping, leading to high-order accuracy in time within a single step~\cite{titarev_ader_2002}. 
In essence, ADER-DG is a predictor-corrector scheme:
First, in each element we predict a solution solely based on the information within the element itself.
In the second step, the predicted solution is corrected using numerical fluxes across element boundaries.
The scheme is explicit in time, which is attractive from a computational perspective, since no global system of equations has to be assembled and solved~\cite{gassner_explicit_2011}
ADER-DG has been successfully used for a broad range of problems, for example, shallow water equations, relativistic magnetohydrodynamics or the Euler equations~\cite{reinarz_exahype_2020}
In particular, ADER-DG is the basis of many seismological applications~\cite{dumbser_arbitrary_2006, kaser_arbitrary_2007, uphoff_extreme_2017,  ulrich_dynamic_2019, palgunadi_dynamic_2020,  reinarz_exahype_2020,  duru_new_2021,  krenz_3d_2021}  \subsection{Spatial discretisation}
We use a DG approach on unstructured tetrahedral elements.
First, we partition the computational domain $\Omega \subset \R^3$ in a set of conforming tetrahedrons $\mathcal{T} = \{T_i\}_{i=1}^n$.
Within each tetrahedron, we expand the solution $q$ in space: $q_p(x,t) = \hat{Q}_{pl}^n(t) \psi^n_l(x)$, for $x \in T_n$ using a set of basis functions $\psi^n_l$.
Here we use Einstein sum convention, i.e. a sum over repeated indices is implied.
The superscript $n$ emphasises that $\hat{Q}$ and $\psi_l$ are specific to the element $T_n$. 
For better readability, however, we will omit that superscript, where it is clear from the context.
We introduce the reference element $\mathcal{E} = \{(x,y,z) \in \R^3: x, y, z > 0 \land x+y+z < 1\}$.
Now, we can construct an affine linear mapping $\Xi^n: T_n \rightarrow \mathcal{E}$, which maps global coordinates to reference coordinates.
Finally, we define the basis functions $\psi_l^n(x) = \phi_l(\Xi^n(x))$, using a set of polynomials $\phi_l$ defined on the reference element.
As $\phi_l$, we choose Dubiner polynomials, which are an orthogonal set of polynomials on tetrahedrons and are based on Jacobi polynomials~\cite{cockburn_discontinuous_2000}.
Any polynomial $p(x,y,z)$ in three spatial variables can be written as $\sum_{i,j,k} \alpha_{ijk} x^i y^j z^k$.
We define the degree of a polynomial as $\deg(p) = \underset{\alpha_{ijk} \neq 0}{\max} i + j + k$.
For example, the polynomial $1$ has degree $0$, the polynomial $x^2yz^2$ has degree $5$ and so on.
In three space dimensions the space of polynomials, which are exactly of degree $N$, can be represented by $\binom{N+2}{2}$ basis functions and the space of polynomials of degree equal or less than $N$ is spanned by $\binom{N+3}{3}$ basis functions.
The Dubiner polynomials $\phi_i$ are ordered such that the degree of $\phi_i$ is less or equal to the degree of $\phi_j$ for $i \leq j$.

We multiply \cref{eq:poroelastic-wave} with a test function $\psi_k^n$ and integrate in space to get a weak formulation of our equation:
\begin{equation}
\label{eq:poroelastic-weak}
    \int_{T_n} \derivative{q}{t} \psi_k^n \dd{V}  + \int_{T_n} \left( A \derivative{q}{x} + B \derivative{q}{y} + C \derivative{q}{z}\right)\psi_k^n \dd{V} = \int_{T_n} E q \psi_k^n \dd{V} 
\end{equation}
We transform the integrals to the reference element and perform integration by parts (c.f.~\cite{kaser_arbitrary_2007}). 
The Jacobians on the reference element can be obtained as $A^* = A \derivative{\xi}{x} + B \derivative{\xi}{y} + C \derivative{\xi}{z}$, $B^* = A \derivative{\eta}{x} + B \derivative{\eta}{y} + C \derivative{\eta}{z}$ and $C^* = A \derivative{\zeta}{x} + B \derivative{\zeta}{y} + C \derivative{\zeta}{z}$.
Here, $\xi$, $\eta$, $\zeta$ are the coordinate components of the transformation $\Xi^n$.
To ease notation, we set 
$\mathcal{A}^1 = A^*$,
$\mathcal{A}^2 = B^*$,
$\mathcal{A}^3 = C^*$ and 
express derivatives in $\xi$ direction as $\partial_1$, 
derivatives in $\eta$ direction as $\partial_2$,  and 
derivatives in $\zeta$ direction as $\partial_3$.
Then we get
\begin{equation}
\label{eq:semi_discrete}
\begin{aligned}
    &\derivative{\hat{Q}^n_{pl}}{t} |J| \int_\mathcal{E} \phi_l \phi_k \dd{V}
    + \sum_{j=1}^4 F^j_{pk}(\hat{Q}^n, \hat{Q}^{n_j}) \\
    &- \sum_{j=1}^3 \mathcal{A}^j_{pq} \hat{Q}^n_{ql} |J| \int_\mathcal{E} \partial_j\phi_k \phi_l \dd{V} 
    = E_{pq} \hat{Q}^n_{ql} |J| \int_\mathcal{E} \phi_k \phi_l \dd{V}. \\
    \end{aligned}
\end{equation}
We assume that the Jacobian matrices $\mathcal{A}$ are constant on each element. 
This has the advantage that the integrals can be precomputed and no quadrature is needed.
Here, we introduced the numerical flux $F_{pk}^j(\hat{Q}^n, \hat{Q}^{n_j})$, which evaluates the exchange of quantities across the $j^{th}$ face of the tetrahedron.
As a numerical flux we choose Godunov's flux method based on the solution of exact Riemann problems at the inter-cell boundaries~\cite{leveque_finite_2002, hesthaven_nodal_2008}.

\subsection{Temporal discretisation with ADER}
We add ADER time stepping~\cite{titarev_ader_2002}.
We replace continuous time with a series of time steps $0 = t_0, t_1, \dots$, for simplicity we assume a regular grid in time: $t_i = i \cdot \Delta t$.
The idea of ADER time stepping is the following: 
Given the local solution $q_p(t_i, x) = \hat{Q}_{pl}^n(t_i)\psi_l(x)$, we predict a solution for upcoming times $t_i + \delta t < t_{i+1}$.
This is classically done using a Taylor series approach,
\begin{equation}
\label{eq:taylor}
    q_p(t_i+\delta t) = \sum_{j=1}^N \derivative{^j q_p}{t_j}\left(t_i\right) \frac{\delta t^j}{j!}.
\end{equation}
The temporal derivatives at $t_i$ are computed from the spatial derivatives using the \ck procedure~\cite{toro_riemann_2009}, where temporal derivatives can be replaced by spatial derivatives.
This Taylor series predicts the time evolution of $q_p$ based on the local information.
If we integrate \cref{eq:semi_discrete} in time over $[t_i, t_{i+1}]$, we get an expression to compute $\hat{Q}^n_{pl}(t_{i+1})$ from $\hat{Q}^n_{pl}(t_i)$.
The integration in time is done using Gaussian quadrature.
This is where the predicted solutions for arbitrary times $t \in [t_i, t_{i+1}]$ is needed.
This corrector step takes the flux information from the neighbouring elements into account.

This scheme has several advantages:
It is a one-step scheme, i.e. no intermediate stages (such as for example, with Runge-Kutta schemes) have to be stored.
The scheme is explicit, in particular, no global system of equations has to be solved.
As a direct consequence, the scheme can easily be parallelised, using mesh partitioning, where only ghost cells at partition boundaries have to be exchanged.
If we choose polynomials up to degree $N$, we achieve a convergence rate $N+1$ in space \textit{and} time.
To ensure stability, a CFL condition in the form 
\begin{equation*}
  \Delta t \leq C(N) \frac{d}{v}
\end{equation*}
has to be fulfilled for each mesh element, where $d$ is the diameter of the element's insphere and $v$ is the maximal wave speed~\cite{dumbser_arbitrary_2007}.
The constant $C$ is set to $C(N) = \frac{c}{2N+1}$ with $c = \frac{1}{2}$.
If we use global time stepping, $\Delta t$ is set to the smallest fraction $\underset{T_i \in \mathcal{T}}{\min} C(N) d(T_i) / v(T_i)$, taking into account all elements.
Material parameters, and thus wave speeds, change over the computational domain.
In addition, meshes are typically refined locally.
So with global time stepping we might impose time steps smaller than actually necessary.
Local time stepping can be added to the ADER-DG scheme, to save computational load where possible~\cite{dumbser_arbitrary_2007}.
As modern supercomputers work best on structured data, clustered local time stepping is used~\cite{breuer_petascale_2016}.

\subsection{Space-Time predictor}
\label{sec:space_time_pred}
The right hand side in \cref{eq:poroelastic-wave}, $Eq$, poses a stiff source term.
As a consequence, the \ck procedure to predict a solution becomes unstable.
In~\cite{de_la_puente_discontinuous_2008}, new time stepping schemes that are stable with a stiff term are analysed -- operator splitting and space-time predictor. 
Since the splitting scheme does not achieve high-order convergence, we focus on the space-time predictor.
For a more general review of this algorithm, we refer to~\cite[sec. 3.3]{gassner_explicit_2011}.
The idea is to express the solution in time using a polynomial expansion: $q_p(x,t) = \tilde{Q}^{ni}_{pls} \psi^n_l(x) \theta^i_s(t)$ for $x \in T_n, t \in [t_i, t_{i+1}]$.
Note that $\tilde{Q}^{ni}$ is not time dependent anymore, but remains constant on the space-time element $\mathcal{E}_n \times [t_i, t_{i+1}]$.
We define $[0,1]$ as a reference element in time. 
On this element, we choose Jacobi polynomials $\chi_s$ as a basis and obtain $\theta^i$ via a transformation onto the reference element.
We chose the same degree for the spatial and temporal basis functions.
Now we multiply \cref{eq:poroelastic-wave} with spatial and temporal basis functions to derive the following system of equations for $\tilde{Q}^{ni}$~\cite{de_la_puente_discontinuous_2008}:
\begin{multline*}
   \int_{t_i}^{t_{i+1}}\int_{T_n}\tilde{Q}^{ni}_{pls}\psi_l\derivative{\theta_s}{t}\psi_k\theta_r\dd{V}\dd{t}\\
   + \int_{t_i}^{t_{i+1}}\int_{T_n} \left(
   A_{pq}\tilde{Q}^{ni}_{qls} \derivative{\psi_l}{x}  \theta_s +
   B_{pq}\tilde{Q}^{ni}_{qls}\derivative{\psi_l}{y}  \theta_s +
   C_{pq}\tilde{Q}^{ni}_{qls}\derivative{\psi_l}{z}  \theta_s 
   \right)\psi_k \theta_r \dd{V}\dd{t}\\
   =\int_{t_i}^{t_{i+1}}\int_{T_n}E_{pq}\tilde{Q}^{ni}_{qls}\psi_l\theta_s\psi_k\theta_r\dd{V}\dd{t}.
\end{multline*} 
We apply integration by parts in time to the first integral and map onto the reference element:
\begin{multline}
\label{eq:stp}
    \delta_{pq}\ips{\chi_r(1)\phi_k}{\chi_s(1)\phi_l}\tilde{Q}^{ni}_{qls} -
    \ips{\chi_r(0)\phi_k}{\phi_l}\tilde{Q}^{n,0}_{pl}
    - \delta_{pq} \ip{\derivative{\chi_r}{\tau
    }\phi_k}{\chi_s\phi_l}\tilde{Q}^{ni}_{qls}\\
    +\sum_{j=1}^3\mathcal{A}^{j*}_{pq}\ip{\chi_r\phi_k}{\chi_s\partial_j\phi_l}\tilde{Q}^{ni}_{qls}
    = E^*_{pq}\ip{\chi_r\phi_k}{\chi_s\phi_l}\tilde{Q}^{ni}_{qls}.
\end{multline}
Here, $\ips{\cdot}{\cdot}$ is a scalar product in space, whereas $\ip{\cdot}{\cdot}$ is a scalar product in space \textit{and} time:
\begin{equation*}
    \begin{aligned}
       \ips{f}{g} &:= \int_{\mathcal{E}} f(\xi, \eta, \zeta) g(\xi, \eta, \zeta) \dd{V(\xi, \eta, \zeta)} \\
       \ip{f}{g} &:= \int_0^1\int_{\mathcal{E}} f(\tau, \xi, \eta, \zeta) g(\tau, \xi, \eta, \zeta) \dd{V(\xi, \eta, \zeta)} \dd{\tau}. \\
    \end{aligned}
\end{equation*}
Note that we use $Q_{pl}^{n,0}$, which collects the spatial degrees only from the previous time step.
We introduce the source matrix $E^* = \Delta t E$ transformed to the reference element, and the Jacobian matrices $\mathcal{A}^{j*} = \Delta t\mathcal{A}^j$, now also scaled with the time step $\Delta t$.
Notice, that we can decompose most of the inner products, e.g.
\begin{equation*}
\begin{aligned}
    \ip{\chi_r\phi_k}{\chi_s\phi_l} &= \int_0^1 \int_\mathcal{E}\chi_r\phi_k\chi_s\phi_l\dd{V}\dd{\tau}
    = \int_0^1 \chi_r\chi_s \left( \int_\mathcal{E} \phi_k\phi_l\dd{V}\right)\dd{\tau} 
    = \int_0^1 \chi_r\chi_s\dd{\tau}\int_\mathcal{E}\phi_k\phi_l\dd{V}.
\end{aligned}
\end{equation*}
We define mass and stiffness matrices in time,
\begin{equation*}
\begin{aligned}
    W_{rs} &= \chi_r(1)\chi_s(1) \\
    w_{r} &= \chi_r(0) \\
    S_{rs} &= \int_{0}^{1} \chi_r\chi_s\dd{\tau} \\
    K^{\tau}_{rs} &= \int_{0}^{1} \derivative{\chi_r}{\tau}\chi_s\dd{\tau}, \\
\end{aligned}
\end{equation*}
and space,
\begin{equation*}
\begin{aligned}
    M_{kl} &= \int_{\mathcal{E}} \phi_k\phi_l\dd{V} \\
    K^{\alpha}_{kl} &= \int_{\mathcal{E}} \phi_k\derivative{\phi_l}{\alpha}\dd{V}, \\
\end{aligned}
\end{equation*}
where $\alpha \in \{\xi, \eta, \zeta\}$ or ${1, 2, 3}$.
We start by inserting our matrix definitions into \cref{eq:stp} and drop the superscript $ni$:
\begin{equation*}
\delta_{pq}W_{rs}M_{kl}\tilde{Q}_{qls} - \delta_{pq}w_rM_{kl}\tilde{Q}^0_{ql}
- \delta_{pq}K^{\tau}_{rs}M_{kl}\tilde{Q}_{qls}
+ \sum_{j=1}^3\mathcal{A}^{j*}_{pq}S_{rs}K_{kl}^{j}\tilde{Q}_{qls}
= E^*_{pq}S_{rs}M_{kl}\tilde{Q}_{qls}.
\end{equation*}
We now collect terms involving $\tilde{Q}_{qlk}$ on the left and the other part on the right:
\begin{equation}
\label{eq:stp-lineq}
    \left(\delta_{pq}W_{rs}M_{kl}
    - \delta_{pq}K^{\tau}_{rs}M_{kl}
    - E^*_{pq}S_{rs}M_{kl}
    + \sum_{j=1}^3\mathcal{A}^{j*}_{pq}S_{rs}K_{kl}^{j} \right)\tilde{Q}_{qls}
    = \delta_{pq}w_rM_{kl}\tilde{Q}^0_{ql}.
\end{equation}
The system of equations can now be stated in the form 
\begin{equation}
    \label{eq:linear_system}
    Y_{pkrqls} \tilde{Q}_{qls} = r_{pkr}.
\end{equation}
This system can be transformed into a matrix-vector form if we map the multi-indices to linear indices.
By solving this system, we get a predicted solution for $q_p(t_i+\delta t)$, which replaces \cref{eq:taylor}.
Again we can combine this with \cref{eq:semi_discrete}, integrate from $t_i$ to $t_{i+1}$ and obtain the solution at time $t_{i+1}$.

%% file: STP.tex
\section{A new efficient inversion of the system matrix}
\label{sec:stp}
In \cref{sec:space_time_pred} we introduce the ADER-DG method with a space-time predictor to effectively treat the stiff source term inherent to poroelastic wave propagation.
What remains is to solve a local system as given in \cref{eq:linear_system}.
We can write this linear system of equations in standard matrix form, $Ax=b$, by unrolling the multi-indices $pkr \rightarrow i$ and $qls \rightarrow j$.
In the approach by ~\citet{de_la_puente_discontinuous_2008}, the inverse of this matrix was precomputed for every element.
During the simulation phase, these systems were simply solved one after the other, which has two main disadvantages:
(1) The operator $Y$ contains information on the material parameters and the shape of the elements, thus, the operator differs for every element.
(2) Since, for example, \num{4368} unknowns are associated with polynomial degree \num{5}, each LU decomposition needs \SI{145.6}{\mega\byte} of storage per element. 
Even on large clusters, this easily poses a severe limitation.
Additionally, precomputing the decomposition demands substantial computational resources.

Here, we present a new, optimised solution approach that relies on a modified back-substitution. 
It does not require explicit unrolling of multi-indices, but makes use of the tensor structure of $\tilde{Q}$.
Thus, no decomposition has to be computed or stored.
In terms of floating-point operations and memory requirement needed for one back-substitution, we outperform an LU decomposition by far.
In addition, our scheme can be implemented using small matrix-matrix multiplications (GEMMs).
For these kinds of tensor operations, the code generator \texttt{YATeTo}~\cite{uphoff_yet_2020} can generate architecture-specific code to achieve high performance.

\subsection{Structure of the system matrix}
\label{sec:stp-structure}
\begin{figure}
    \centering
    \input{figures/sparsity/sparsity.pgf}
    \caption{Sparsity patterns of the system matrices of \cref{eq:linear_system} with basis functions of maximum degree $1$ and $2$, respectively.
    The white blocks contain only zeros.
    We identify $(N+1)$ blue blocks of increasing size at the diagonal.
    We further distinguish these blocks in a light and a dark blue.
    The light-blue blocks also contain only zeros.
    The dark-blue blocks are all upper block-triangular with blocks of size $(N+1) \times (N+1)$ along the diagonal.
    In total, there are $B$ dark-blue blocks, all of size $\mathcal{Q}(N+1) \times \mathcal{Q}(N+1)$.
    Finally, we recognise a re-occurring sparsity pattern in the green blocks. See the text for a detailed explanation on the origin of the blue and green block structure.
}
    \label{fig:sparsity}
\end{figure}
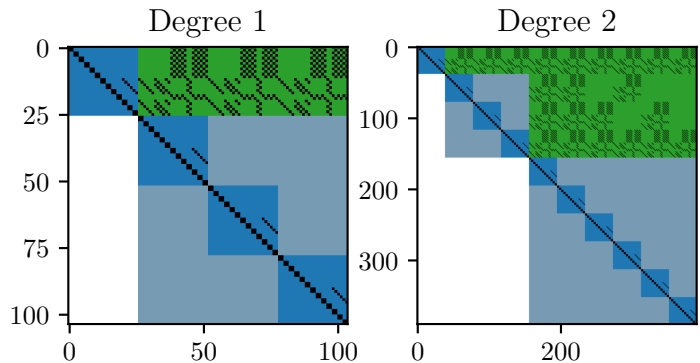

In \cref{eq:linear_system,eq:stp-lineq}, the indices $p$ and $q$ range from $1$ to $\mathcal{Q}$ (number of quantities), $r$ and $s$ range from $1$ to $N+1$ (number of temporal basis functions), and $k$ and $l$ range from $1$ to $B = B(N)$ (number of spatial basis functions with degrees up to $N$).
There are several ways to unroll the multi-indices $pkr \rightarrow i$ and $qls \rightarrow j$.
If we choose $k/l$ as the slowest and $r/s$ as the fastest-running indices, we observe a sparsity pattern and block-structure of the system matrix $A_{ij}$, as shown in \cref{fig:sparsity}.
Overall, the system \eqref{eq:linear_system} is in upper block-triangular form. 
The blocks on the diagonal hinder us from using standard row-wise back-substitution (such as the triangular solver from Level 2 BLAS~\cite{blackford_updated_2001}). 
Nonetheless, we can make use of this property to derive a block-wise back-substitution algorithm in \cref{sec:stp-backsubstitution}.

The block-triangular structure of the matrix stems from our choice of basis functions, which are orthogonal polynomials.
\begin{lem}
If the basis functions $\phi_l$ are orthogonal and ordered such that $\deg(\phi_k) \leq \deg(\phi_l)$ for $k \leq l$, the stiffness matrix $K^\alpha$ is upper triangular.
In particular, we see larger blocks of zeros: 
Let $B_n = \binom{n+3}{3}$. 
Then for our choice of basis functions, we have:
\begin{equation*}
\forall n \in [1,N+1] : \forall i \in (B_{n-1}, B_{n}] : \forall j \in [1, B_{n}]: K_{ij}^\alpha = 0
\end{equation*}
\end{lem}
\begin{proof}
The basis functions are numbered with increasing degree, where each degree adds $\binom{n+2}{2}$ basis functions.
That is, the first basis function is of degree 0, the next 3 basis functions have degree 1, the next 6 basis functions have degree 2, and so forth. 
Let $i \in (B_{n-1}, B_n]$ and $j \in [1, B_n]$.
Denote the degree of $\phi_i$ with $k_i$ and the degree of $\phi_j$ with $k_j$. 
Then $k_j \leq k_i$.
We obtain the stiffness matrices by an inner product $K_{ij}^\alpha= \ips{\phi_i}{\derivative{\phi_j}{\alpha}}$.
Taking the derivative of a basis function of degree $k_j$ yields a polynomial of degree $k_j-1$, which we can write as a linear combination of the basis functions up to degree $k_j-1$.
Since all basis functions of degree $k_i$ are orthogonal to the polynomials of degree up to $k_j-1$, it follows that $K^\alpha_{ij} = 0$
\end{proof}
Now, we can see how the blue and green blocks in \cref{fig:sparsity} arise from \cref{eq:stp-lineq}:
The overall shape of the green blocks resembles the sparsity pattern of the stiffness matrices.
Therefore, we identify the green block with the pattern $\sum_{j=1 }^3 \mathcal{A}_{pq}^{j*} S_{rs} K_{kl}^j$.
Within each block, we see a replication of the sparsity pattern of the Jacobian matrices ($\mathcal{A}^{j*}_{pq}$), where each entry on the diagonal is replaced by the $(N+1)\times(N+1)$ diagonal matrix $S_{rs}$.
The blue blocks on the diagonal resemble the other part of \cref{eq:stp-lineq}: $\left(\delta_{pq}W_{rs} - \delta_{pq}K^\tau_{rs} - E^*_{pq}S_{rs}\right) M_{kl}$.
The mass matrix $M_{kl}$ is diagonal, which explains the location of the dark-blue blocks and the existence of the light-blue blocks. 
Within each dark-blue block, we see smaller blocks of size $(N+1)\times(N+1)$ on the diagonal ($\delta_{pq}W_{rs} - \delta_{pq}K^\tau_{rs}$).
Above the diagonal (in particular on a side diagonal), we see the entries $E_{pq}S_{rs}$.
We can group the dark-blue blocks on the diagonal to $(N+1)$ blocks of size $(N+1) \mathcal{Q} \binom{n+2}{2} \times (N+1) \mathcal{Q} \binom{n+2}{2}$.
Each of these blocks corresponds to basis functions of a certain polynomial degree.

\subsection{Block-wise back-substitution}
\label{sec:stp-backsubstitution}
We continue with \cref{eq:stp-lineq} and derive the block-wise back-substitution procedure.
First, we multiply with $M_{mk}^{-1}S_{ur}^{-1}$:
\begin{multline*}
    \left(\delta_{pq}S_{ur}^{-1}W_{rs}\delta_{ml}
    - \delta_{pq}S_{ur}^{-1}K^{\tau}_{rs}\delta_{ml}
    - E^*_{pq}\delta_{us}\delta_{ml}
    + \sum_{j=1}^3\mathcal{A}^{j*}_{pq}\delta_{us}M_{mk}^{-1}K_{kl}^{j}
    \right)Q_{qls} \\
    = \delta_{pq}S_{ur}^{-1}w_r\delta_{ml}Q^0_{ql} = S_{ur}^{-1}w_rQ^0_{pm}.
\end{multline*}
Now, we move the parts containing the spatial stiffness matrices onto the right-hand side, since we know this part vanishes for some index combinations:
\begin{equation*}
\left(\delta_{pq}S_{ur}^{-1}W_{rs}\delta_{ml}
- \delta_{pq}S_{ur}^{-1}K^{\tau}_{rs}\delta_{ml}
- E^*_{pq}\delta_{us}\delta_{ml}\right)Q_{qls} 
= S_{ur}^{-1}w_rQ^0_{pm} - \left(\sum_{j=1}^3\mathcal{A}^{j*}_{pq}\delta_{us}M_{mk}^{-1}K_{kl}^{j}
\right)Q_{qls}.
\end{equation*}
Next, we can factor out some of the $\delta$ functions.
Furthermore, we introduce $\hat{K}^\alpha_{ml} = M_{mr}^{-1}K^\alpha_{rl}$, which renders the equations simpler:
\begin{equation*}
    \left(\delta_{pq}S_{ur}^{-1}W_{rs} - \delta_{pq}S_{ur}^{-1}K^{\tau}_{rs}
    - E^*_{pq}\delta_{us}\right)Q_{qms}=
    S_{ur}^{-1}w_rQ^0_{pm} - \left( 
      \sum_{j=1}^m \mathcal{A}^{j*}_{pq}\hat{K}_{ml}^{j}
    \right)Q_{qlu}.
\end{equation*}
Since the mass matrix is diagonal due to the choice of orthogonal basis functions, $\hat{K}^{\alpha}$ has the same sparsity pattern as $K^\alpha$.
We can write the sum over $l$ on the right-hand side explicitly now and neglect all parts which are $0$.
Then:
\begin{equation}
\label{eq:b_pmu}
b_{pmu} := S_{ur}^{-1}w_rQ^0_{pm}
- \sum_{l=m+1}^B\left( A^*_{pq}\hat{K}_{ml}^{\xi}
+ B^*_{pq}\hat{K}_{ml}^{\eta}
+ C^*_{pq}\hat{K}_{ml}^{\zeta}\right)Q_{qlu},
\end{equation}
Thus, $b_{pmu}$ only depends on $Q_{:l:}$ with $l > m$.
In particular, for $m=B$ we obtain
\begin{equation*}
b_{pBu} := S_{ur}^{-1}w_rQ^0_{pB}.
\end{equation*}
Recall that we want to solve the system for every $m$:
\begin{equation}\label{eq:system}
\left(\delta_{pq}S_{ur}^{-1}W_{rs} - \delta_{pq}S_{ur}^{-1}K^{\tau}_{rs} - E^*_{pq}\delta_{us}\right)Q_{qms}
= b_{pmu}.
\end{equation}
For $m=B$, the right-hand side $b$ does not depend on $Q$.
For $m < B$, the right-hand side $b$ depends on $Q_{:l:}$ with $l > m$.
Hence we can solve the system of equations for $Q_{:m:}$ backwards in the order $m = B, \dots, 1$ and update the right-hand side with the already computed values of $Q$, as shown in \cref{alg:back-substition-1}.
\begin{algorithm}
\label{alg:back-substition-1}
\SetAlgoLined
$b_{pBu} \gets S_{ur}^{-1}w_r Q_{pB}^0$\;
\For{$m \gets B$ \KwDownto $1$}{
    \tcp{dark-blue blocks}
    Solve $\left(\delta_{pq}S_{ur}^{-1}W_{rs} - \delta_{pq}S_{ur}^{-1}K^{\tau}_{rs} - E^*_{pq}\delta_{us}\right)Q_{qms} = b_{pmu}$\;
    \tcp{green blocks}
    Update $b_{p(m-1)u}$ using \cref{eq:b_pmu}\;
}
\caption{First simple block-wise back-substitution algorithm.}
\end{algorithm}
Here we still need to solve a system of size $\mathcal{Q}(N+1) \times \mathcal{Q}(N+1)$ for $B$ iterations.

We can further optimise the algorithm by using the sparsity pattern of $E^*$, which is upper triangular and repeat what we have done earlier.
We collect the matrices on the left-hand side of \cref{eq:system} further:
\begin{equation*}
\left(\delta_{pq}\underbrace{S_{ur}^{-1}\left(W_{rs}
- K^{\tau}_{rs}\right)}_{=:Z_{us}}
- E^*_{pq}\delta_{us}\right)Q_{qms}
= b_{pmu}.
\end{equation*}
Then we split $E^*$ in a diagonal and in a strictly upper triangular part, i.e. $E^* = F+G$, where $F = \diag(E^*)$, and $G=E^*-F$.
We put $G$ on the right-hand side and explicitly write the sum over $q$:
\begin{equation*}
\left(\delta_{pq}Z_{us} - F_{pq}\delta_{us}\right)Q_{qms}
= b_{pmu} + \sum_{o=p+1}^\mathcal{Q}G_{po}Q_{omu} =: \hat{b}_{pmu}.
\end{equation*}
Again, we see that the right-hand side does not depend on $Q$ for  $p=\mathcal{Q}$:
\begin{equation*}
\hat{b}_{\mathcal{Q}mu} = b_{\mathcal{Q}mu}.
\end{equation*}
$F$ is diagonal, i.e.\ $F_{pq} = E^*_{PP}\delta_{pq}$.
Note that the upper-case $P$ has the same value as the lower-case $p$ but no summation is implied.
We can further simplify the equations by pulling out $\delta_{pq}$ and obtain the final system that needs to be solved for all $p$ and $m$:
\begin{equation*}
\left(Z_{us} - E^{*}_{PP}\delta_{us}\right)Q_{pms} = \hat{b}_{pmu}.
\end{equation*}
In summary, we obtain \cref{alg:algorithm-2}, where $\cdot$ stands for a matrix multiplication and $\circ$ for a tensor product. 
In comparison to \cref{alg:back-substition-1}, we only have to solve $B\mathcal{Q}$ systems of size $(N+1)\times(N+1)$.
\begin{algorithm}
\label{alg:algorithm-2} 
\SetAlgoLined
$b \gets Q^0 \circ (S^{-1}w)$\;
\For{$m \gets B$ \KwDownto $1$}{
    \tcp{use sparsity pattern of dark-blue blocks}
    \For{$p \gets \mathcal{Q}$ \KwDownto $1$}{
        $Q_{pm:} \gets (Z-E^*_{pp}I)^{-1}\cdot b_{pm:}$\;
        \For{$o \gets 1$ \KwTo $p-1$}{
            $b_{om:} \gets b_{om:} + G_{op}\cdot Q_{pm:}$\;
        }
    }
    \tcp{green blocks}
    \For{$n \gets 1$ \KwTo $m-1$}{
        $b_{:n:} \gets b_{:n:} - \hat{K}_{nm}^\xi A^*\cdot Q_{:m:} - \hat{K}_{nm}^\eta B^*\cdot Q_{:m:} - \hat{K}_{nm}^\zeta C^*\cdot Q_{:m:}$\;
    }
}
\caption{Back-substitution algorithm unrolled over $m$ and $p$.}
\end{algorithm}

We can even further optimise the algorithm by using the internal structure of the blue and green blocks.
Until now, we have only used the block structure given by the dark-blue blocks, but we also see that the system matrix contains larger blocks of zeros (light-blue in \cref{fig:sparsity}).
Thus, for $m \leq B_n$ the right-hand side $b_{pmu}$ depends on $Q_{:l:}$ for $l \in (B_n, B]$.
We can use this information to construct \cref{alg:algorithm-3}, in which we fuse iterations within the loop over $m$.
\begin{algorithm}
\label{alg:algorithm-3}
\SetAlgoLined
$b \gets Q^0 \circ (S^{-1}w)$\;
\For{$n \gets N+1$ \KwDownto $1$}{
    \tcp{dark and light-blue blocks together}
    $m \gets (B_{n-1}, B_n]$\;
    \For{$p \gets \mathcal{Q}$ \KwDownto $1$}{
        $Q_{pm:} \gets b_{pm:}\cdot (Z-E^*_{pp}I)^{-T}$ \tcp*{$2d_n(N+1)^2$ flop}
        \For{$o \gets 1$ \KwTo $p-1$}{
            $b_{om:} \gets b_{om:} + G_{op}\cdot Q_{pm:}$ \tcp*{$2d_n(N+1)$ flop}
        }
    }
    \tcp{green blocks}
    \If{$n > 1$}{
        $b \gets b - A^*\times_1 Q_{:m:} \times_2 \hat{K}_{:m}^\xi$ \;
        $b \gets b - B^*\times_1 Q_{:m:} \times_2 \hat{K}_{:m}^\eta$\;
        $b \gets b - C^*\times_1 Q_{:m:} \times_2 \hat{K}_{:m}^\zeta$\;
        \tcp{each: $\mathcal{Q}B(N+1) + 2(N+1)\mathcal{Q}^2d_n + 2(N+1)\mathcal{Q}d_nB$ flop}
    }
}
\caption{Back-substitution algorithm with iterations fused over $m$ to better match the sparsity pattern. The number of floating-point operations for each tensor contraction is provided as comments. The product $\times_n$ denotes the $n$-mode product, c.f. \cite{kolda_tensor_2009}.}
\end{algorithm}

Next, we compare the number of floating-point operations needed to solve one system of equations using an LU decomposition and our \cref{alg:algorithm-3}.
As material parameters do not change over time, we can reuse the same decomposition in each time step.
The LU decomposition can be computed in advance for each element and stored. 
During the simulation, we have to perform back-substitution twice, thus, the number of floating-point operations is $2s^2$, where $s$ is the number of unknowns.
Now, consider the newly proposed block-wise back-substitution approach: Let $d_n = \binom{n+2}{2}$.
The number of floating-point operations for each tensor contraction is provided in \cref{alg:algorithm-3}. 
Now we only have to sum over the loops.
Let us note here that $G$ has $3$ non-zero entries, thus, we execute line $8$ only three times.
The number of floating-point operations for the $n^{th}$ execution of the outermost loop is bounded by 
\begin{multline*}
\underset{\texttt{line 6}}{\underbrace{\mathcal{Q} \cdot 2d_n(N+1)^2}}
+ \underset{\texttt{line 8}}{\underbrace{3\cdot 2d_n(N+1)}}
+ \underset{\texttt{lines 13-16}}{\underbrace{3 \left(\mathcal{Q}B(N+1) + 2 (N+1)\mathcal{Q}^2d_N + 2 (N+1)\mathcal{Q}d_NB \right)}} \\
= 2d_n \left( \mathcal{Q} (N+1)^2 + 3(N+1) + 3(N+1)\mathcal{Q}^2 + 3(N+1)\mathcal{Q}B \right) + 3 \mathcal{Q}B(N+1).
\end{multline*}
As $\sum_{n=1}^{N}d_n = B$ and the other terms are independent of $n$, we obtain
\begin{equation*}
2B \left( \mathcal{Q} (N+1)^2 + 3(N+1) + 3(N+1)\mathcal{Q}^2 + 3(N+1)\mathcal{Q}B \right) + 3 \mathcal{Q}B(N+1)^2 
\end{equation*}
as an upper bound for \cref{alg:algorithm-3}.
Comparison of the number of floating-point operations in \cref{tab:flop-count} shows significant speed-up comparing our \cref{alg:algorithm-3} to the LU decomposition for all considered orders of accuracy. 
The speed-up factor increases approximately linearly with the polynomial degree.
Specifically, we see a reduction of computational effort by a factor of $\approx 25$ for polynomial degree \num{6}.
Our \cref{alg:algorithm-3} outperforms LU decomposition also in terms of memory requirements.
For the LU decomposition, two triangular matrices of size $B\mathcal{Q}(N+1) \times B\mathcal{Q}(N+1)$ have to be stored.
For our back-substitution algorithm, we only need to store the matrices $(Z - E^*_{pp}I) ^{-1}$ (\num{13} matrices of size $(N+1)\times(N+1)$) and the matrices $A^*, B^*, C^*$ (each of size $\mathcal{Q} \times \mathcal{Q}$) and $E^*$ (\num{6} non-zero entries).
\begin{table}
  \centering
\pgfplotstabletypeset[
multicolumn names=l,
string type,
col sep=&,row sep=\\,
header=false,
every head row/.style={before row=\toprule,after row=\midrule},
every last row/.style={after row=\bottomrule},
display columns/0/.style={column name=N, column type={l}},
display columns/1/.style={column name=$2$, column type={S[table-format=1.2e2, round-mode=figures, round-precision=3, scientific-notation=true]}},
display columns/2/.style={column name=$3$, column type={S[table-format=1.2e2, round-mode=figures, round-precision=3, scientific-notation=true]}},
display columns/3/.style={column name=$4$, column type={S[table-format=1.2e2, round-mode=figures, round-precision=3, scientific-notation=true]}},
display columns/4/.style={column name=$5$, column type={S[table-format=1.2e2, round-mode=figures, round-precision=3, scientific-notation=true]}},
display columns/5/.style={column name=$6$, column type={S[table-format=1.2e2, round-mode=figures, round-precision=3, scientific-notation=true]}},
]{
\#unknowns & 390 & 1040 & 2275 & 4368 & 7644 \\
\hline
\#FLOP LU  & 304200 & 2163200 & 10351250 & 38158848 & 116861472 \\
\#FLOP STP &  59850 &  227200 &   713125 &  1941408 &   4719876 \\
reduction & 5.08270677 & 9.52112676 & 14.51533742 & 19.65524403 & 24.7594369 \\
\hline
storage LU [\si{\mega\byte}] & 1.1634063720703125 & 8.2598876953125 & 39.504241943359375 & 145.5977783203125 & 445.8494567871094\\
storage STP [\si{\mega\byte}] & 0.0048065185546875 & 0.00550079345703125 & 0.0063934326171875 & 0.00748443603515625 & 0.0087738037109375 \\
reduction & 242.04761905 & 1501.58113731 & 6178.87828162 & 19453.40672783 & 50815.9826087 \\
  }
  \caption{Comparison of the computational effort (absolute count of floating-point operations) and memory requirements for the standard LU decomposition and our newly proposed approach (STP, \cref{alg:algorithm-3}).}
  \label{tab:flop-count}
\end{table}

%% file: figures/sparsity/sparsity.pgf
\begingroup%
\makeatletter%
\begin{pgfpicture}%
\pgfpathrectangle{\pgfpointorigin}{\pgfqpoint{3.760556in}{2.066419in}}%
\pgfusepath{use as bounding box, clip}%
\begin{pgfscope}%
\pgfsetbuttcap%
\pgfsetmiterjoin%
\definecolor{currentfill}{rgb}{1.000000,1.000000,1.000000}%
\pgfsetfillcolor{currentfill}%
\pgfsetlinewidth{0.000000pt}%
\definecolor{currentstroke}{rgb}{1.000000,1.000000,1.000000}%
\pgfsetstrokecolor{currentstroke}%
\pgfsetdash{}{0pt}%
\pgfpathmoveto{\pgfqpoint{0.000000in}{0.000000in}}%
\pgfpathlineto{\pgfqpoint{3.760556in}{0.000000in}}%
\pgfpathlineto{\pgfqpoint{3.760556in}{2.066419in}}%
\pgfpathlineto{\pgfqpoint{0.000000in}{2.066419in}}%
\pgfpathclose%
\pgfusepath{fill}%
\end{pgfscope}%
\begin{pgfscope}%
\pgfsetbuttcap%
\pgfsetmiterjoin%
\definecolor{currentfill}{rgb}{1.000000,1.000000,1.000000}%
\pgfsetfillcolor{currentfill}%
\pgfsetlinewidth{0.000000pt}%
\definecolor{currentstroke}{rgb}{0.000000,0.000000,0.000000}%
\pgfsetstrokecolor{currentstroke}%
\pgfsetstrokeopacity{0.000000}%
\pgfsetdash{}{0pt}%
\pgfpathmoveto{\pgfqpoint{0.405556in}{0.320679in}}%
\pgfpathlineto{\pgfqpoint{1.852223in}{0.320679in}}%
\pgfpathlineto{\pgfqpoint{1.852223in}{1.767346in}}%
\pgfpathlineto{\pgfqpoint{0.405556in}{1.767346in}}%
\pgfpathclose%
\pgfusepath{fill}%
\end{pgfscope}%
\begin{pgfscope}%
\pgfpathrectangle{\pgfqpoint{0.405556in}{0.320679in}}{\pgfqpoint{1.446667in}{1.446667in}}%
\pgfusepath{clip}%
\pgfsys@transformshift{0.405556in}{0.320679in}%
\pgftext[left,bottom]{\includegraphics[interpolate=true,width=1.446667in,height=1.446667in]{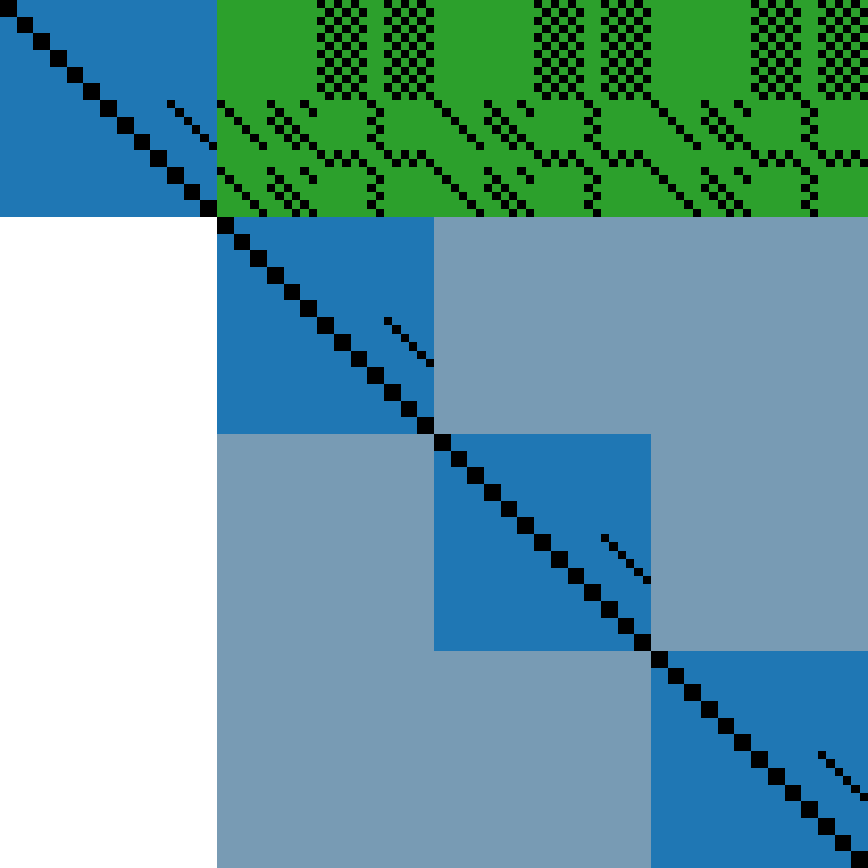}}%
\end{pgfscope}%
\begin{pgfscope}%
\pgfsetbuttcap%
\pgfsetroundjoin%
\definecolor{currentfill}{rgb}{0.000000,0.000000,0.000000}%
\pgfsetfillcolor{currentfill}%
\pgfsetlinewidth{0.803000pt}%
\definecolor{currentstroke}{rgb}{0.000000,0.000000,0.000000}%
\pgfsetstrokecolor{currentstroke}%
\pgfsetdash{}{0pt}%
\pgfsys@defobject{currentmarker}{\pgfqpoint{0.000000in}{-0.048611in}}{\pgfqpoint{0.000000in}{0.000000in}}{%
\pgfpathmoveto{\pgfqpoint{0.000000in}{0.000000in}}%
\pgfpathlineto{\pgfqpoint{0.000000in}{-0.048611in}}%
\pgfusepath{stroke,fill}%
}%
\begin{pgfscope}%
\pgfsys@transformshift{0.412511in}{0.320679in}%
\pgfsys@useobject{currentmarker}{}%
\end{pgfscope}%
\end{pgfscope}%
\begin{pgfscope}%
\definecolor{textcolor}{rgb}{0.000000,0.000000,0.000000}%
\pgfsetstrokecolor{textcolor}%
\pgfsetfillcolor{textcolor}%
\pgftext[x=0.412511in,y=0.223457in,,top]{\color{textcolor}\rmfamily\fontsize{10.000000}{12.000000}\selectfont \(\displaystyle {0}\)}%
\end{pgfscope}%
\begin{pgfscope}%
\pgfsetbuttcap%
\pgfsetroundjoin%
\definecolor{currentfill}{rgb}{0.000000,0.000000,0.000000}%
\pgfsetfillcolor{currentfill}%
\pgfsetlinewidth{0.803000pt}%
\definecolor{currentstroke}{rgb}{0.000000,0.000000,0.000000}%
\pgfsetstrokecolor{currentstroke}%
\pgfsetdash{}{0pt}%
\pgfsys@defobject{currentmarker}{\pgfqpoint{0.000000in}{-0.048611in}}{\pgfqpoint{0.000000in}{0.000000in}}{%
\pgfpathmoveto{\pgfqpoint{0.000000in}{0.000000in}}%
\pgfpathlineto{\pgfqpoint{0.000000in}{-0.048611in}}%
\pgfusepath{stroke,fill}%
}%
\begin{pgfscope}%
\pgfsys@transformshift{1.108024in}{0.320679in}%
\pgfsys@useobject{currentmarker}{}%
\end{pgfscope}%
\end{pgfscope}%
\begin{pgfscope}%
\definecolor{textcolor}{rgb}{0.000000,0.000000,0.000000}%
\pgfsetstrokecolor{textcolor}%
\pgfsetfillcolor{textcolor}%
\pgftext[x=1.108024in,y=0.223457in,,top]{\color{textcolor}\rmfamily\fontsize{10.000000}{12.000000}\selectfont \(\displaystyle {50}\)}%
\end{pgfscope}%
\begin{pgfscope}%
\pgfsetbuttcap%
\pgfsetroundjoin%
\definecolor{currentfill}{rgb}{0.000000,0.000000,0.000000}%
\pgfsetfillcolor{currentfill}%
\pgfsetlinewidth{0.803000pt}%
\definecolor{currentstroke}{rgb}{0.000000,0.000000,0.000000}%
\pgfsetstrokecolor{currentstroke}%
\pgfsetdash{}{0pt}%
\pgfsys@defobject{currentmarker}{\pgfqpoint{0.000000in}{-0.048611in}}{\pgfqpoint{0.000000in}{0.000000in}}{%
\pgfpathmoveto{\pgfqpoint{0.000000in}{0.000000in}}%
\pgfpathlineto{\pgfqpoint{0.000000in}{-0.048611in}}%
\pgfusepath{stroke,fill}%
}%
\begin{pgfscope}%
\pgfsys@transformshift{1.803537in}{0.320679in}%
\pgfsys@useobject{currentmarker}{}%
\end{pgfscope}%
\end{pgfscope}%
\begin{pgfscope}%
\definecolor{textcolor}{rgb}{0.000000,0.000000,0.000000}%
\pgfsetstrokecolor{textcolor}%
\pgfsetfillcolor{textcolor}%
\pgftext[x=1.803537in,y=0.223457in,,top]{\color{textcolor}\rmfamily\fontsize{10.000000}{12.000000}\selectfont \(\displaystyle {100}\)}%
\end{pgfscope}%
\begin{pgfscope}%
\pgfsetbuttcap%
\pgfsetroundjoin%
\definecolor{currentfill}{rgb}{0.000000,0.000000,0.000000}%
\pgfsetfillcolor{currentfill}%
\pgfsetlinewidth{0.803000pt}%
\definecolor{currentstroke}{rgb}{0.000000,0.000000,0.000000}%
\pgfsetstrokecolor{currentstroke}%
\pgfsetdash{}{0pt}%
\pgfsys@defobject{currentmarker}{\pgfqpoint{-0.048611in}{0.000000in}}{\pgfqpoint{-0.000000in}{0.000000in}}{%
\pgfpathmoveto{\pgfqpoint{-0.000000in}{0.000000in}}%
\pgfpathlineto{\pgfqpoint{-0.048611in}{0.000000in}}%
\pgfusepath{stroke,fill}%
}%
\begin{pgfscope}%
\pgfsys@transformshift{0.405556in}{1.760390in}%
\pgfsys@useobject{currentmarker}{}%
\end{pgfscope}%
\end{pgfscope}%
\begin{pgfscope}%
\definecolor{textcolor}{rgb}{0.000000,0.000000,0.000000}%
\pgfsetstrokecolor{textcolor}%
\pgfsetfillcolor{textcolor}%
\pgftext[x=0.238889in, y=1.712165in, left, base]{\color{textcolor}\rmfamily\fontsize{10.000000}{12.000000}\selectfont \(\displaystyle {0}\)}%
\end{pgfscope}%
\begin{pgfscope}%
\pgfsetbuttcap%
\pgfsetroundjoin%
\definecolor{currentfill}{rgb}{0.000000,0.000000,0.000000}%
\pgfsetfillcolor{currentfill}%
\pgfsetlinewidth{0.803000pt}%
\definecolor{currentstroke}{rgb}{0.000000,0.000000,0.000000}%
\pgfsetstrokecolor{currentstroke}%
\pgfsetdash{}{0pt}%
\pgfsys@defobject{currentmarker}{\pgfqpoint{-0.048611in}{0.000000in}}{\pgfqpoint{-0.000000in}{0.000000in}}{%
\pgfpathmoveto{\pgfqpoint{-0.000000in}{0.000000in}}%
\pgfpathlineto{\pgfqpoint{-0.048611in}{0.000000in}}%
\pgfusepath{stroke,fill}%
}%
\begin{pgfscope}%
\pgfsys@transformshift{0.405556in}{1.412634in}%
\pgfsys@useobject{currentmarker}{}%
\end{pgfscope}%
\end{pgfscope}%
\begin{pgfscope}%
\definecolor{textcolor}{rgb}{0.000000,0.000000,0.000000}%
\pgfsetstrokecolor{textcolor}%
\pgfsetfillcolor{textcolor}%
\pgftext[x=0.169445in, y=1.364409in, left, base]{\color{textcolor}\rmfamily\fontsize{10.000000}{12.000000}\selectfont \(\displaystyle {25}\)}%
\end{pgfscope}%
\begin{pgfscope}%
\pgfsetbuttcap%
\pgfsetroundjoin%
\definecolor{currentfill}{rgb}{0.000000,0.000000,0.000000}%
\pgfsetfillcolor{currentfill}%
\pgfsetlinewidth{0.803000pt}%
\definecolor{currentstroke}{rgb}{0.000000,0.000000,0.000000}%
\pgfsetstrokecolor{currentstroke}%
\pgfsetdash{}{0pt}%
\pgfsys@defobject{currentmarker}{\pgfqpoint{-0.048611in}{0.000000in}}{\pgfqpoint{-0.000000in}{0.000000in}}{%
\pgfpathmoveto{\pgfqpoint{-0.000000in}{0.000000in}}%
\pgfpathlineto{\pgfqpoint{-0.048611in}{0.000000in}}%
\pgfusepath{stroke,fill}%
}%
\begin{pgfscope}%
\pgfsys@transformshift{0.405556in}{1.064878in}%
\pgfsys@useobject{currentmarker}{}%
\end{pgfscope}%
\end{pgfscope}%
\begin{pgfscope}%
\definecolor{textcolor}{rgb}{0.000000,0.000000,0.000000}%
\pgfsetstrokecolor{textcolor}%
\pgfsetfillcolor{textcolor}%
\pgftext[x=0.169445in, y=1.016652in, left, base]{\color{textcolor}\rmfamily\fontsize{10.000000}{12.000000}\selectfont \(\displaystyle {50}\)}%
\end{pgfscope}%
\begin{pgfscope}%
\pgfsetbuttcap%
\pgfsetroundjoin%
\definecolor{currentfill}{rgb}{0.000000,0.000000,0.000000}%
\pgfsetfillcolor{currentfill}%
\pgfsetlinewidth{0.803000pt}%
\definecolor{currentstroke}{rgb}{0.000000,0.000000,0.000000}%
\pgfsetstrokecolor{currentstroke}%
\pgfsetdash{}{0pt}%
\pgfsys@defobject{currentmarker}{\pgfqpoint{-0.048611in}{0.000000in}}{\pgfqpoint{-0.000000in}{0.000000in}}{%
\pgfpathmoveto{\pgfqpoint{-0.000000in}{0.000000in}}%
\pgfpathlineto{\pgfqpoint{-0.048611in}{0.000000in}}%
\pgfusepath{stroke,fill}%
}%
\begin{pgfscope}%
\pgfsys@transformshift{0.405556in}{0.717121in}%
\pgfsys@useobject{currentmarker}{}%
\end{pgfscope}%
\end{pgfscope}%
\begin{pgfscope}%
\definecolor{textcolor}{rgb}{0.000000,0.000000,0.000000}%
\pgfsetstrokecolor{textcolor}%
\pgfsetfillcolor{textcolor}%
\pgftext[x=0.169445in, y=0.668896in, left, base]{\color{textcolor}\rmfamily\fontsize{10.000000}{12.000000}\selectfont \(\displaystyle {75}\)}%
\end{pgfscope}%
\begin{pgfscope}%
\pgfsetbuttcap%
\pgfsetroundjoin%
\definecolor{currentfill}{rgb}{0.000000,0.000000,0.000000}%
\pgfsetfillcolor{currentfill}%
\pgfsetlinewidth{0.803000pt}%
\definecolor{currentstroke}{rgb}{0.000000,0.000000,0.000000}%
\pgfsetstrokecolor{currentstroke}%
\pgfsetdash{}{0pt}%
\pgfsys@defobject{currentmarker}{\pgfqpoint{-0.048611in}{0.000000in}}{\pgfqpoint{-0.000000in}{0.000000in}}{%
\pgfpathmoveto{\pgfqpoint{-0.000000in}{0.000000in}}%
\pgfpathlineto{\pgfqpoint{-0.048611in}{0.000000in}}%
\pgfusepath{stroke,fill}%
}%
\begin{pgfscope}%
\pgfsys@transformshift{0.405556in}{0.369365in}%
\pgfsys@useobject{currentmarker}{}%
\end{pgfscope}%
\end{pgfscope}%
\begin{pgfscope}%
\definecolor{textcolor}{rgb}{0.000000,0.000000,0.000000}%
\pgfsetstrokecolor{textcolor}%
\pgfsetfillcolor{textcolor}%
\pgftext[x=0.100000in, y=0.321140in, left, base]{\color{textcolor}\rmfamily\fontsize{10.000000}{12.000000}\selectfont \(\displaystyle {100}\)}%
\end{pgfscope}%
\begin{pgfscope}%
\pgfsetrectcap%
\pgfsetmiterjoin%
\pgfsetlinewidth{0.803000pt}%
\definecolor{currentstroke}{rgb}{0.000000,0.000000,0.000000}%
\pgfsetstrokecolor{currentstroke}%
\pgfsetdash{}{0pt}%
\pgfpathmoveto{\pgfqpoint{0.405556in}{0.320679in}}%
\pgfpathlineto{\pgfqpoint{0.405556in}{1.767346in}}%
\pgfusepath{stroke}%
\end{pgfscope}%
\begin{pgfscope}%
\pgfsetrectcap%
\pgfsetmiterjoin%
\pgfsetlinewidth{0.803000pt}%
\definecolor{currentstroke}{rgb}{0.000000,0.000000,0.000000}%
\pgfsetstrokecolor{currentstroke}%
\pgfsetdash{}{0pt}%
\pgfpathmoveto{\pgfqpoint{1.852223in}{0.320679in}}%
\pgfpathlineto{\pgfqpoint{1.852223in}{1.767346in}}%
\pgfusepath{stroke}%
\end{pgfscope}%
\begin{pgfscope}%
\pgfsetrectcap%
\pgfsetmiterjoin%
\pgfsetlinewidth{0.803000pt}%
\definecolor{currentstroke}{rgb}{0.000000,0.000000,0.000000}%
\pgfsetstrokecolor{currentstroke}%
\pgfsetdash{}{0pt}%
\pgfpathmoveto{\pgfqpoint{0.405556in}{0.320679in}}%
\pgfpathlineto{\pgfqpoint{1.852223in}{0.320679in}}%
\pgfusepath{stroke}%
\end{pgfscope}%
\begin{pgfscope}%
\pgfsetrectcap%
\pgfsetmiterjoin%
\pgfsetlinewidth{0.803000pt}%
\definecolor{currentstroke}{rgb}{0.000000,0.000000,0.000000}%
\pgfsetstrokecolor{currentstroke}%
\pgfsetdash{}{0pt}%
\pgfpathmoveto{\pgfqpoint{0.405556in}{1.767346in}}%
\pgfpathlineto{\pgfqpoint{1.852223in}{1.767346in}}%
\pgfusepath{stroke}%
\end{pgfscope}%
\begin{pgfscope}%
\definecolor{textcolor}{rgb}{0.000000,0.000000,0.000000}%
\pgfsetstrokecolor{textcolor}%
\pgfsetfillcolor{textcolor}%
\pgftext[x=1.128890in,y=1.850679in,,base]{\color{textcolor}\rmfamily\fontsize{12.000000}{14.400000}\selectfont Degree 1}%
\end{pgfscope}%
\begin{pgfscope}%
\pgfsetbuttcap%
\pgfsetmiterjoin%
\definecolor{currentfill}{rgb}{1.000000,1.000000,1.000000}%
\pgfsetfillcolor{currentfill}%
\pgfsetlinewidth{0.000000pt}%
\definecolor{currentstroke}{rgb}{0.000000,0.000000,0.000000}%
\pgfsetstrokecolor{currentstroke}%
\pgfsetstrokeopacity{0.000000}%
\pgfsetdash{}{0pt}%
\pgfpathmoveto{\pgfqpoint{2.213890in}{0.320679in}}%
\pgfpathlineto{\pgfqpoint{3.660556in}{0.320679in}}%
\pgfpathlineto{\pgfqpoint{3.660556in}{1.767346in}}%
\pgfpathlineto{\pgfqpoint{2.213890in}{1.767346in}}%
\pgfpathclose%
\pgfusepath{fill}%
\end{pgfscope}%
\begin{pgfscope}%
\pgfpathrectangle{\pgfqpoint{2.213890in}{0.320679in}}{\pgfqpoint{1.446667in}{1.446667in}}%
\pgfusepath{clip}%
\pgfsys@transformshift{2.213890in}{0.320679in}%
\pgftext[left,bottom]{\includegraphics[interpolate=true,width=1.446667in,height=1.446667in]{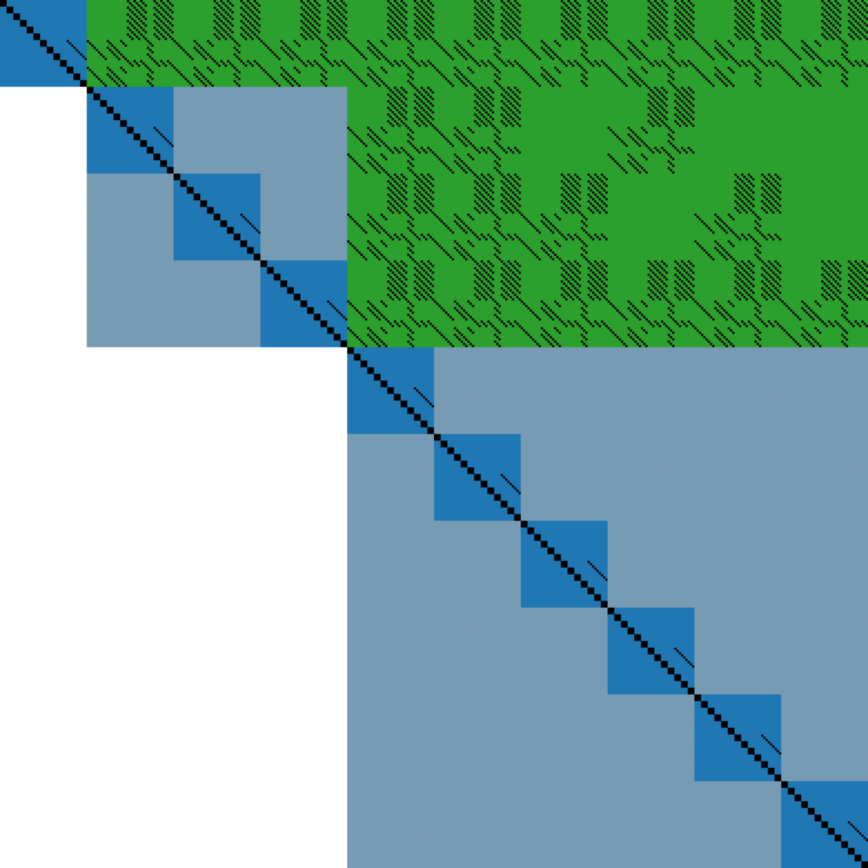}}%
\end{pgfscope}%
\begin{pgfscope}%
\pgfsetbuttcap%
\pgfsetroundjoin%
\definecolor{currentfill}{rgb}{0.000000,0.000000,0.000000}%
\pgfsetfillcolor{currentfill}%
\pgfsetlinewidth{0.803000pt}%
\definecolor{currentstroke}{rgb}{0.000000,0.000000,0.000000}%
\pgfsetstrokecolor{currentstroke}%
\pgfsetdash{}{0pt}%
\pgfsys@defobject{currentmarker}{\pgfqpoint{0.000000in}{-0.048611in}}{\pgfqpoint{0.000000in}{0.000000in}}{%
\pgfpathmoveto{\pgfqpoint{0.000000in}{0.000000in}}%
\pgfpathlineto{\pgfqpoint{0.000000in}{-0.048611in}}%
\pgfusepath{stroke,fill}%
}%
\begin{pgfscope}%
\pgfsys@transformshift{2.215744in}{0.320679in}%
\pgfsys@useobject{currentmarker}{}%
\end{pgfscope}%
\end{pgfscope}%
\begin{pgfscope}%
\definecolor{textcolor}{rgb}{0.000000,0.000000,0.000000}%
\pgfsetstrokecolor{textcolor}%
\pgfsetfillcolor{textcolor}%
\pgftext[x=2.215744in,y=0.223457in,,top]{\color{textcolor}\rmfamily\fontsize{10.000000}{12.000000}\selectfont \(\displaystyle {0}\)}%
\end{pgfscope}%
\begin{pgfscope}%
\pgfsetbuttcap%
\pgfsetroundjoin%
\definecolor{currentfill}{rgb}{0.000000,0.000000,0.000000}%
\pgfsetfillcolor{currentfill}%
\pgfsetlinewidth{0.803000pt}%
\definecolor{currentstroke}{rgb}{0.000000,0.000000,0.000000}%
\pgfsetstrokecolor{currentstroke}%
\pgfsetdash{}{0pt}%
\pgfsys@defobject{currentmarker}{\pgfqpoint{0.000000in}{-0.048611in}}{\pgfqpoint{0.000000in}{0.000000in}}{%
\pgfpathmoveto{\pgfqpoint{0.000000in}{0.000000in}}%
\pgfpathlineto{\pgfqpoint{0.000000in}{-0.048611in}}%
\pgfusepath{stroke,fill}%
}%
\begin{pgfscope}%
\pgfsys@transformshift{2.957625in}{0.320679in}%
\pgfsys@useobject{currentmarker}{}%
\end{pgfscope}%
\end{pgfscope}%
\begin{pgfscope}%
\definecolor{textcolor}{rgb}{0.000000,0.000000,0.000000}%
\pgfsetstrokecolor{textcolor}%
\pgfsetfillcolor{textcolor}%
\pgftext[x=2.957625in,y=0.223457in,,top]{\color{textcolor}\rmfamily\fontsize{10.000000}{12.000000}\selectfont \(\displaystyle {200}\)}%
\end{pgfscope}%
\begin{pgfscope}%
\pgfsetbuttcap%
\pgfsetroundjoin%
\definecolor{currentfill}{rgb}{0.000000,0.000000,0.000000}%
\pgfsetfillcolor{currentfill}%
\pgfsetlinewidth{0.803000pt}%
\definecolor{currentstroke}{rgb}{0.000000,0.000000,0.000000}%
\pgfsetstrokecolor{currentstroke}%
\pgfsetdash{}{0pt}%
\pgfsys@defobject{currentmarker}{\pgfqpoint{-0.048611in}{0.000000in}}{\pgfqpoint{-0.000000in}{0.000000in}}{%
\pgfpathmoveto{\pgfqpoint{-0.000000in}{0.000000in}}%
\pgfpathlineto{\pgfqpoint{-0.048611in}{0.000000in}}%
\pgfusepath{stroke,fill}%
}%
\begin{pgfscope}%
\pgfsys@transformshift{2.213890in}{1.765491in}%
\pgfsys@useobject{currentmarker}{}%
\end{pgfscope}%
\end{pgfscope}%
\begin{pgfscope}%
\definecolor{textcolor}{rgb}{0.000000,0.000000,0.000000}%
\pgfsetstrokecolor{textcolor}%
\pgfsetfillcolor{textcolor}%
\pgftext[x=2.047223in, y=1.717266in, left, base]{\color{textcolor}\rmfamily\fontsize{10.000000}{12.000000}\selectfont \(\displaystyle {0}\)}%
\end{pgfscope}%
\begin{pgfscope}%
\pgfsetbuttcap%
\pgfsetroundjoin%
\definecolor{currentfill}{rgb}{0.000000,0.000000,0.000000}%
\pgfsetfillcolor{currentfill}%
\pgfsetlinewidth{0.803000pt}%
\definecolor{currentstroke}{rgb}{0.000000,0.000000,0.000000}%
\pgfsetstrokecolor{currentstroke}%
\pgfsetdash{}{0pt}%
\pgfsys@defobject{currentmarker}{\pgfqpoint{-0.048611in}{0.000000in}}{\pgfqpoint{-0.000000in}{0.000000in}}{%
\pgfpathmoveto{\pgfqpoint{-0.000000in}{0.000000in}}%
\pgfpathlineto{\pgfqpoint{-0.048611in}{0.000000in}}%
\pgfusepath{stroke,fill}%
}%
\begin{pgfscope}%
\pgfsys@transformshift{2.213890in}{1.394551in}%
\pgfsys@useobject{currentmarker}{}%
\end{pgfscope}%
\end{pgfscope}%
\begin{pgfscope}%
\definecolor{textcolor}{rgb}{0.000000,0.000000,0.000000}%
\pgfsetstrokecolor{textcolor}%
\pgfsetfillcolor{textcolor}%
\pgftext[x=1.908333in, y=1.346325in, left, base]{\color{textcolor}\rmfamily\fontsize{10.000000}{12.000000}\selectfont \(\displaystyle {100}\)}%
\end{pgfscope}%
\begin{pgfscope}%
\pgfsetbuttcap%
\pgfsetroundjoin%
\definecolor{currentfill}{rgb}{0.000000,0.000000,0.000000}%
\pgfsetfillcolor{currentfill}%
\pgfsetlinewidth{0.803000pt}%
\definecolor{currentstroke}{rgb}{0.000000,0.000000,0.000000}%
\pgfsetstrokecolor{currentstroke}%
\pgfsetdash{}{0pt}%
\pgfsys@defobject{currentmarker}{\pgfqpoint{-0.048611in}{0.000000in}}{\pgfqpoint{-0.000000in}{0.000000in}}{%
\pgfpathmoveto{\pgfqpoint{-0.000000in}{0.000000in}}%
\pgfpathlineto{\pgfqpoint{-0.048611in}{0.000000in}}%
\pgfusepath{stroke,fill}%
}%
\begin{pgfscope}%
\pgfsys@transformshift{2.213890in}{1.023611in}%
\pgfsys@useobject{currentmarker}{}%
\end{pgfscope}%
\end{pgfscope}%
\begin{pgfscope}%
\definecolor{textcolor}{rgb}{0.000000,0.000000,0.000000}%
\pgfsetstrokecolor{textcolor}%
\pgfsetfillcolor{textcolor}%
\pgftext[x=1.908333in, y=0.975385in, left, base]{\color{textcolor}\rmfamily\fontsize{10.000000}{12.000000}\selectfont \(\displaystyle {200}\)}%
\end{pgfscope}%
\begin{pgfscope}%
\pgfsetbuttcap%
\pgfsetroundjoin%
\definecolor{currentfill}{rgb}{0.000000,0.000000,0.000000}%
\pgfsetfillcolor{currentfill}%
\pgfsetlinewidth{0.803000pt}%
\definecolor{currentstroke}{rgb}{0.000000,0.000000,0.000000}%
\pgfsetstrokecolor{currentstroke}%
\pgfsetdash{}{0pt}%
\pgfsys@defobject{currentmarker}{\pgfqpoint{-0.048611in}{0.000000in}}{\pgfqpoint{-0.000000in}{0.000000in}}{%
\pgfpathmoveto{\pgfqpoint{-0.000000in}{0.000000in}}%
\pgfpathlineto{\pgfqpoint{-0.048611in}{0.000000in}}%
\pgfusepath{stroke,fill}%
}%
\begin{pgfscope}%
\pgfsys@transformshift{2.213890in}{0.652670in}%
\pgfsys@useobject{currentmarker}{}%
\end{pgfscope}%
\end{pgfscope}%
\begin{pgfscope}%
\definecolor{textcolor}{rgb}{0.000000,0.000000,0.000000}%
\pgfsetstrokecolor{textcolor}%
\pgfsetfillcolor{textcolor}%
\pgftext[x=1.908333in, y=0.604445in, left, base]{\color{textcolor}\rmfamily\fontsize{10.000000}{12.000000}\selectfont \(\displaystyle {300}\)}%
\end{pgfscope}%
\begin{pgfscope}%
\pgfsetrectcap%
\pgfsetmiterjoin%
\pgfsetlinewidth{0.803000pt}%
\definecolor{currentstroke}{rgb}{0.000000,0.000000,0.000000}%
\pgfsetstrokecolor{currentstroke}%
\pgfsetdash{}{0pt}%
\pgfpathmoveto{\pgfqpoint{2.213890in}{0.320679in}}%
\pgfpathlineto{\pgfqpoint{2.213890in}{1.767346in}}%
\pgfusepath{stroke}%
\end{pgfscope}%
\begin{pgfscope}%
\pgfsetrectcap%
\pgfsetmiterjoin%
\pgfsetlinewidth{0.803000pt}%
\definecolor{currentstroke}{rgb}{0.000000,0.000000,0.000000}%
\pgfsetstrokecolor{currentstroke}%
\pgfsetdash{}{0pt}%
\pgfpathmoveto{\pgfqpoint{3.660556in}{0.320679in}}%
\pgfpathlineto{\pgfqpoint{3.660556in}{1.767346in}}%
\pgfusepath{stroke}%
\end{pgfscope}%
\begin{pgfscope}%
\pgfsetrectcap%
\pgfsetmiterjoin%
\pgfsetlinewidth{0.803000pt}%
\definecolor{currentstroke}{rgb}{0.000000,0.000000,0.000000}%
\pgfsetstrokecolor{currentstroke}%
\pgfsetdash{}{0pt}%
\pgfpathmoveto{\pgfqpoint{2.213890in}{0.320679in}}%
\pgfpathlineto{\pgfqpoint{3.660556in}{0.320679in}}%
\pgfusepath{stroke}%
\end{pgfscope}%
\begin{pgfscope}%
\pgfsetrectcap%
\pgfsetmiterjoin%
\pgfsetlinewidth{0.803000pt}%
\definecolor{currentstroke}{rgb}{0.000000,0.000000,0.000000}%
\pgfsetstrokecolor{currentstroke}%
\pgfsetdash{}{0pt}%
\pgfpathmoveto{\pgfqpoint{2.213890in}{1.767346in}}%
\pgfpathlineto{\pgfqpoint{3.660556in}{1.767346in}}%
\pgfusepath{stroke}%
\end{pgfscope}%
\begin{pgfscope}%
\definecolor{textcolor}{rgb}{0.000000,0.000000,0.000000}%
\pgfsetstrokecolor{textcolor}%
\pgfsetfillcolor{textcolor}%
\pgftext[x=2.937223in,y=1.850679in,,base]{\color{textcolor}\rmfamily\fontsize{12.000000}{14.400000}\selectfont Degree 2}%
\end{pgfscope}%
\end{pgfpicture}%
\makeatother%
\endgroup%

%% file: Benchmarks.tex
\section{Verification}
\label{sec:verification}
\label{sec:benchmarks}
To verify the implementation of the block-wise back-substitution approach (\cref{alg:algorithm-3} in \cref{sec:space_time_pred}) in SeisSol, we perform a series of numerical verification tests with canonical models of different complexity:
\begin{enumerate}
    \item  We impose a planar wave as initial condition and let it evolve over time. 
    On a cascade of finer and finer meshes, we verify convergence of all $13$ unknowns (stresses, solid particle velocities, pore pressure, relative fluid velocities) against an analytical reference solution.
    \item We consider an explosive point source in a homogeneous full space with either a viscous or an inviscid fluid.
    We compare the time histories of the solid particle velocities and relative fluid velocities at selected receiver positions using the semi-analytical solution by~\citet{karpfinger_greens_2009}.
    \item We consider a contact of two half-spaces with an inviscid fluid to assess how well SeisSol resolves reflection and transmission of waves at internal interfaces.
    We compare the solid particle velocities to the semi-analytical solution by~\citet{diaz_analytical_2008}.
    \item We verify the reflection of waves at a free surface of a half-space filled with an inviscid fluid.
    Again, the reference for the solid particle velocities is given by~\cite{diaz_analytical_2008}.
    \item As final verification and demonstration example, we define a new layer over half-space problem. For isotropic elastic materials, the layer over half-space community benchmark is widely accepted (c.f.~\cite{moczo_comparison_2006}).
    Here, we combine verification setups 2, 3 and 4 to simulate wave propagation with a free surface and an internal interface in a poroelastic medium filled with a viscous fluid.
    To the best of our knowledge, there is no analytical or semi-analytical solution available for such model configurations.
    Therefore, we compare our solution against an independent numerical simulation using the FD method~\cite{moczo_discrete_2019, gregor_subcell-resolution_2021}.
\end{enumerate}
The first three verification setups follow~\cite{de_la_puente_discontinuous_2008}, the two last test cases are inspired by similar 2D examples in~\cite{gregor_subcell-resolution_2021}.
\subsection{Planar wave convergence analysis}
\label{sec:convergence}
We consider a cube $[-1,1]^3$ with periodic boundary conditions.
The cube is refined into $4^3$, $8^3$, $16^3$, $32^3$ and $64^3$ equally sized sub-cubes and every cube is partitioned into $5$ tetrahedrons.
We express the reference solution in the form of a planar wave: 
\begin{equation*}
    q(x,t) = \sum_{n=1}^{13} \alpha_n r_n e^{i (\omega_n  t - k \cdot x)}.
\end{equation*}
Here $\omega_n, r_n$ are eigenpairs of the matrix $k_x A + k_y B + k_z C - i E$ and $i$ denotes the imaginary unit.
It is easy to verify that the real part of $\hat{Q}$ is a solution of \cref{eq:poroelastic-wave}.
Each eigenpair corresponds to a wave mode.
The vector $k = \left(\pi,\, \pi,\, \pi\right)$ describes the direction in which the waves propagate.
The scaling factors $\alpha_n$ define the relative amplitudes of these wave modes.
We set: $\alpha_1, \alpha_{2}, \alpha_{3}, \alpha_{4} = 100.0$ and all others to $0$.
The first mode corresponds to a fast P-wave, the second and third modes correspond to different polarisations of the S-wave, and the fourth mode is the slow P-wave.
The material parameters are given in \cref{tab:material_convergence}.
As an initial condition we set $\hat{Q}(x, 0)$.
The final time of the simulation is \SI{1e-4}{\second}.
The fast P-wave has a velocity of \SI{2715.6}{\meter \per \second}, thus, the wave travels \SI{0.27}{\meter} during the simulation time.
We compare the simulation result at the final time step of our simulation with the analytic reference solution using the $L^1$, $L^2$ and $L^\infty$ norms within the cube $[-1,1]^3$.
\begin{table}
  \centering
  \caption{Material parameters used for the planar wave convergence analysis.}
  \label{tab:material_convergence}
  \pgfplotstabletypeset[
    multicolumn names=l,
    string type,
    col sep=&,row sep=\\,
    header=false,
    every head row/.style={before row=\toprule,after row=\midrule},
    every last row/.style={after row=\bottomrule},
    display columns/0/.style={column name=Parameter, column type={l}},
    display columns/1/.style={column name=, column type={l}},
    display columns/2/.style={column name=Value, column type={S[table-format=1.2e2]}},
    display columns/3/.style={column name=, column type={l}},
   ]{
    Solid Bulk modulus                & {$K_S$}       &  4.00e10  & \si{\Pa}\\
    Solid density                     & {$\rho_S$}    &  2.50e3   & \si{\kg \per \meter \tothe{3}}\\
    Matrix $1^{st}$ Lam\'e parameter  & {$\lambda_M$} &  1.20e10  & \si{\Pa}\\
    Matrix $2^{nd}$ Lam\'e parameter  & {$\mu_M$}     &  1.00e10  & \si{\Pa}\\
    Matrix porosity                   & {$\phi$}      &  0.20     & \\
    Matrix permeability               & {$\kappa$}    &  6.00e-13 & \si{\meter \tothe{2}}\\
    Matrix tortuosity                 & {$T$}         &  3.00     & \\
    Fluid bulk modulus                & {$K_F$}       &  2.5e9    & \si{\Pa}\\
    Fluid density                     & {$\rho_F$}    &  1.04e3   & \si{\kg \per \meter \tothe{3}}\\
    Fluid viscosity                   & {$\nu$}       &  1.00e-3  & \si{\Pa \second} \\
  }
\end{table}

\begin{figure}
  \center{
    \input{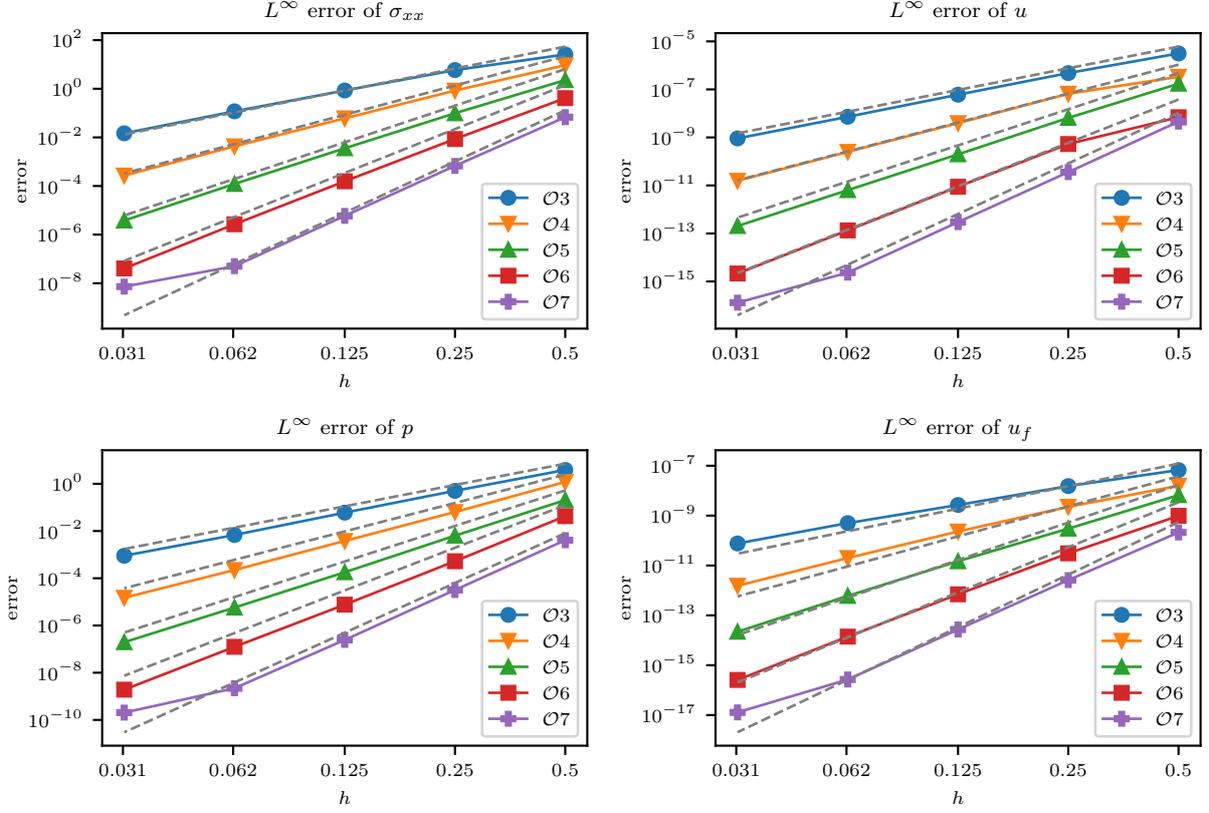}
  }
  \caption{Convergence plots for selected unknowns of the planar wave convergence setup in the $L^\infty$ norm. The expected convergence order is plotted in grey dashed lines. For $\mathcal{O}7$ we are close to machine precision on the finest mesh.}
  \label{fig:conv}
\end{figure}

In \cref{fig:conv} we show the dependence of the $L^\infty$ error between simulated results and analytic solution on the characteristic edge length $h$ for stress $\sigma_{xx}$, solid particle velocity $u$, pore pressure $p$ and relative fluid velocity $u_f$.
\cref{tab:mesh_spacing} summarises the characteristic element edge length $h$ and the number of elements for the considered meshes.
When we use polynomial basis functions up to degree $N$, we expect convergence of the order $\mathcal{O} = N+1$~\cite{gassner_explicit_2011}.
In \cref{fig:conv}, we clearly see the expected convergence behaviour for the different choices of basis functions.
For convergence order $7$ and the finest mesh, the convergence slows down, since we reach machine precision for the solid particle velocities and the relative fluid velocities.
In other norms and for other quantities we find the same expected convergence behaviour, see \cref{sec:convergence_other_norms}.
\begin{table}
  \centering
  \caption{Characteristic length in meter and number of elements for the different meshes used in the planar wave convergence analysis.}
  \label{tab:mesh_spacing}
  \pgfplotstabletypeset[
    multicolumn names,
    string type,
    col sep=&,row sep=\\,
    header=false,
    every head row/.style={before row=\toprule,after row=\midrule},
    every last row/.style={after row=\bottomrule},
    display columns/0/.style={column name=$N$, column type={c}},
    display columns/1/.style={column name=$4$, column type={l}},
    display columns/2/.style={column name=$8$, column type={l}},
    display columns/3/.style={column name=$16$, column type={l}},
    display columns/4/.style={column name=$32$, column type={l}},
    display columns/5/.style={column name=$64$, column type={l}},
   ]{
   {$h$}       & 0.5 & 0.25  & 0.125   & 0.0625  & 0.03125 \\
   \#elem & 320 & 2560  & 20480   & 163840  & 1310720 \\
  }
\end{table}


\subsection{Homogeneous full-space}
\label{sec:homogeneous-fullspace}
Next, we verify the wave propagation excited by a point source by modifying a similar problem setup in~\cite{de_la_puente_discontinuous_2008}.
We consider a homogeneous full-space with an explosive point source at the origin.
The time history is a Ricker wavelet with dominant frequency $f_0 = \SI{16}{\Hz}$ and time delay $t_0 = \SI{0.07}{\second}$:
\begin{equation*}
    s(t) = \left(1-2\left(\pi f_0\left(t-t_0\right)\right)^2\right) \cdot \exp\left(-\left(\pi f_0\left(t-t_0\right)\right)^2\right).
\end{equation*}
The source acts on all three diagonal parts of the stress tensor, $\sigma_{xx}, \sigma_{yy}, \sigma_{zz}$, and the pore pressure $p$.
A set of four receivers is placed along the x-axis, while another set of four receivers is placed along the body diagonal.
The coordinates of all receivers are stated in \cref{tab:receivers_homogeneous_fullspace}.
\begin{table}
    \centering
    \caption{Receiver positions (in meter) for the homogeneous full space problem.}
    \label{tab:receivers_homogeneous_fullspace}
    \pgfplotstabletypeset[
    multicolumn names,
    string type,
    col sep=&,row sep=\\,
    header=false,
    every head row/.style={before row=\toprule,after row=\midrule},
    every last row/.style={after row=\bottomrule},
    display columns/0/.style={column name=, column type={c}},
    display columns/1/.style={column name=\texttt{x1}, column type={S[table-format=4.0]}},
    display columns/2/.style={column name=\texttt{x2}, column type={S[table-format=4.0]}},
    display columns/3/.style={column name=\texttt{x3}, column type={S[table-format=4.0]}},
    display columns/4/.style={column name=\texttt{x4}, column type={S[table-format=4.0]}},
    display columns/5/.style={column name=\texttt{d1}, column type={S[table-format=4.0]}},
    display columns/6/.style={column name=\texttt{d2}, column type={S[table-format=4.0]}},
    display columns/7/.style={column name=\texttt{d3}, column type={S[table-format=4.0]}},
    display columns/8/.style={column name=\texttt{d4}, column type={S[table-format=4.0]}}]
    {
        {$x$} & -1000  &  -600  &   600  &  1000  & -575  &  -345  &  345  &  575 \\
        {$y$} &     0  &     0  &     0  &     0  & -575  &  -345  &  345  &  575 \\
        {$z$} &     0  &     0  &     0  &     0  & -575  &  -345  &  345  &  575 \\
    }
\end{table}
The material parameters are stated in \cref{tab:material_homogeneous}.
We consider the viscous case ($\nu = \SI{0.001}{\pascal\second}$) and the inviscid case ($\nu = \SI{0.0}{\pascal\second}$).
\begin{table}
  \caption{Material parameters for the homogeneous full-space problem.}
  \label{tab:material_homogeneous}
  \centering
  \pgfplotstabletypeset[
    multicolumn names=l,
    string type,
    col sep=&,row sep=\\,
    header=false,
    every head row/.style={before row=\toprule,after row=\midrule},
    every last row/.style={after row=\bottomrule},
    display columns/0/.style={column name=Parameter, column type={l}},
    display columns/1/.style={column name=Value, column type={S[table-format=1.2e2]}},
    display columns/2/.style={column name=, column type={l}},
   ]{
    {$K_S$}       &  2.00e10 & \si{\Pa}\\
    {$\rho_S$}    &  2.08e3  & \si{\kg \per \meter \tothe{3}}\\
    {$\lambda_M$} &  5.28e9  & \si{\Pa}\\
    {$\mu_M$}     &  6.40e9  & \si{\Pa}\\
    {$\phi$}      &  0.40    & \\
    {$\kappa$}    &  6.00e-13& \si{\meter \tothe{2}}\\
    {$T$}         &  2.00    & \\
    {$K_F$}       &  2.50e9  & \si{\Pa}\\
    {$\rho_F$}    &  1.04e3  & \si{\kg \per \meter \tothe{3}}\\
    {$\nu$}       &  1.00e-3 & \si{\Pa \second} \\
  }
\end{table}

The computational domain is a box $[-3000, 3000]^3$.
To simulate the full space, we impose absorbing boundary conditions at the outer faces.
SeisSol supports absorbing boundary conditions as detailed in \cite{dumbser_arbitrary_2006}.
These boundary conditions absorb waves with normal incidence very well, but in corners or in the case of grazing incidence spurious reflections can occur.
We typically enlarge the computational domain, such that these artifacts are not observable in the region of interest.
In the cube $[-1500, 1500]^3$ the mesh has a resolution of \SI{30}{\meter}.
The mesh is further refined around the origin with element edge size down to \SI{3}{\meter} to be able to resolve the explosive source accurately.
Outside of the refined cube, the mesh is coarsened towards the mesh borders.
In total, the mesh consists of \num{6.27e6} elements.
In the \SI{1.5}{\second} simulation window the generated waves pass through all receivers. 
The simulation is carried out with convergence order \num{6}.

\begin{figure}
  \begin{subfigure}{0.9\textwidth}
    \center{
      \input{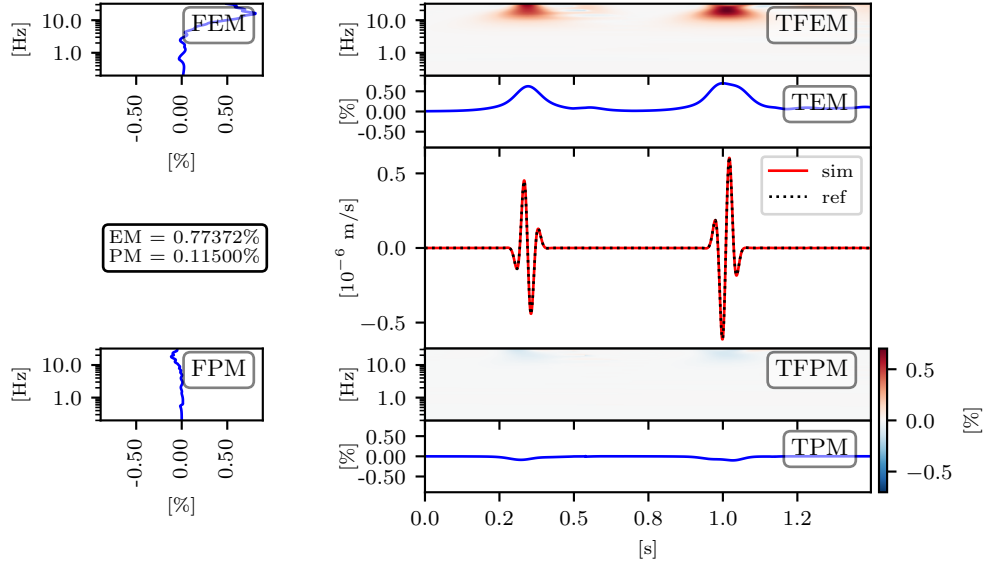}
    }
    \caption{Inviscid fluid}
    \label{fig:explosive-misfits-d4-solid-x-inviscid}
  \end{subfigure}\vspace{5pt}
  \begin{subfigure}{0.9\textwidth}
    \center{
      \input{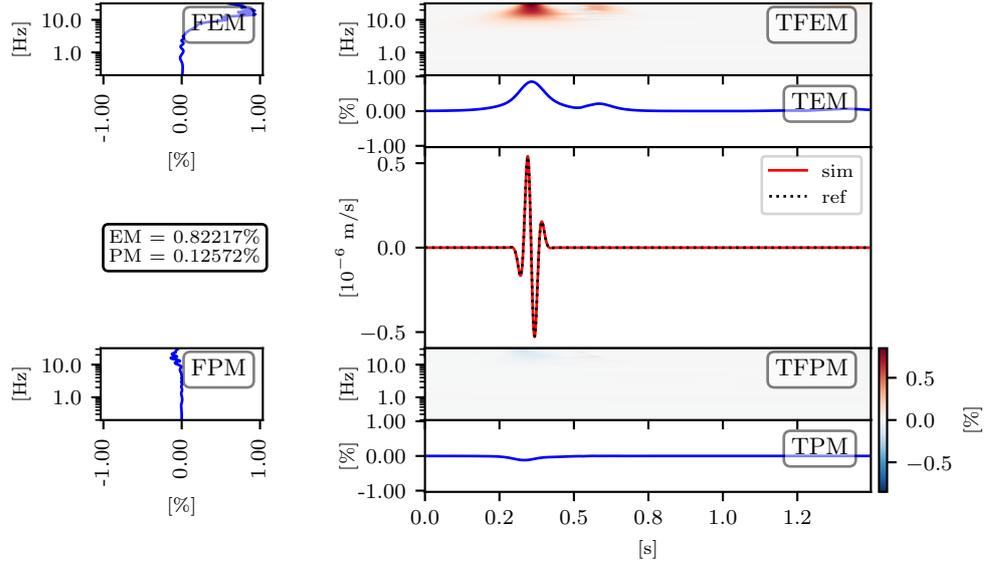}
    }
    \caption{Viscous fluid}
    \label{fig:explosive-misfits-d4-solid-x-viscous}
  \end{subfigure}\vspace{5pt}
  \caption{Detailed misfit plots for the solid particle velocity component $u$ at receiver \texttt{d4} for the homogeneous full-space problem. The wave is excited by an explosive point source in a homogeneous full space. We plot Frequency Envelope Misfit (FEM), Time Frequency Envelope Misfit (TFEM), Time Envelope Misfit (TEM), single-valued Envelope Misfit (EM), single-valued Phase Misfit (PM), the simulated and reference signal, Frequency Phase Misfit (FPM), Time Frequency Phase Misfit (TFPM) and Time Phase Misfit (TPM).}
  \label{fig:explosive-misfits-d4-solid-x}
\end{figure}

We compare the simulation results obtained from SeisSol with semi-analytical solutions calculated using a Green's function approach~\cite{karpfinger_greens_2009}.
Throughout this section we use time-frequency misfits~\cite{kristekova_misfit_2006, kristekova_time-frequency_2009} to quantify differences between SeisSol and the reference solution.
A detailed comparison for the x-component of the solid particle velocity ($u$) at receiver \texttt{d4} in \cref{fig:explosive-misfits-d4-solid-x} illustrates the excellent agreement between both solutions.
In the inviscid case (\cref{fig:explosive-misfits-d4-solid-x-inviscid}), we observe both the fast P-wave (the first pulse) and the slow P-wave (the second pulse).
In the viscous case (\cref{fig:explosive-misfits-d4-solid-x-viscous}), we only observe the fast P-wave since the slow P-wave is strongly diffusive and attenuates very quickly with distance from the source.

\begin{figure}
  \center{
    \input{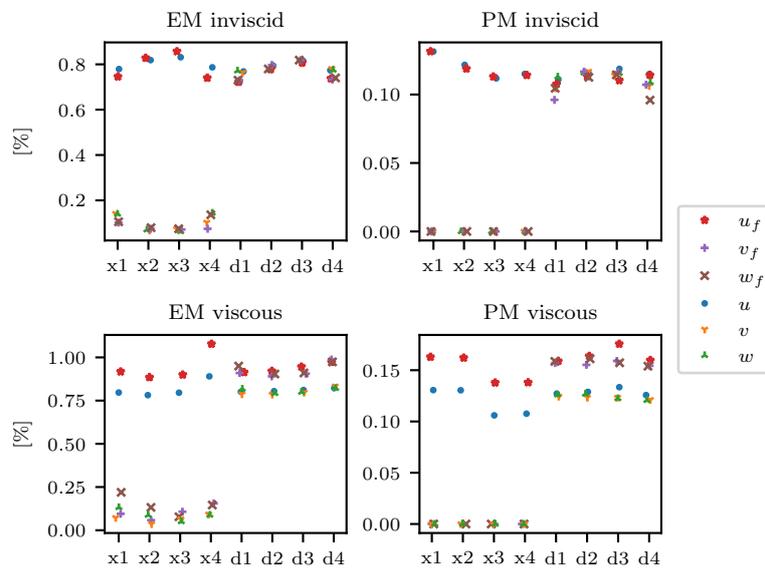}
  }
  \caption{Plot of the envelope and phase misfits of all receivers for the homogeneous full-space problem. We distinguish between an inviscid and a viscous fluid. For both cases, we achieve excellent agreement between simulation and reference.}
  \label{fig:explosive-misfits-overview}
\end{figure}

In \cref{fig:explosive-misfits-overview}, we compare single-valued envelope misfits (EM) and phase misfits (PM) for all components of the solid particle velocity and relative fluid velocity for all considered receivers. 
This summary comparison reveals consistently excellent agreement between SeisSol and semi-analytic solutions with PM below \SI{0.2}{\percent} and EM below $\approx$ \SI{1}{\percent}, which is is slightly exceeded only for $u_f$ at receiver \texttt{x4}. 
We note that small variations at individual receivers are expected due to the unstructured mesh.



\subsection{Contact of two half-spaces}
\label{sec:heterogeneous-fullspace}
To verify whether SeisSol correctly simulates waves reflected from and transmitted through an internal interface, we consider a contact of two half-spaces benchmark problem.
We modify a 2D version proposed in~\cite{de_la_puente_discontinuous_2008} to a 3D volume with an adapted source-receiver configuration.
Material properties of the half-spaces are given in \cref{tab:material_heterogeneous}.
We place an explosive point source at $\left( \SI{0}{\meter} ,\, \SI{0}{\meter} ,\, \SI{500}{\meter} \right)$.
As in \cref{sec:homogeneous-fullspace} the time history is a Ricker wavelet with $f_0 = \SI{16}{\Hz}$ and a time delay of \SI{0.07}{\second}.
We consider two receivers:
Receiver \texttt{r} $\left( \SI{500}{\meter},\, \SI{400}{\meter},\, \SI{500}{\meter} \right)$ at the same side of the interface as the source, to see the reflected waves, and
Receiver \texttt{t} $\left( \SI{500}{\meter},\, \SI{400}{\meter},\, \SI{-500}{\meter} \right)$ across the material interface to record the transmitted waves.

\begin{figure}
  \centering
  \begin{subfigure}{0.9\textwidth}
    \center{
      \input{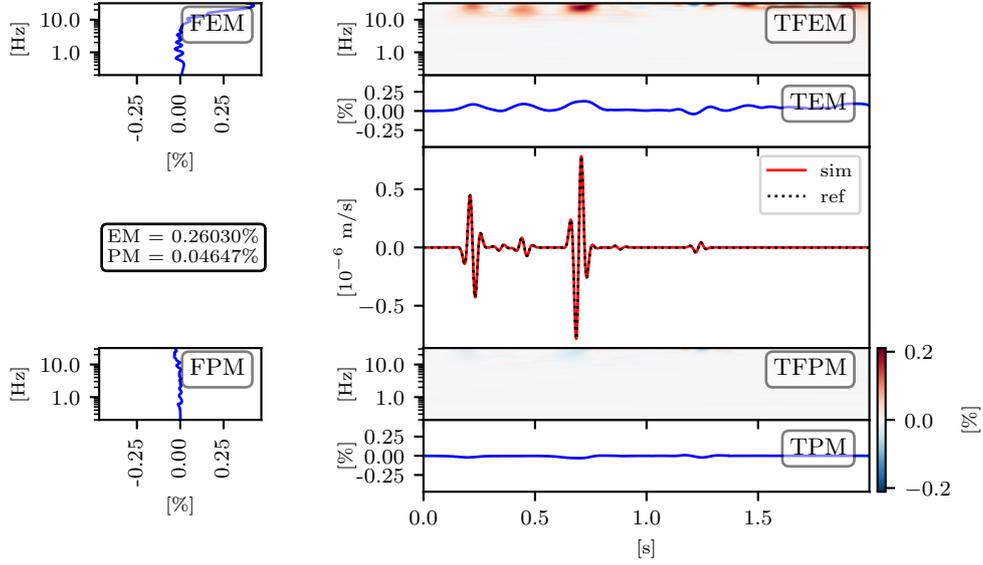}
    }
    \caption{Primary and reflected waves at receiver \texttt{r}.}
    \label{fig:heterogeneous-misfits-r}
  \end{subfigure}\vspace{5pt}
  \begin{subfigure}{0.9\textwidth}
    \center{
      \input{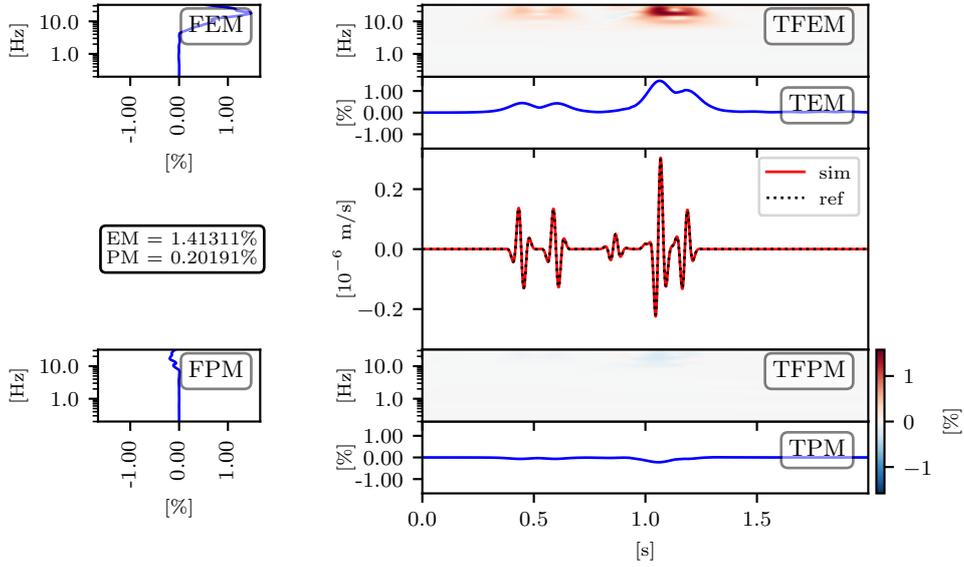}
    }
    \caption{Transmitted waves at receiver \texttt{t}.}
    \label{fig:heterogeneous-misfits-t}
  \end{subfigure}\vspace{5pt}
  \caption{Detailed misfit plots for the solid particle velocity component $u$ at the receivers \texttt{r} and \texttt{t} for the contact of two half-spaces problem.
  For an explanation of the abbreviations, see \cref{fig:explosive-misfits-d4-solid-x}.
  The materials for each half-space have different wave speeds, but the unstructured mesh has the same resolution everywhere, which explains the slightly increased misfit at receiver \texttt{t}.}
  \label{fig:heterogeneous-misfits}
\end{figure}

The computational domain is $[-5000, 5000]^3$ with absorbing boundary conditions.
We refine in a cube with edge length \SI{2}{\km} around the origin up to a characteristic length of \SI{40}{\meter} with further refinement around the source and coarsening towards the boundary.
The mesh has \num{2.08e6} elements.
\begin{table}
\centering
\caption{Material parameters we used for the contact of two half-spaces problem.}
\label{tab:material_heterogeneous}
  \pgfplotstabletypeset[
    multicolumn names=l,
    string type,
    col sep=&,row sep=\\,
    header=false,
    every head row/.style={before row=\toprule,after row=\midrule},
    every last row/.style={after row=\bottomrule},
    display columns/0/.style={column name=Parameter, column type={l}},
    display columns/1/.style={column name=$z>0$, column type={S[table-format=1.2e2]}},
    display columns/2/.style={column name=$z<0$, column type={S[table-format=1.2e2]}},
    display columns/3/.style={column name=, column type={l}},
   ]{
    {$K_S$}       &  4.00e10 &  7.60e9  & \si{\Pa}\\
    {$\rho_S$}    &  2.50e3  &  2.21e3  & \si{\kg \per \meter \tothe{3}}\\
    {$\lambda_M$} &  1.20e10 &  3.96e9  & \si{\Pa}\\
    {$\mu_M$}     &  1.20e10 &  3.96e9  & \si{\Pa}\\
    {$\phi$}      &  0.20    &  0.16    & \\
    {$\kappa$}    &  6.00e-13&  1.00e-13& \si{\meter \tothe{2}}\\
    {$T$}         &  2.00    &  2.00    & \\
    {$K_F$}       &  2.50e9  &  2.50e9  & \si{\Pa}\\
    {$\rho_F$}    &  1.04e3  &  1.04e3  & \si{\kg \per \meter \tothe{3}}\\
    {$\nu$}       &  0.00    &  0.00    & \si{\Pa \second} \\
  }
\end{table}
The method by~\cite{karpfinger_greens_2009}, used to calculate the reference solution in the previous setup, is applicable only in full spaces. 
Therefore, we use the semi-analytical code Gar6More3D ~\cite{diaz_analytical_2008} that allows to include planar interfaces between two half-spaces. 
However, Gar6More3D only supports inviscid fluids ($\nu = \SI{0}{\pascal \second}$) and provides solutions for solid particle velocities. 
\Cref{fig:heterogeneous-misfits} shows a detailed comparison for the x-component of the solid particle velocity between simulated results and the reference solution for both receivers. 
\cref{tab:heterogenous-misfits} summarises all EM and PM values for all components of the solid particle velocity. 
Both comparisons illustrate excellent agreement between the SeisSol and the reference solutions. 
However, we note that the agreement at receiver \texttt{t} is slightly worse. 
The speed of the fast P-wave is \SI[round-mode=places,round-precision=1]{2480.66384633}{\meter \per \second} for $z < 0$ and \SI[round-mode=places,round-precision=1]{4246.85151203}{\meter \per \second} for $z  > 0$.
With a lower wave speed the wavelength decreases, hence a finer mesh is needed for the same accuracy.
We used the same characteristic edge length on both sides of the interface, which explains the slightly increased misfit at receiver \texttt{t}.

\begin{table}
  \centering
  \caption{Misfits (in \si{\percent}) for the contact of two half-spaces problem.}
  \label{tab:heterogenous-misfits}
  \begin{filecontents*}{misfits-heterogeneous-fullspace.csv}
r,0.260295003287,0.046467829217999995,0.220030612162,0.0364999576195,0.174972928394,0.019196193835800002
t,1.41310774945,0.20190887248,1.12427745892,0.161085823498,1.53755203355,0.277134152185
\end{filecontents*}
\pgfplotstabletypeset[
    multicolumn names=l,
    string type,
    col sep=comma,
    header=false,
    every head row/.style={before row=\toprule,after row=\midrule},
    every last row/.style={after row=\bottomrule},
    display columns/0/.style={column name=Receiver, column type={>{\ttfamily}l}},
    display columns/1/.style={column name=EM $u$, column type={S[table-format=1.2, round-mode=places, round-precision=2]}},
    display columns/2/.style={column name=PM $u$, column type={S[table-format=1.2, round-mode=places, round-precision=2]}},
    display columns/3/.style={column name=EM $v$, column type={S[table-format=1.2, round-mode=places, round-precision=2]}},
    display columns/4/.style={column name=PM $v$, column type={S[table-format=1.2, round-mode=places, round-precision=2]}},
    display columns/5/.style={column name=EM $w$, column type={S[table-format=1.2, round-mode=places, round-precision=2]}},
    display columns/6/.style={column name=PM $w$, column type={S[table-format=1.2, round-mode=places, round-precision=2]}},
    ]{misfits-heterogeneous-fullspace.csv}
\end{table}
\clearpage
\subsection{Free surface}
\label{sec:free-surface}
Including a free surface is of special interest in seismic simulations, since in most seismological configurations the traction-free condition is a sufficient approximation of the Earth's surface.
Furthermore, seismic motion is commonly recorded at seismic stations located at Earth's surface. 
A benchmark setup with a homogeneous half-space is reported in~\cite{gregor_subcell-resolution_2021} for a 2D geometry.
We here modify this to 3D, use different material parameters and another source-receiver configuration.
We consider a half-space with a free surface boundary condition ($\sigma \cdot n = 0, p=0$) at $z=0$.
The half-space is homogeneous with material parameters in \cref{tab:material_heterogeneous} for $z>0$.
We place an explosive point source (Ricker wavelet, $f_0$ = \SI{5}{\Hz}, $t_0$ = \SI{0.25}{\s}) at $\left(\SI{0}{\meter},\, \SI{0}{\meter},\, \SI{500}{\meter}\right)$.
We consider a receiver at a shallow depth at $\left(\SI{500}{\meter},\, \SI{400}{\meter},\, \SI{0.5}{\meter}\right)$ due to slow convergence of the reference solution using Gar6More3D~\cite{diaz_analytical_2008} directly at the free surface. 
Additionally, we consider a receiver at depth $\left(\SI{500}{\meter},\, \SI{400}{\meter},\, \SI{500}{\meter}\right)$ to record the reflected waves.
The computational domain is $[-5000, 5000] \times [-5000, 5000] \times [0, 5000]$.
The mesh has a characteristic length of \SI{30}{\m} in the cuboid $[-1000, 1000] \times [-1000, 1000] \times [0, 1000]$.
As before, the mesh is coarsened towards the boundary and refined towards the source. It consists of \num{2.17e6} elements.

\begin{figure}
  \centering
  \begin{subfigure}{0.9\textwidth}
    \center{
      \input{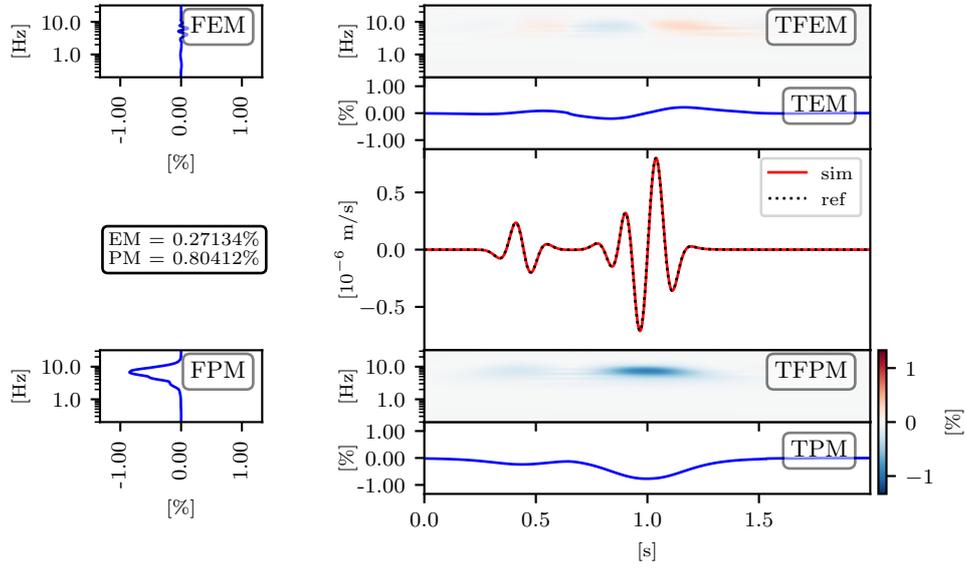}
    }
    \caption{Waveform at the free surface (receiver \texttt{0.5m}).}
    \label{fig:freesurface-misfits-0.5}
  \end{subfigure}\vspace{5pt}
  \begin{subfigure}{0.9\textwidth}
    \center{
      \input{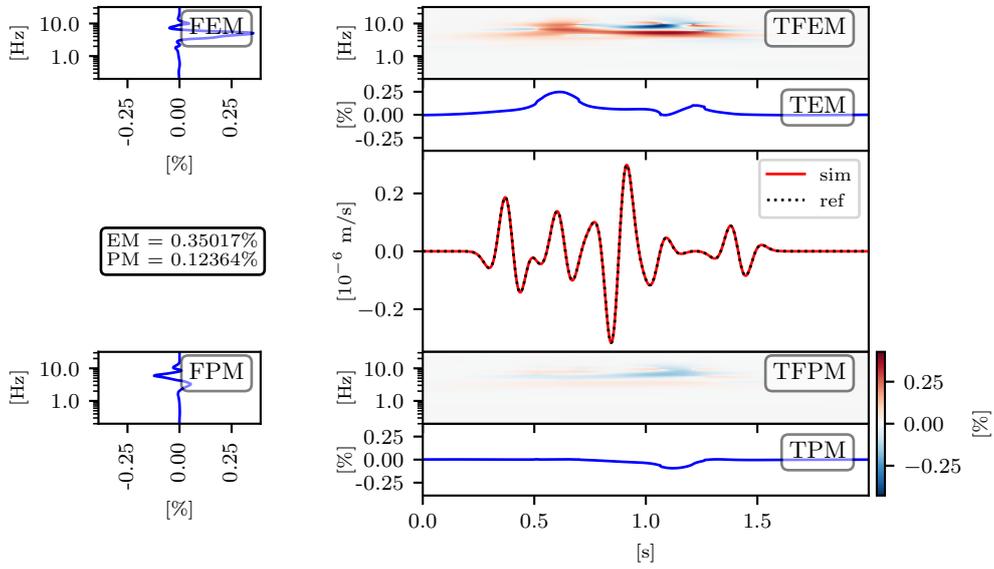}
    }
    \caption{Waveform at depth (receiver \texttt{500m}).}
    \label{fig:freesurface-misfits-100}
  \end{subfigure}\vspace{5pt}
  \caption{Detailed misfit plots for the solid particle velocity $u$ at the receiver at the surface and at depth for the free surface problem.
  For an explanation of the abbreviations, see \cref{fig:explosive-misfits-d4-solid-x}.}
  \label{fig:freesurface-misfits}
\end{figure}

The agreement of the SeisSol and reference solution is excellent as documented by all misfit values below \SI{1}{\percent} (\cref{tab:freesurface-misfits}).
\cref{fig:freesurface-misfits} shows a detailed comparison between the SeisSol and reference solutions for the $x$ component of the solid particle velocity at both receivers.
We note that the reference solution contains, in the plots barely visible, a wave at approximately \SI{1.5}{\second}.
The amplitude of this wave slowly decreases with a finer resolution of the numerical quadrature scheme used in Gar6more3D.
In agreement with one of the Gar6more3D authors (Diaz, 2021, personal communication), we conclude that this wave is unphysical. 
By shifting the receiver to a shallow depth, we were able to obtain a reliable and sufficiently accurate reference solution. 
\begin{table}
  \centering
  \caption{Misfits (in \si{\percent}) for the free surface problem.}
  \label{tab:freesurface-misfits}
  \begin{filecontents*}{misfits-free-surface.csv}
0.5m,0.271337074713,0.804117082473,0.21711374699599997,0.643320297407,0.400986783434,1.21645163097
500m,0.350165995902,0.123639239402,0.280931257361,0.098945162935,0.295408074029,0.0835877304191
\end{filecontents*}
\pgfplotstabletypeset[
    multicolumn names=l,
    string type,
    col sep=comma,
    header=false,
    every head row/.style={before row=\toprule,after row=\midrule},
    every last row/.style={after row=\bottomrule},
    display columns/0/.style={column name=Receiver, column type={>{\ttfamily}l}},
    display columns/1/.style={column name=EM $u$, column type={S[table-format=1.2, round-mode=places, round-precision=2]}},
    display columns/2/.style={column name=PM $u$, column type={S[table-format=1.2, round-mode=places, round-precision=2]}},
    display columns/3/.style={column name=EM $v$, column type={S[table-format=1.2, round-mode=places, round-precision=2]}},
    display columns/4/.style={column name=PM $v$, column type={S[table-format=1.2, round-mode=places, round-precision=2]}},
    display columns/5/.style={column name=EM $w$, column type={S[table-format=1.2, round-mode=places, round-precision=2]}},
    display columns/6/.style={column name=PM $w$, column type={S[table-format=1.2, round-mode=places, round-precision=2]}},
    ]{misfits-free-surface.csv}
\end{table}


\subsection{Layer over half-space}
\label{sec:lohp}
Our final verification benchmark is inspired by the SISMOWINE LOH1 configuration~\cite{moczo_comparison_2006}. 
To emphasise that we consider a poroelastic layer over half-space scenario, we will call this setup LOHp.
In~\cite{gregor_subcell-resolution_2021}, a similar setup has been included using poroelastic materials for 2D geometries.
This model configuration is more complex compared to the previous ones since it contains both the internal material interface and the free surface at the same time. 
Since there is no analytical or semi-analytical solution for this model, we compare our SeisSol solution with a reference solution computed by the FD method~\cite{gregor_subcell-resolution_2021}.

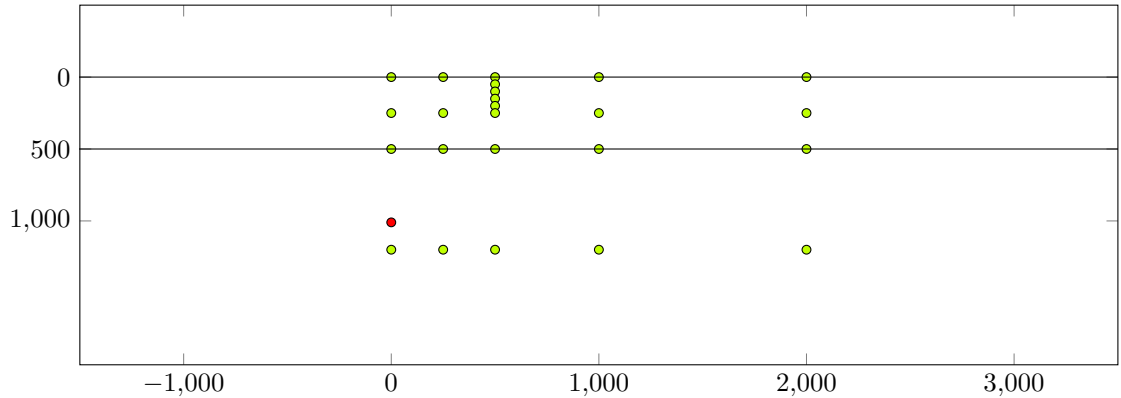
\begin{figure}
    \centering
    \input{figures/loh_plots/loh.tikz}
    \caption{Geometry of the LOHp problem: Source in red and receivers in green.
    We place receivers at the free surface, in the middle of the layer, \SI{10}{\meter} above and below the interface and \SI{200}{\meter} below the source.
    In the horizontal direction, receivers are \SI{0}{\meter}, \SI{250}{\meter}, \SI{500}{\meter}, \SI{1000}{\meter} and \SI{2000}{\meter} away from the source.
    Instead of two receivers at the interface, only one is plotted. 
    Under the receiver at $x=500$, we place a line of receivers from the free surface to the middle of the layer at \SI{5}{\meter} distance to see how the solution varies from the free surface into the layer.
    For this line, most receivers are omitted for readability.}
    \label{fig:loh_geometry}
\end{figure}

\cref{fig:loh_geometry} shows the geometry and the source--receiver configuration.
We consider a \SI{500}{\m}-thick layer atop a homogeneous half-space.
We slightly modify the material parameters from \cref{tab:material_heterogeneous} ($z>0$ for the half-space and $z<0$ for the layer).
Since we are interested in a realistic scenario, we consider a fluid with \textit{non-zero viscosity}: $\nu = \SI{0.001}{\Pa \second}$.
Additionally, we set the permeabilities to \SI{6e-12}{\meter \tothe{2}} and \SI{1e-12}{\meter \tothe{2}} in the half-space and the layer, respectively.
We place an explosive source at $\left(\SI{0}{\meter},\, \SI{0}{\meter},\, \SI{1010}{\meter}\right)$.
We use a Gabor-type source time function:
\begin{equation*}
      s(t) = \cos(w \cdot T + \psi) \cdot \exp( -(\omega \cdot T / \gamma)^2 ),
    \end{equation*}
where $T = t - t_0$,  $\omega = 2 \cdot \pi \cdot f_0$ with parameters: $f_0 = 0.5$, $\gamma = 0.25$, $\psi = 0$, $t_0 = 0.25$.
With this choice of parameters, we excite waves with a flat amplitude spectrum up to $\approx$ \SI{10}{\Hz}.
To be able to compare to a 2D solution, we approximate a line source in SeisSol by \num{401} point sources at $(0, k \cdot 50, 1010), k \in \{-200, \dots 200\}$.
The simulated time window is \SI{5}{\second} long to allow waves to pass through all receivers.

\begin{figure}
    \center{\input{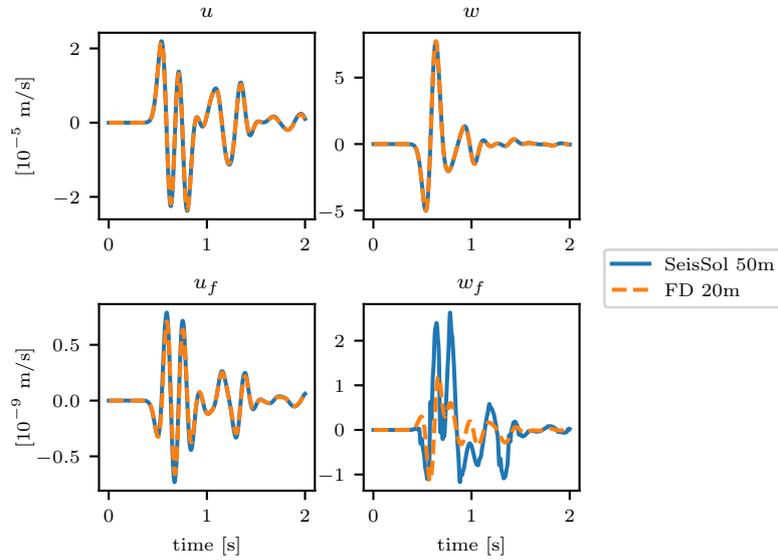}}
    \caption{Horizontal and vertical component of the solid and relative fluid velocities at $\left( \SI{500}{\meter},\, \SI{0}{\meter},\, \SI{0}{\meter}\right)$ for the LOHp problem calculated by SeisSol with characteristic edge length \SI{50}{\meter} method and FD method with grid spacing \SI{20}{\meter}.}
    \label{fig:lohp_wave_form_rec3}
\end{figure}

\begin{figure}
    \center{
      \input{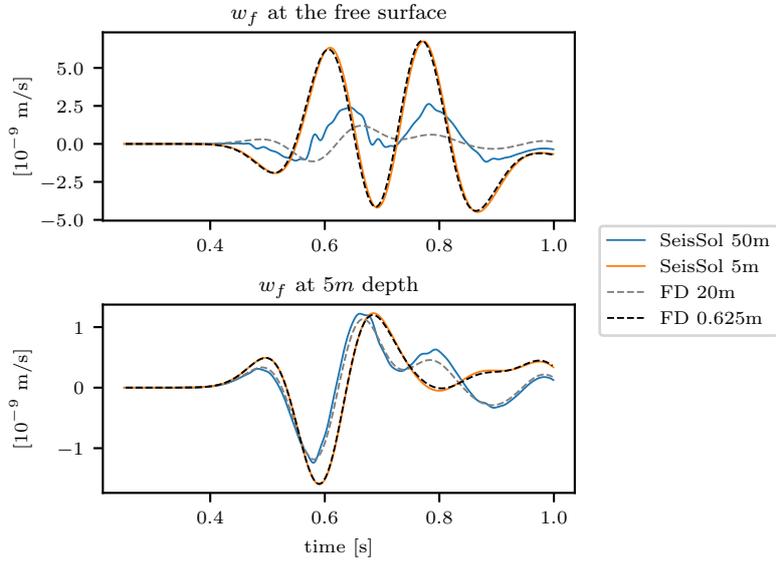}
    }
    \caption{Comparison of the waveform of $w_f$ for the LOHp problem directly at the free surface and \SI{5}{\meter} below it.
    We plot SeisSol solutions with characteristic edge lengths of \SI{5}{\meter} and \SI{50}{\meter} and FD solutions with grid spacings of \SI{20}{\meter} and \SI{0.625}{\meter}.
    SeisSol and FD solutions for the coarse resolutions do not match, but the ones obtained on refined meshes match.}
    
    \label{fig:loh_conv_surface}
\end{figure}

First, we use a standard resolution, i.e. a mesh with characteristic edge length determined according to \citet{kaser_quantitative_2008}, who thoroughly analysed the accuracy of ADER-DG for elastic wave propagation.
Meshes built according to these rules provided reliable results in the previous benchmarks (\cref{sec:homogeneous-fullspace,sec:heterogeneous-fullspace,sec:free-surface}).
For the LOHp model, the standard resolution yields excellent agreement between the SeisSol and the FD solutions only for the solid particle velocities.
However, we find discrepancies with respect to the relative fluid velocities near the free surface and near the interface (c.f. \cref{fig:lohp_wave_form_rec3}).
Note that in the previous tests with an internal material interface or a free surface  (\cref{sec:free-surface,sec:heterogeneous-fullspace}), we could not evaluate the accuracy of the relative fluid velocities due to limitations in the reference solution.
Therefore, now we use finer resolutions to verify that both methods converge to the same solution even for the relative fluid velocity at and near the free surface (c.f. \cref{fig:loh_conv_surface}).
We first discuss the results with standard resolution and then with the fine resolution.

First, we consider SeisSol solutions for a characteristic edge length of \SI{50}{\meter} in the cuboid $[-500, 2500] \times [-1500, 1500] \times [0, 1500]$.
Furthermore, the mesh is refined towards the source and coarsened towards the boundary.
As the reference solution, we consider FD solution for grid spacing of \SI{20}{\meter}.
Because we consider a 2D problem, $y$-components of the solid and relative fluid velocities ($v, v_f$) are zero.
In \cref{fig:lohp_wave_form_rec3} we compare the waveforms obtained with SeisSol and the FD code for a selected receiver at the free surface.
A visual comparison reveals good agreement for $u$, $w$ and $u_f$, however, it also reveals substantial differences for $w_f$.

\begin{figure}
  \center{
    \input{figures/loh_plots/misfit_matrix_em.pgf}
  }
  \caption{Comparison of the EM between the SeisSol and FD solutions at the grid of receivers for the LOHp problem.
  The first row in each matrix represents the receivers at the free surface. 
  The second row has the receivers in the middle of the layer.
  In the next two rows are the receivers a little above and below the interface.
  The last row gathers all receivers below the source.}
  \label{fig:lohp_misfit_matrix}
\end{figure}
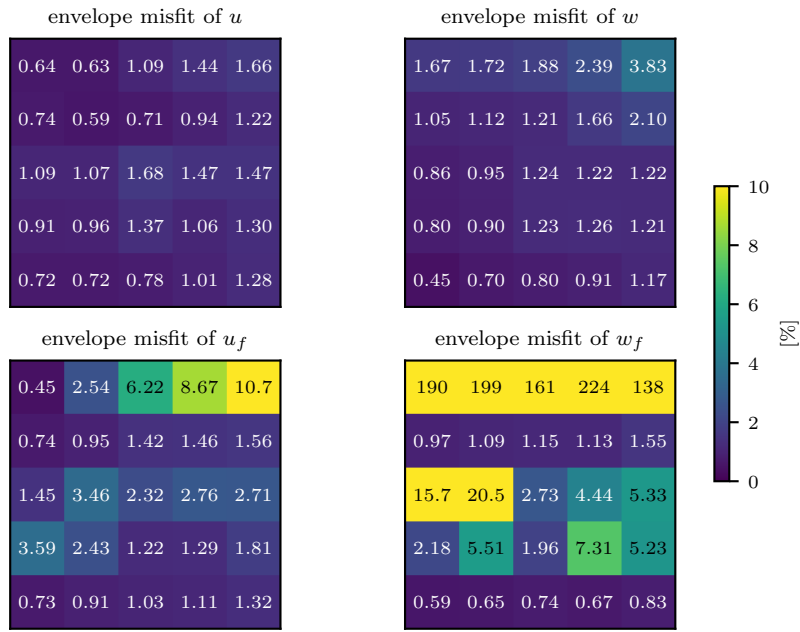

Next, we present a quantitative error analysis.
Since we want to compare a large number of receivers, we concentrate on the single-valued envelope misfit (EM) only, while considering the FD solution as a reference.
\cref{fig:lohp_misfit_matrix} summarises the EM values for the vertical and horizontal components of the solid and relative fluid velocities at the grid of receivers, shown in \cref{fig:loh_geometry}.
In the case of solid particle velocities, we see excellent agreement (EM below \SI{4}{\percent}) between the SeisSol and FD solutions.
We note that a slightly larger error is expected for receivers further away from the source due to the cumulative character of numerical errors. 
We also note that the receivers directly at the surface and close to the material interface have a slightly worse match than the ones in the middle of the layer or below the source independent of their distance to the source.
However, in the case of the relative fluid velocities, we observe agreement comparable to that for solid particle velocities only in the middle of the layer and below the source.
The misfits for the vertical component near the material interface and the free surface are unacceptably large.
The mismatch at the free surface is very high, such that the waveforms do qualitatively differ (c.f. \cref{fig:lohp_wave_form_rec3}).

We assume that the above-mentioned disagreement between the FD and SeisSol solutions is due to the fact that the slow P-wave is poorly resolved in the close vicinity of the free surface and material interface in one or both methods. 
In the low-frequency regime, the slow P-wave behaves as a diffusive wave, having very small wavelength compared to the fast P-wave and S-wave, and attenuates very quickly with distance from its origin.
\citet{dutta_seismic_1983} show that when a fast P- or S-wave impinges on a material interface or free surface, as part of the partitioning of energy, mode conversion to a slow P-wave occurs, and its generation draws energy from the propagating wave process.
However, the relative fluid velocities are $\approx 5$ orders of magnitude smaller than the solid particle velocities and thus this inaccuracy does not affect the accuracy of the solid particle velocities.
Moreover, because of the diffusive character of the slow P-wave, the slow P-wave remains in the vicinity of the interface during time scales for seismic wave propagation. 
Therefore, we do not observe differences at the receivers at depths much greater than the characteristic diffusion length~\cite{wenzlau_finite-difference_2009}, which is $\approx \SI{0.7}{\meter}$ for our configuration. 

\begin{figure}
  \center{
    \input{figures/loh_plots/misfit_depth.pgf}
  }
  \caption{Envelope misfit between the SeisSol solutions with \SI{50}{\meter} resolution and SeisSol solutions with \SI{5}{\meter} refinement towards the free surface for the LOHp problem.}
  \label{fig:lohp_misfit_depth}
\end{figure}
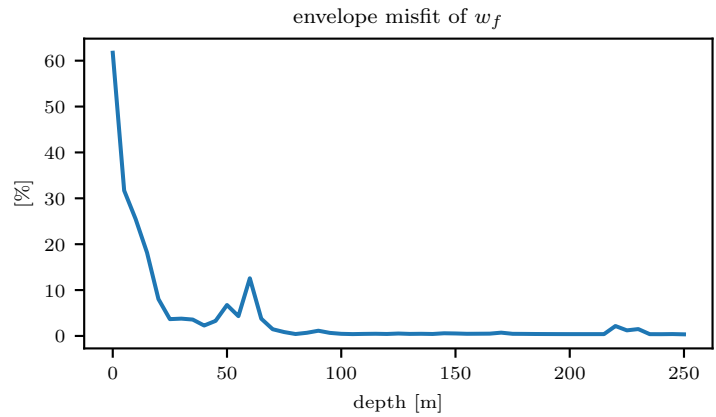

To overcome the observed discrepancies, we refined both the SeisSol and the FD resolutions to see, if we can also resolve the slow P-wave accurately.
For FD, we use a grid spacing of \SI{0.625}{\m}, based on a series of simulations with gradually finer grid resolution to verify that the solution converged (see \cref{fig:convergence_fd} in \cref{sec:convergence_fd}).
For SeisSol we refined the mesh at the free surface to a characteristic edge length of only \SI{5}{\meter}. 
For computational reasons, we restricted the computation time to \SI{2}{\second}.
\cref{fig:loh_conv_surface} reveals that the qualitative behaviour of the waveform changes considerably with grid spacing/mesh refinement.
If we compare the refined solutions, we conclude that both numerical solvers converge to the same solution.
Finally, we would like to remark that the qualitative behaviour changes rapidly between the receiver at the free surface and \SI{5}{\meter} below the free surface, which indicates that the diffusive P-wave plays a significant role here.
To further assess how the slow P-wave affects the quality of the solution near the free surface, we compare the EM between the SeisSol solutions for characteristic element lengths \SI{50}{\m} and \SI{5}{\m} for the vertical component of the relative fluid velocity ($w_f$) at various depths (\cref{fig:lohp_misfit_depth}).
We observe large EM values only above \SI{70}{\meter} from the free surface.
This means that the large differences are, in fact, concentrated in a layer with a thickness comparable with characteristic element length (\SI{50}{\meter}). 
This justifies our assumption that the slow P-wave was the source of mismatch at the free surface.


To conclude, we remark that the slow P-wave at the free surface can be simulated accurately with both, the here presented SeisSol extension and the 2D FD approach, but both methods need an extremely fine grid to accurately resolve the relative fluid velocity at the free surface.
For most practical scenarios, the simulation at the coarse level will suffice, because:
\begin{figure}
    \center{
      \input{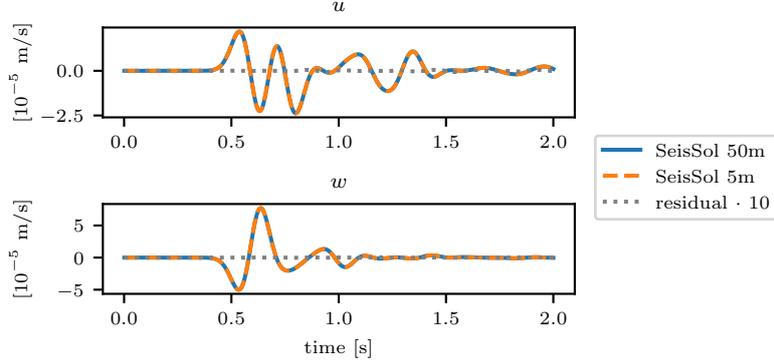}
    }
    \caption{Comparison of the solid particle velocities at  $\left( \SI{500}{\meter},\, \SI{0}{\meter},\, \SI{0}{\meter}\right)$ for the LOHp problem on meshes with different characteristic edge lengths with SeisSol.}
    \label{fig:loh_solid_velocities}
\end{figure}
\begin{itemize}
    \item If we compare the solid particle velocities at the free surface for different mesh resolutions (c.f. \cref{fig:loh_solid_velocities}), we do not see a difference.
    We conclude that at the free surface, the solid particle velocities are not affected by the slow P-wave.
    This means for a seismic simulation, where one is interested in ground motion, the direct effect from the slow P-wave at the free surface is negligible.
    \item  The slow P-wave only affects the solution at small spatial scales.
    Within the volume, all reflected waves are correctly simulated, even with coarse resolutions.
    If the relative fluid velocities are needed at a high resolution, another approach, e.g. by solving a diffusion equation, might be better suited.
\end{itemize}

%% file: figures/loh_plots/loh.tikz
\begin{tikzpicture}
\begin{axis}[
  xmin=-1500,xmax=3500,ymin=-500, ymax=2000,
  width  = 6in,
  height = 2.5in,
  xtick = {-1000, 0, 1000, 2000, 3000},
  ytick = {1000, 500, 0},
  y dir = reverse
]
 	\node (0) at (axis cs: -5000,     0) {};
 	\node (1) at (axis cs:  5000,     0) {};
 	\node (2) at (axis cs: -5000, 5000) {};
 	\node (3) at (axis cs:  5000, 5000) {};
 	\node (4) at (axis cs: -5000,  500) {};
 	\node (5) at (axis cs:  5000,  500) {};
 	\node [style=large red dot]   (6)  at (axis cs:    0, 1010) {}; 
 	\node [style=large green dot] (7)  at (axis cs:    0,     0) {};
 	\node [style=large green dot] (8)  at (axis cs:    0,  250) {};
 	\node [style=large green dot] (9)  at (axis cs:    0,  500) {};
 	\node [style=large green dot] (10) at (axis cs:    0, 1200) {};
 	\node [style=large green dot] (11) at (axis cs:  250,     0) {};
 	\node [style=large green dot] (12) at (axis cs:  250,  250) {};
 	\node [style=large green dot] (13) at (axis cs:  250,  500) {};
 	\node [style=large green dot] (12) at (axis cs:  250, 1200) {};
 	\node [style=large green dot] (17) at (axis cs:  500,     0) {};
 	\node [style=large green dot] (24) at (axis cs:  500,   50) {};
 	\node [style=large green dot] (25) at (axis cs:  500,  100) {};
 	\node [style=large green dot] (26) at (axis cs:  500,  150) {};
 	\node [style=large green dot] (27) at (axis cs:  500,  200) {};
 	\node [style=large green dot] (17) at (axis cs:  500,  250) {};
 	\node [style=large green dot] (17) at (axis cs:  500,  500) {};
 	\node [style=large green dot] (15) at (axis cs:  500, 1200) {};
 	\node [style=large green dot] (16) at (axis cs: 1000,     0) {};
 	\node [style=large green dot] (17) at (axis cs: 1000,  250) {};
 	\node [style=large green dot] (18) at (axis cs: 1000,  500) {};
 	\node [style=large green dot] (19) at (axis cs: 1000, 1200) {};
 	\node [style=large green dot] (20) at (axis cs: 2000,     0) {};
 	\node [style=large green dot] (21) at (axis cs: 2000,  250) {};
 	\node [style=large green dot] (22) at (axis cs: 2000,  500) {};
 	\node [style=large green dot] (23) at (axis cs: 2000, 1200) {};
 	\draw (0.center) to (4.center);
 	\draw (4.center) to (5.center);
 	\draw (5.center) to (1.center);
 	\draw (1.center) to (0.center);
 	\draw (4.center) to (2.center);
 	\draw (2.center) to (3.center);
 	\draw (3.center) to (5.center);
\end{axis}
\end{tikzpicture}

%% file: figures/loh_plots/misfit_matrix_em.pgf
\begingroup%
\makeatletter%
\begin{pgfpicture}%
\pgfpathrectangle{\pgfpointorigin}{\pgfqpoint{4.285680in}{3.440494in}}%
\pgfusepath{use as bounding box, clip}%
\begin{pgfscope}%
\pgfsetbuttcap%
\pgfsetmiterjoin%
\definecolor{currentfill}{rgb}{1.000000,1.000000,1.000000}%
\pgfsetfillcolor{currentfill}%
\pgfsetlinewidth{0.000000pt}%
\definecolor{currentstroke}{rgb}{1.000000,1.000000,1.000000}%
\pgfsetstrokecolor{currentstroke}%
\pgfsetdash{}{0pt}%
\pgfpathmoveto{\pgfqpoint{0.000000in}{0.000000in}}%
\pgfpathlineto{\pgfqpoint{4.285680in}{0.000000in}}%
\pgfpathlineto{\pgfqpoint{4.285680in}{3.440494in}}%
\pgfpathlineto{\pgfqpoint{0.000000in}{3.440494in}}%
\pgfpathclose%
\pgfusepath{fill}%
\end{pgfscope}%
\begin{pgfscope}%
\pgfsetbuttcap%
\pgfsetmiterjoin%
\definecolor{currentfill}{rgb}{1.000000,1.000000,1.000000}%
\pgfsetfillcolor{currentfill}%
\pgfsetlinewidth{0.000000pt}%
\definecolor{currentstroke}{rgb}{0.000000,0.000000,0.000000}%
\pgfsetstrokecolor{currentstroke}%
\pgfsetstrokeopacity{0.000000}%
\pgfsetdash{}{0pt}%
\pgfpathmoveto{\pgfqpoint{0.100000in}{1.780000in}}%
\pgfpathlineto{\pgfqpoint{1.500000in}{1.780000in}}%
\pgfpathlineto{\pgfqpoint{1.500000in}{3.180000in}}%
\pgfpathlineto{\pgfqpoint{0.100000in}{3.180000in}}%
\pgfpathclose%
\pgfusepath{fill}%
\end{pgfscope}%
\begin{pgfscope}%
\pgfpathrectangle{\pgfqpoint{0.100000in}{1.780000in}}{\pgfqpoint{1.400000in}{1.400000in}}%
\pgfusepath{clip}%
\pgfsys@transformshift{0.100000in}{1.780000in}%
\pgftext[left,bottom]{\includegraphics[interpolate=true,width=1.400000in,height=1.400000in]{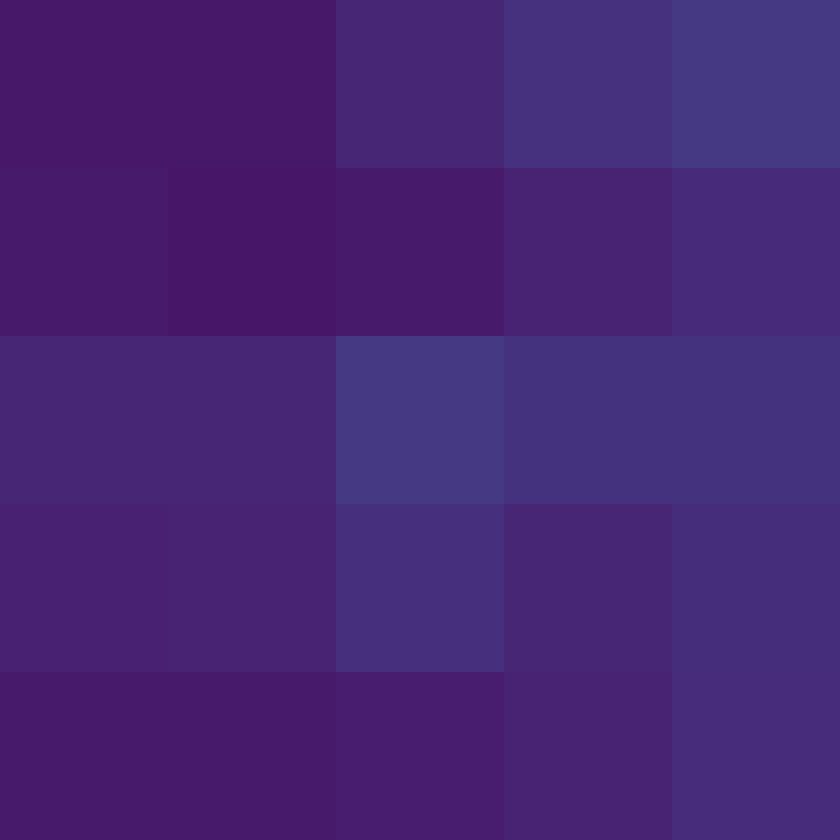}}%
\end{pgfscope}%
\begin{pgfscope}%
\pgfsetrectcap%
\pgfsetmiterjoin%
\pgfsetlinewidth{0.803000pt}%
\definecolor{currentstroke}{rgb}{0.000000,0.000000,0.000000}%
\pgfsetstrokecolor{currentstroke}%
\pgfsetdash{}{0pt}%
\pgfpathmoveto{\pgfqpoint{0.100000in}{1.780000in}}%
\pgfpathlineto{\pgfqpoint{0.100000in}{3.180000in}}%
\pgfusepath{stroke}%
\end{pgfscope}%
\begin{pgfscope}%
\pgfsetrectcap%
\pgfsetmiterjoin%
\pgfsetlinewidth{0.803000pt}%
\definecolor{currentstroke}{rgb}{0.000000,0.000000,0.000000}%
\pgfsetstrokecolor{currentstroke}%
\pgfsetdash{}{0pt}%
\pgfpathmoveto{\pgfqpoint{1.500000in}{1.780000in}}%
\pgfpathlineto{\pgfqpoint{1.500000in}{3.180000in}}%
\pgfusepath{stroke}%
\end{pgfscope}%
\begin{pgfscope}%
\pgfsetrectcap%
\pgfsetmiterjoin%
\pgfsetlinewidth{0.803000pt}%
\definecolor{currentstroke}{rgb}{0.000000,0.000000,0.000000}%
\pgfsetstrokecolor{currentstroke}%
\pgfsetdash{}{0pt}%
\pgfpathmoveto{\pgfqpoint{0.100000in}{1.780000in}}%
\pgfpathlineto{\pgfqpoint{1.500000in}{1.780000in}}%
\pgfusepath{stroke}%
\end{pgfscope}%
\begin{pgfscope}%
\pgfsetrectcap%
\pgfsetmiterjoin%
\pgfsetlinewidth{0.803000pt}%
\definecolor{currentstroke}{rgb}{0.000000,0.000000,0.000000}%
\pgfsetstrokecolor{currentstroke}%
\pgfsetdash{}{0pt}%
\pgfpathmoveto{\pgfqpoint{0.100000in}{3.180000in}}%
\pgfpathlineto{\pgfqpoint{1.500000in}{3.180000in}}%
\pgfusepath{stroke}%
\end{pgfscope}%
\begin{pgfscope}%
\definecolor{textcolor}{rgb}{1.000000,1.000000,1.000000}%
\pgfsetstrokecolor{textcolor}%
\pgfsetfillcolor{textcolor}%
\pgftext[x=0.240000in,y=3.040000in,,]{\color{textcolor}\rmfamily\fontsize{7.000000}{8.400000}\selectfont 0.64}%
\end{pgfscope}%
\begin{pgfscope}%
\definecolor{textcolor}{rgb}{1.000000,1.000000,1.000000}%
\pgfsetstrokecolor{textcolor}%
\pgfsetfillcolor{textcolor}%
\pgftext[x=0.520000in,y=3.040000in,,]{\color{textcolor}\rmfamily\fontsize{7.000000}{8.400000}\selectfont 0.63}%
\end{pgfscope}%
\begin{pgfscope}%
\definecolor{textcolor}{rgb}{1.000000,1.000000,1.000000}%
\pgfsetstrokecolor{textcolor}%
\pgfsetfillcolor{textcolor}%
\pgftext[x=0.800000in,y=3.040000in,,]{\color{textcolor}\rmfamily\fontsize{7.000000}{8.400000}\selectfont 1.09}%
\end{pgfscope}%
\begin{pgfscope}%
\definecolor{textcolor}{rgb}{1.000000,1.000000,1.000000}%
\pgfsetstrokecolor{textcolor}%
\pgfsetfillcolor{textcolor}%
\pgftext[x=1.080000in,y=3.040000in,,]{\color{textcolor}\rmfamily\fontsize{7.000000}{8.400000}\selectfont 1.44}%
\end{pgfscope}%
\begin{pgfscope}%
\definecolor{textcolor}{rgb}{1.000000,1.000000,1.000000}%
\pgfsetstrokecolor{textcolor}%
\pgfsetfillcolor{textcolor}%
\pgftext[x=1.360000in,y=3.040000in,,]{\color{textcolor}\rmfamily\fontsize{7.000000}{8.400000}\selectfont 1.66}%
\end{pgfscope}%
\begin{pgfscope}%
\definecolor{textcolor}{rgb}{1.000000,1.000000,1.000000}%
\pgfsetstrokecolor{textcolor}%
\pgfsetfillcolor{textcolor}%
\pgftext[x=0.240000in,y=2.760000in,,]{\color{textcolor}\rmfamily\fontsize{7.000000}{8.400000}\selectfont 0.74}%
\end{pgfscope}%
\begin{pgfscope}%
\definecolor{textcolor}{rgb}{1.000000,1.000000,1.000000}%
\pgfsetstrokecolor{textcolor}%
\pgfsetfillcolor{textcolor}%
\pgftext[x=0.520000in,y=2.760000in,,]{\color{textcolor}\rmfamily\fontsize{7.000000}{8.400000}\selectfont 0.59}%
\end{pgfscope}%
\begin{pgfscope}%
\definecolor{textcolor}{rgb}{1.000000,1.000000,1.000000}%
\pgfsetstrokecolor{textcolor}%
\pgfsetfillcolor{textcolor}%
\pgftext[x=0.800000in,y=2.760000in,,]{\color{textcolor}\rmfamily\fontsize{7.000000}{8.400000}\selectfont 0.71}%
\end{pgfscope}%
\begin{pgfscope}%
\definecolor{textcolor}{rgb}{1.000000,1.000000,1.000000}%
\pgfsetstrokecolor{textcolor}%
\pgfsetfillcolor{textcolor}%
\pgftext[x=1.080000in,y=2.760000in,,]{\color{textcolor}\rmfamily\fontsize{7.000000}{8.400000}\selectfont 0.94}%
\end{pgfscope}%
\begin{pgfscope}%
\definecolor{textcolor}{rgb}{1.000000,1.000000,1.000000}%
\pgfsetstrokecolor{textcolor}%
\pgfsetfillcolor{textcolor}%
\pgftext[x=1.360000in,y=2.760000in,,]{\color{textcolor}\rmfamily\fontsize{7.000000}{8.400000}\selectfont 1.22}%
\end{pgfscope}%
\begin{pgfscope}%
\definecolor{textcolor}{rgb}{1.000000,1.000000,1.000000}%
\pgfsetstrokecolor{textcolor}%
\pgfsetfillcolor{textcolor}%
\pgftext[x=0.240000in,y=2.480000in,,]{\color{textcolor}\rmfamily\fontsize{7.000000}{8.400000}\selectfont 1.09}%
\end{pgfscope}%
\begin{pgfscope}%
\definecolor{textcolor}{rgb}{1.000000,1.000000,1.000000}%
\pgfsetstrokecolor{textcolor}%
\pgfsetfillcolor{textcolor}%
\pgftext[x=0.520000in,y=2.480000in,,]{\color{textcolor}\rmfamily\fontsize{7.000000}{8.400000}\selectfont 1.07}%
\end{pgfscope}%
\begin{pgfscope}%
\definecolor{textcolor}{rgb}{1.000000,1.000000,1.000000}%
\pgfsetstrokecolor{textcolor}%
\pgfsetfillcolor{textcolor}%
\pgftext[x=0.800000in,y=2.480000in,,]{\color{textcolor}\rmfamily\fontsize{7.000000}{8.400000}\selectfont 1.68}%
\end{pgfscope}%
\begin{pgfscope}%
\definecolor{textcolor}{rgb}{1.000000,1.000000,1.000000}%
\pgfsetstrokecolor{textcolor}%
\pgfsetfillcolor{textcolor}%
\pgftext[x=1.080000in,y=2.480000in,,]{\color{textcolor}\rmfamily\fontsize{7.000000}{8.400000}\selectfont 1.47}%
\end{pgfscope}%
\begin{pgfscope}%
\definecolor{textcolor}{rgb}{1.000000,1.000000,1.000000}%
\pgfsetstrokecolor{textcolor}%
\pgfsetfillcolor{textcolor}%
\pgftext[x=1.360000in,y=2.480000in,,]{\color{textcolor}\rmfamily\fontsize{7.000000}{8.400000}\selectfont 1.47}%
\end{pgfscope}%
\begin{pgfscope}%
\definecolor{textcolor}{rgb}{1.000000,1.000000,1.000000}%
\pgfsetstrokecolor{textcolor}%
\pgfsetfillcolor{textcolor}%
\pgftext[x=0.240000in,y=2.200000in,,]{\color{textcolor}\rmfamily\fontsize{7.000000}{8.400000}\selectfont 0.91}%
\end{pgfscope}%
\begin{pgfscope}%
\definecolor{textcolor}{rgb}{1.000000,1.000000,1.000000}%
\pgfsetstrokecolor{textcolor}%
\pgfsetfillcolor{textcolor}%
\pgftext[x=0.520000in,y=2.200000in,,]{\color{textcolor}\rmfamily\fontsize{7.000000}{8.400000}\selectfont 0.96}%
\end{pgfscope}%
\begin{pgfscope}%
\definecolor{textcolor}{rgb}{1.000000,1.000000,1.000000}%
\pgfsetstrokecolor{textcolor}%
\pgfsetfillcolor{textcolor}%
\pgftext[x=0.800000in,y=2.200000in,,]{\color{textcolor}\rmfamily\fontsize{7.000000}{8.400000}\selectfont 1.37}%
\end{pgfscope}%
\begin{pgfscope}%
\definecolor{textcolor}{rgb}{1.000000,1.000000,1.000000}%
\pgfsetstrokecolor{textcolor}%
\pgfsetfillcolor{textcolor}%
\pgftext[x=1.080000in,y=2.200000in,,]{\color{textcolor}\rmfamily\fontsize{7.000000}{8.400000}\selectfont 1.06}%
\end{pgfscope}%
\begin{pgfscope}%
\definecolor{textcolor}{rgb}{1.000000,1.000000,1.000000}%
\pgfsetstrokecolor{textcolor}%
\pgfsetfillcolor{textcolor}%
\pgftext[x=1.360000in,y=2.200000in,,]{\color{textcolor}\rmfamily\fontsize{7.000000}{8.400000}\selectfont 1.30}%
\end{pgfscope}%
\begin{pgfscope}%
\definecolor{textcolor}{rgb}{1.000000,1.000000,1.000000}%
\pgfsetstrokecolor{textcolor}%
\pgfsetfillcolor{textcolor}%
\pgftext[x=0.240000in,y=1.920000in,,]{\color{textcolor}\rmfamily\fontsize{7.000000}{8.400000}\selectfont 0.72}%
\end{pgfscope}%
\begin{pgfscope}%
\definecolor{textcolor}{rgb}{1.000000,1.000000,1.000000}%
\pgfsetstrokecolor{textcolor}%
\pgfsetfillcolor{textcolor}%
\pgftext[x=0.520000in,y=1.920000in,,]{\color{textcolor}\rmfamily\fontsize{7.000000}{8.400000}\selectfont 0.72}%
\end{pgfscope}%
\begin{pgfscope}%
\definecolor{textcolor}{rgb}{1.000000,1.000000,1.000000}%
\pgfsetstrokecolor{textcolor}%
\pgfsetfillcolor{textcolor}%
\pgftext[x=0.800000in,y=1.920000in,,]{\color{textcolor}\rmfamily\fontsize{7.000000}{8.400000}\selectfont 0.78}%
\end{pgfscope}%
\begin{pgfscope}%
\definecolor{textcolor}{rgb}{1.000000,1.000000,1.000000}%
\pgfsetstrokecolor{textcolor}%
\pgfsetfillcolor{textcolor}%
\pgftext[x=1.080000in,y=1.920000in,,]{\color{textcolor}\rmfamily\fontsize{7.000000}{8.400000}\selectfont 1.01}%
\end{pgfscope}%
\begin{pgfscope}%
\definecolor{textcolor}{rgb}{1.000000,1.000000,1.000000}%
\pgfsetstrokecolor{textcolor}%
\pgfsetfillcolor{textcolor}%
\pgftext[x=1.360000in,y=1.920000in,,]{\color{textcolor}\rmfamily\fontsize{7.000000}{8.400000}\selectfont 1.28}%
\end{pgfscope}%
\begin{pgfscope}%
\definecolor{textcolor}{rgb}{0.000000,0.000000,0.000000}%
\pgfsetstrokecolor{textcolor}%
\pgfsetfillcolor{textcolor}%
\pgftext[x=0.800000in,y=3.263333in,,base]{\color{textcolor}\rmfamily\fontsize{8.400000}{10.080000}\selectfont envelope misfit of \(\displaystyle u\)}%
\end{pgfscope}%
\begin{pgfscope}%
\pgfsetbuttcap%
\pgfsetmiterjoin%
\definecolor{currentfill}{rgb}{1.000000,1.000000,1.000000}%
\pgfsetfillcolor{currentfill}%
\pgfsetlinewidth{0.000000pt}%
\definecolor{currentstroke}{rgb}{0.000000,0.000000,0.000000}%
\pgfsetstrokecolor{currentstroke}%
\pgfsetstrokeopacity{0.000000}%
\pgfsetdash{}{0pt}%
\pgfpathmoveto{\pgfqpoint{2.145455in}{1.780000in}}%
\pgfpathlineto{\pgfqpoint{3.545455in}{1.780000in}}%
\pgfpathlineto{\pgfqpoint{3.545455in}{3.180000in}}%
\pgfpathlineto{\pgfqpoint{2.145455in}{3.180000in}}%
\pgfpathclose%
\pgfusepath{fill}%
\end{pgfscope}%
\begin{pgfscope}%
\pgfpathrectangle{\pgfqpoint{2.145455in}{1.780000in}}{\pgfqpoint{1.400000in}{1.400000in}}%
\pgfusepath{clip}%
\pgfsys@transformshift{2.145455in}{1.780000in}%
\pgftext[left,bottom]{\includegraphics[interpolate=true,width=1.401667in,height=1.400000in]{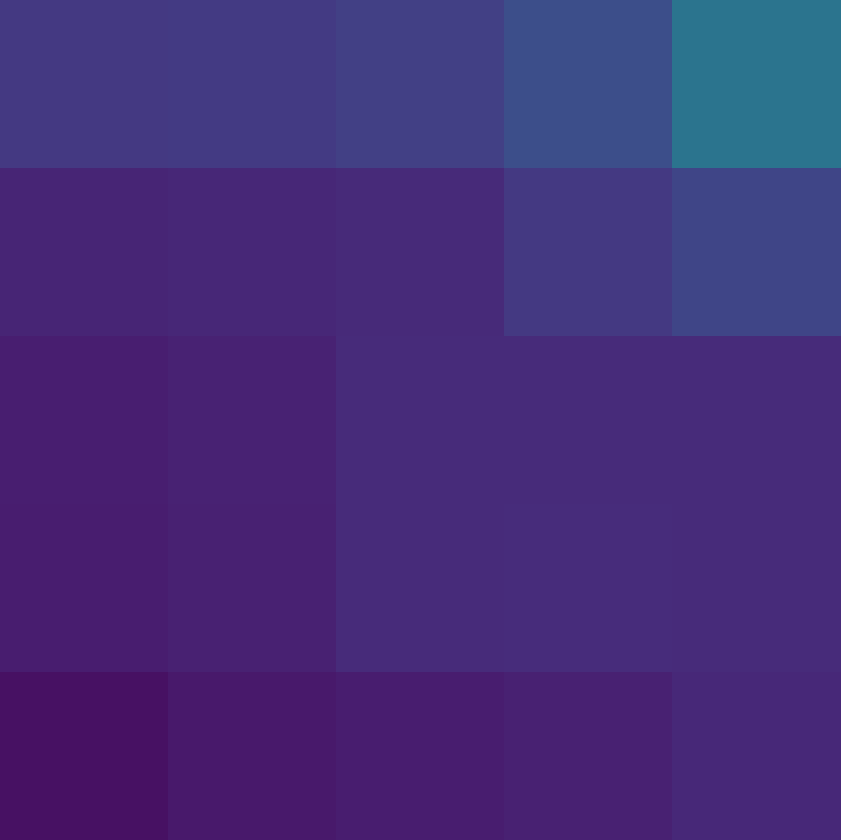}}%
\end{pgfscope}%
\begin{pgfscope}%
\pgfsetrectcap%
\pgfsetmiterjoin%
\pgfsetlinewidth{0.803000pt}%
\definecolor{currentstroke}{rgb}{0.000000,0.000000,0.000000}%
\pgfsetstrokecolor{currentstroke}%
\pgfsetdash{}{0pt}%
\pgfpathmoveto{\pgfqpoint{2.145455in}{1.780000in}}%
\pgfpathlineto{\pgfqpoint{2.145455in}{3.180000in}}%
\pgfusepath{stroke}%
\end{pgfscope}%
\begin{pgfscope}%
\pgfsetrectcap%
\pgfsetmiterjoin%
\pgfsetlinewidth{0.803000pt}%
\definecolor{currentstroke}{rgb}{0.000000,0.000000,0.000000}%
\pgfsetstrokecolor{currentstroke}%
\pgfsetdash{}{0pt}%
\pgfpathmoveto{\pgfqpoint{3.545455in}{1.780000in}}%
\pgfpathlineto{\pgfqpoint{3.545455in}{3.180000in}}%
\pgfusepath{stroke}%
\end{pgfscope}%
\begin{pgfscope}%
\pgfsetrectcap%
\pgfsetmiterjoin%
\pgfsetlinewidth{0.803000pt}%
\definecolor{currentstroke}{rgb}{0.000000,0.000000,0.000000}%
\pgfsetstrokecolor{currentstroke}%
\pgfsetdash{}{0pt}%
\pgfpathmoveto{\pgfqpoint{2.145455in}{1.780000in}}%
\pgfpathlineto{\pgfqpoint{3.545455in}{1.780000in}}%
\pgfusepath{stroke}%
\end{pgfscope}%
\begin{pgfscope}%
\pgfsetrectcap%
\pgfsetmiterjoin%
\pgfsetlinewidth{0.803000pt}%
\definecolor{currentstroke}{rgb}{0.000000,0.000000,0.000000}%
\pgfsetstrokecolor{currentstroke}%
\pgfsetdash{}{0pt}%
\pgfpathmoveto{\pgfqpoint{2.145455in}{3.180000in}}%
\pgfpathlineto{\pgfqpoint{3.545455in}{3.180000in}}%
\pgfusepath{stroke}%
\end{pgfscope}%
\begin{pgfscope}%
\definecolor{textcolor}{rgb}{1.000000,1.000000,1.000000}%
\pgfsetstrokecolor{textcolor}%
\pgfsetfillcolor{textcolor}%
\pgftext[x=2.285455in,y=3.040000in,,]{\color{textcolor}\rmfamily\fontsize{7.000000}{8.400000}\selectfont 1.67}%
\end{pgfscope}%
\begin{pgfscope}%
\definecolor{textcolor}{rgb}{1.000000,1.000000,1.000000}%
\pgfsetstrokecolor{textcolor}%
\pgfsetfillcolor{textcolor}%
\pgftext[x=2.565455in,y=3.040000in,,]{\color{textcolor}\rmfamily\fontsize{7.000000}{8.400000}\selectfont 1.72}%
\end{pgfscope}%
\begin{pgfscope}%
\definecolor{textcolor}{rgb}{1.000000,1.000000,1.000000}%
\pgfsetstrokecolor{textcolor}%
\pgfsetfillcolor{textcolor}%
\pgftext[x=2.845455in,y=3.040000in,,]{\color{textcolor}\rmfamily\fontsize{7.000000}{8.400000}\selectfont 1.88}%
\end{pgfscope}%
\begin{pgfscope}%
\definecolor{textcolor}{rgb}{1.000000,1.000000,1.000000}%
\pgfsetstrokecolor{textcolor}%
\pgfsetfillcolor{textcolor}%
\pgftext[x=3.125455in,y=3.040000in,,]{\color{textcolor}\rmfamily\fontsize{7.000000}{8.400000}\selectfont 2.39}%
\end{pgfscope}%
\begin{pgfscope}%
\definecolor{textcolor}{rgb}{1.000000,1.000000,1.000000}%
\pgfsetstrokecolor{textcolor}%
\pgfsetfillcolor{textcolor}%
\pgftext[x=3.405455in,y=3.040000in,,]{\color{textcolor}\rmfamily\fontsize{7.000000}{8.400000}\selectfont 3.83}%
\end{pgfscope}%
\begin{pgfscope}%
\definecolor{textcolor}{rgb}{1.000000,1.000000,1.000000}%
\pgfsetstrokecolor{textcolor}%
\pgfsetfillcolor{textcolor}%
\pgftext[x=2.285455in,y=2.760000in,,]{\color{textcolor}\rmfamily\fontsize{7.000000}{8.400000}\selectfont 1.05}%
\end{pgfscope}%
\begin{pgfscope}%
\definecolor{textcolor}{rgb}{1.000000,1.000000,1.000000}%
\pgfsetstrokecolor{textcolor}%
\pgfsetfillcolor{textcolor}%
\pgftext[x=2.565455in,y=2.760000in,,]{\color{textcolor}\rmfamily\fontsize{7.000000}{8.400000}\selectfont 1.12}%
\end{pgfscope}%
\begin{pgfscope}%
\definecolor{textcolor}{rgb}{1.000000,1.000000,1.000000}%
\pgfsetstrokecolor{textcolor}%
\pgfsetfillcolor{textcolor}%
\pgftext[x=2.845455in,y=2.760000in,,]{\color{textcolor}\rmfamily\fontsize{7.000000}{8.400000}\selectfont 1.21}%
\end{pgfscope}%
\begin{pgfscope}%
\definecolor{textcolor}{rgb}{1.000000,1.000000,1.000000}%
\pgfsetstrokecolor{textcolor}%
\pgfsetfillcolor{textcolor}%
\pgftext[x=3.125455in,y=2.760000in,,]{\color{textcolor}\rmfamily\fontsize{7.000000}{8.400000}\selectfont 1.66}%
\end{pgfscope}%
\begin{pgfscope}%
\definecolor{textcolor}{rgb}{1.000000,1.000000,1.000000}%
\pgfsetstrokecolor{textcolor}%
\pgfsetfillcolor{textcolor}%
\pgftext[x=3.405455in,y=2.760000in,,]{\color{textcolor}\rmfamily\fontsize{7.000000}{8.400000}\selectfont 2.10}%
\end{pgfscope}%
\begin{pgfscope}%
\definecolor{textcolor}{rgb}{1.000000,1.000000,1.000000}%
\pgfsetstrokecolor{textcolor}%
\pgfsetfillcolor{textcolor}%
\pgftext[x=2.285455in,y=2.480000in,,]{\color{textcolor}\rmfamily\fontsize{7.000000}{8.400000}\selectfont 0.86}%
\end{pgfscope}%
\begin{pgfscope}%
\definecolor{textcolor}{rgb}{1.000000,1.000000,1.000000}%
\pgfsetstrokecolor{textcolor}%
\pgfsetfillcolor{textcolor}%
\pgftext[x=2.565455in,y=2.480000in,,]{\color{textcolor}\rmfamily\fontsize{7.000000}{8.400000}\selectfont 0.95}%
\end{pgfscope}%
\begin{pgfscope}%
\definecolor{textcolor}{rgb}{1.000000,1.000000,1.000000}%
\pgfsetstrokecolor{textcolor}%
\pgfsetfillcolor{textcolor}%
\pgftext[x=2.845455in,y=2.480000in,,]{\color{textcolor}\rmfamily\fontsize{7.000000}{8.400000}\selectfont 1.24}%
\end{pgfscope}%
\begin{pgfscope}%
\definecolor{textcolor}{rgb}{1.000000,1.000000,1.000000}%
\pgfsetstrokecolor{textcolor}%
\pgfsetfillcolor{textcolor}%
\pgftext[x=3.125455in,y=2.480000in,,]{\color{textcolor}\rmfamily\fontsize{7.000000}{8.400000}\selectfont 1.22}%
\end{pgfscope}%
\begin{pgfscope}%
\definecolor{textcolor}{rgb}{1.000000,1.000000,1.000000}%
\pgfsetstrokecolor{textcolor}%
\pgfsetfillcolor{textcolor}%
\pgftext[x=3.405455in,y=2.480000in,,]{\color{textcolor}\rmfamily\fontsize{7.000000}{8.400000}\selectfont 1.22}%
\end{pgfscope}%
\begin{pgfscope}%
\definecolor{textcolor}{rgb}{1.000000,1.000000,1.000000}%
\pgfsetstrokecolor{textcolor}%
\pgfsetfillcolor{textcolor}%
\pgftext[x=2.285455in,y=2.200000in,,]{\color{textcolor}\rmfamily\fontsize{7.000000}{8.400000}\selectfont 0.80}%
\end{pgfscope}%
\begin{pgfscope}%
\definecolor{textcolor}{rgb}{1.000000,1.000000,1.000000}%
\pgfsetstrokecolor{textcolor}%
\pgfsetfillcolor{textcolor}%
\pgftext[x=2.565455in,y=2.200000in,,]{\color{textcolor}\rmfamily\fontsize{7.000000}{8.400000}\selectfont 0.90}%
\end{pgfscope}%
\begin{pgfscope}%
\definecolor{textcolor}{rgb}{1.000000,1.000000,1.000000}%
\pgfsetstrokecolor{textcolor}%
\pgfsetfillcolor{textcolor}%
\pgftext[x=2.845455in,y=2.200000in,,]{\color{textcolor}\rmfamily\fontsize{7.000000}{8.400000}\selectfont 1.23}%
\end{pgfscope}%
\begin{pgfscope}%
\definecolor{textcolor}{rgb}{1.000000,1.000000,1.000000}%
\pgfsetstrokecolor{textcolor}%
\pgfsetfillcolor{textcolor}%
\pgftext[x=3.125455in,y=2.200000in,,]{\color{textcolor}\rmfamily\fontsize{7.000000}{8.400000}\selectfont 1.26}%
\end{pgfscope}%
\begin{pgfscope}%
\definecolor{textcolor}{rgb}{1.000000,1.000000,1.000000}%
\pgfsetstrokecolor{textcolor}%
\pgfsetfillcolor{textcolor}%
\pgftext[x=3.405455in,y=2.200000in,,]{\color{textcolor}\rmfamily\fontsize{7.000000}{8.400000}\selectfont 1.21}%
\end{pgfscope}%
\begin{pgfscope}%
\definecolor{textcolor}{rgb}{1.000000,1.000000,1.000000}%
\pgfsetstrokecolor{textcolor}%
\pgfsetfillcolor{textcolor}%
\pgftext[x=2.285455in,y=1.920000in,,]{\color{textcolor}\rmfamily\fontsize{7.000000}{8.400000}\selectfont 0.45}%
\end{pgfscope}%
\begin{pgfscope}%
\definecolor{textcolor}{rgb}{1.000000,1.000000,1.000000}%
\pgfsetstrokecolor{textcolor}%
\pgfsetfillcolor{textcolor}%
\pgftext[x=2.565455in,y=1.920000in,,]{\color{textcolor}\rmfamily\fontsize{7.000000}{8.400000}\selectfont 0.70}%
\end{pgfscope}%
\begin{pgfscope}%
\definecolor{textcolor}{rgb}{1.000000,1.000000,1.000000}%
\pgfsetstrokecolor{textcolor}%
\pgfsetfillcolor{textcolor}%
\pgftext[x=2.845455in,y=1.920000in,,]{\color{textcolor}\rmfamily\fontsize{7.000000}{8.400000}\selectfont 0.80}%
\end{pgfscope}%
\begin{pgfscope}%
\definecolor{textcolor}{rgb}{1.000000,1.000000,1.000000}%
\pgfsetstrokecolor{textcolor}%
\pgfsetfillcolor{textcolor}%
\pgftext[x=3.125455in,y=1.920000in,,]{\color{textcolor}\rmfamily\fontsize{7.000000}{8.400000}\selectfont 0.91}%
\end{pgfscope}%
\begin{pgfscope}%
\definecolor{textcolor}{rgb}{1.000000,1.000000,1.000000}%
\pgfsetstrokecolor{textcolor}%
\pgfsetfillcolor{textcolor}%
\pgftext[x=3.405455in,y=1.920000in,,]{\color{textcolor}\rmfamily\fontsize{7.000000}{8.400000}\selectfont 1.17}%
\end{pgfscope}%
\begin{pgfscope}%
\definecolor{textcolor}{rgb}{0.000000,0.000000,0.000000}%
\pgfsetstrokecolor{textcolor}%
\pgfsetfillcolor{textcolor}%
\pgftext[x=2.845455in,y=3.263333in,,base]{\color{textcolor}\rmfamily\fontsize{8.400000}{10.080000}\selectfont envelope misfit of \(\displaystyle w\)}%
\end{pgfscope}%
\begin{pgfscope}%
\pgfsetbuttcap%
\pgfsetmiterjoin%
\definecolor{currentfill}{rgb}{1.000000,1.000000,1.000000}%
\pgfsetfillcolor{currentfill}%
\pgfsetlinewidth{0.000000pt}%
\definecolor{currentstroke}{rgb}{0.000000,0.000000,0.000000}%
\pgfsetstrokecolor{currentstroke}%
\pgfsetstrokeopacity{0.000000}%
\pgfsetdash{}{0pt}%
\pgfpathmoveto{\pgfqpoint{0.100000in}{0.100000in}}%
\pgfpathlineto{\pgfqpoint{1.500000in}{0.100000in}}%
\pgfpathlineto{\pgfqpoint{1.500000in}{1.500000in}}%
\pgfpathlineto{\pgfqpoint{0.100000in}{1.500000in}}%
\pgfpathclose%
\pgfusepath{fill}%
\end{pgfscope}%
\begin{pgfscope}%
\pgfpathrectangle{\pgfqpoint{0.100000in}{0.100000in}}{\pgfqpoint{1.400000in}{1.400000in}}%
\pgfusepath{clip}%
\pgfsys@transformshift{0.100000in}{0.100000in}%
\pgftext[left,bottom]{\includegraphics[interpolate=true,width=1.400000in,height=1.400000in]{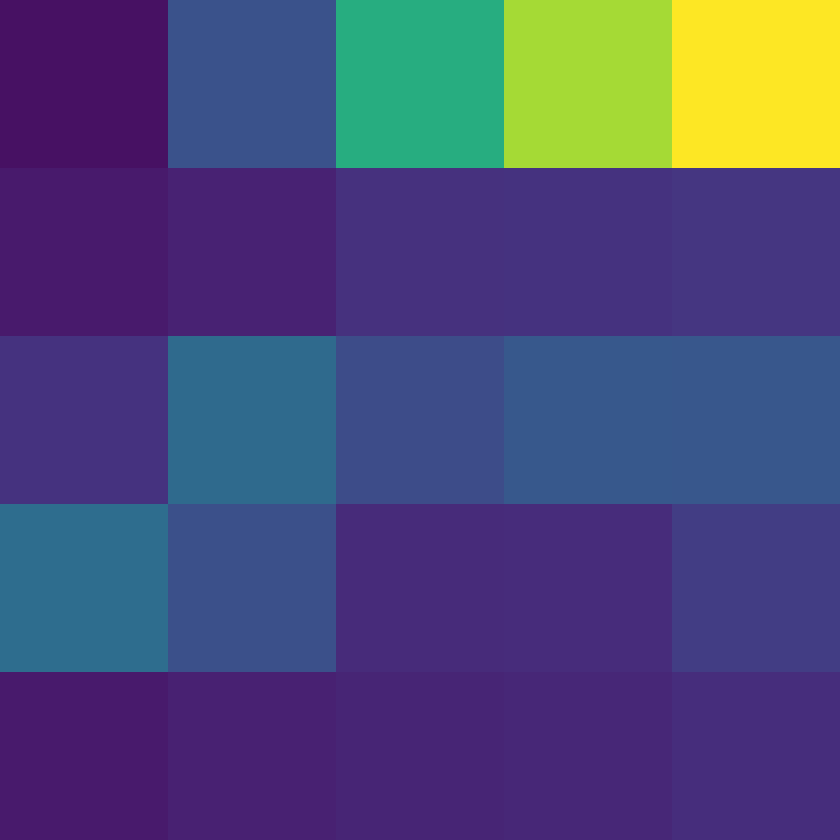}}%
\end{pgfscope}%
\begin{pgfscope}%
\pgfsetrectcap%
\pgfsetmiterjoin%
\pgfsetlinewidth{0.803000pt}%
\definecolor{currentstroke}{rgb}{0.000000,0.000000,0.000000}%
\pgfsetstrokecolor{currentstroke}%
\pgfsetdash{}{0pt}%
\pgfpathmoveto{\pgfqpoint{0.100000in}{0.100000in}}%
\pgfpathlineto{\pgfqpoint{0.100000in}{1.500000in}}%
\pgfusepath{stroke}%
\end{pgfscope}%
\begin{pgfscope}%
\pgfsetrectcap%
\pgfsetmiterjoin%
\pgfsetlinewidth{0.803000pt}%
\definecolor{currentstroke}{rgb}{0.000000,0.000000,0.000000}%
\pgfsetstrokecolor{currentstroke}%
\pgfsetdash{}{0pt}%
\pgfpathmoveto{\pgfqpoint{1.500000in}{0.100000in}}%
\pgfpathlineto{\pgfqpoint{1.500000in}{1.500000in}}%
\pgfusepath{stroke}%
\end{pgfscope}%
\begin{pgfscope}%
\pgfsetrectcap%
\pgfsetmiterjoin%
\pgfsetlinewidth{0.803000pt}%
\definecolor{currentstroke}{rgb}{0.000000,0.000000,0.000000}%
\pgfsetstrokecolor{currentstroke}%
\pgfsetdash{}{0pt}%
\pgfpathmoveto{\pgfqpoint{0.100000in}{0.100000in}}%
\pgfpathlineto{\pgfqpoint{1.500000in}{0.100000in}}%
\pgfusepath{stroke}%
\end{pgfscope}%
\begin{pgfscope}%
\pgfsetrectcap%
\pgfsetmiterjoin%
\pgfsetlinewidth{0.803000pt}%
\definecolor{currentstroke}{rgb}{0.000000,0.000000,0.000000}%
\pgfsetstrokecolor{currentstroke}%
\pgfsetdash{}{0pt}%
\pgfpathmoveto{\pgfqpoint{0.100000in}{1.500000in}}%
\pgfpathlineto{\pgfqpoint{1.500000in}{1.500000in}}%
\pgfusepath{stroke}%
\end{pgfscope}%
\begin{pgfscope}%
\definecolor{textcolor}{rgb}{1.000000,1.000000,1.000000}%
\pgfsetstrokecolor{textcolor}%
\pgfsetfillcolor{textcolor}%
\pgftext[x=0.240000in,y=1.360000in,,]{\color{textcolor}\rmfamily\fontsize{7.000000}{8.400000}\selectfont 0.45}%
\end{pgfscope}%
\begin{pgfscope}%
\definecolor{textcolor}{rgb}{1.000000,1.000000,1.000000}%
\pgfsetstrokecolor{textcolor}%
\pgfsetfillcolor{textcolor}%
\pgftext[x=0.520000in,y=1.360000in,,]{\color{textcolor}\rmfamily\fontsize{7.000000}{8.400000}\selectfont 2.54}%
\end{pgfscope}%
\begin{pgfscope}%
\definecolor{textcolor}{rgb}{0.000000,0.000000,0.000000}%
\pgfsetstrokecolor{textcolor}%
\pgfsetfillcolor{textcolor}%
\pgftext[x=0.800000in,y=1.360000in,,]{\color{textcolor}\rmfamily\fontsize{7.000000}{8.400000}\selectfont 6.22}%
\end{pgfscope}%
\begin{pgfscope}%
\definecolor{textcolor}{rgb}{0.000000,0.000000,0.000000}%
\pgfsetstrokecolor{textcolor}%
\pgfsetfillcolor{textcolor}%
\pgftext[x=1.080000in,y=1.360000in,,]{\color{textcolor}\rmfamily\fontsize{7.000000}{8.400000}\selectfont 8.67}%
\end{pgfscope}%
\begin{pgfscope}%
\definecolor{textcolor}{rgb}{0.000000,0.000000,0.000000}%
\pgfsetstrokecolor{textcolor}%
\pgfsetfillcolor{textcolor}%
\pgftext[x=1.360000in,y=1.360000in,,]{\color{textcolor}\rmfamily\fontsize{7.000000}{8.400000}\selectfont 10.7}%
\end{pgfscope}%
\begin{pgfscope}%
\definecolor{textcolor}{rgb}{1.000000,1.000000,1.000000}%
\pgfsetstrokecolor{textcolor}%
\pgfsetfillcolor{textcolor}%
\pgftext[x=0.240000in,y=1.080000in,,]{\color{textcolor}\rmfamily\fontsize{7.000000}{8.400000}\selectfont 0.74}%
\end{pgfscope}%
\begin{pgfscope}%
\definecolor{textcolor}{rgb}{1.000000,1.000000,1.000000}%
\pgfsetstrokecolor{textcolor}%
\pgfsetfillcolor{textcolor}%
\pgftext[x=0.520000in,y=1.080000in,,]{\color{textcolor}\rmfamily\fontsize{7.000000}{8.400000}\selectfont 0.95}%
\end{pgfscope}%
\begin{pgfscope}%
\definecolor{textcolor}{rgb}{1.000000,1.000000,1.000000}%
\pgfsetstrokecolor{textcolor}%
\pgfsetfillcolor{textcolor}%
\pgftext[x=0.800000in,y=1.080000in,,]{\color{textcolor}\rmfamily\fontsize{7.000000}{8.400000}\selectfont 1.42}%
\end{pgfscope}%
\begin{pgfscope}%
\definecolor{textcolor}{rgb}{1.000000,1.000000,1.000000}%
\pgfsetstrokecolor{textcolor}%
\pgfsetfillcolor{textcolor}%
\pgftext[x=1.080000in,y=1.080000in,,]{\color{textcolor}\rmfamily\fontsize{7.000000}{8.400000}\selectfont 1.46}%
\end{pgfscope}%
\begin{pgfscope}%
\definecolor{textcolor}{rgb}{1.000000,1.000000,1.000000}%
\pgfsetstrokecolor{textcolor}%
\pgfsetfillcolor{textcolor}%
\pgftext[x=1.360000in,y=1.080000in,,]{\color{textcolor}\rmfamily\fontsize{7.000000}{8.400000}\selectfont 1.56}%
\end{pgfscope}%
\begin{pgfscope}%
\definecolor{textcolor}{rgb}{1.000000,1.000000,1.000000}%
\pgfsetstrokecolor{textcolor}%
\pgfsetfillcolor{textcolor}%
\pgftext[x=0.240000in,y=0.800000in,,]{\color{textcolor}\rmfamily\fontsize{7.000000}{8.400000}\selectfont 1.45}%
\end{pgfscope}%
\begin{pgfscope}%
\definecolor{textcolor}{rgb}{1.000000,1.000000,1.000000}%
\pgfsetstrokecolor{textcolor}%
\pgfsetfillcolor{textcolor}%
\pgftext[x=0.520000in,y=0.800000in,,]{\color{textcolor}\rmfamily\fontsize{7.000000}{8.400000}\selectfont 3.46}%
\end{pgfscope}%
\begin{pgfscope}%
\definecolor{textcolor}{rgb}{1.000000,1.000000,1.000000}%
\pgfsetstrokecolor{textcolor}%
\pgfsetfillcolor{textcolor}%
\pgftext[x=0.800000in,y=0.800000in,,]{\color{textcolor}\rmfamily\fontsize{7.000000}{8.400000}\selectfont 2.32}%
\end{pgfscope}%
\begin{pgfscope}%
\definecolor{textcolor}{rgb}{1.000000,1.000000,1.000000}%
\pgfsetstrokecolor{textcolor}%
\pgfsetfillcolor{textcolor}%
\pgftext[x=1.080000in,y=0.800000in,,]{\color{textcolor}\rmfamily\fontsize{7.000000}{8.400000}\selectfont 2.76}%
\end{pgfscope}%
\begin{pgfscope}%
\definecolor{textcolor}{rgb}{1.000000,1.000000,1.000000}%
\pgfsetstrokecolor{textcolor}%
\pgfsetfillcolor{textcolor}%
\pgftext[x=1.360000in,y=0.800000in,,]{\color{textcolor}\rmfamily\fontsize{7.000000}{8.400000}\selectfont 2.71}%
\end{pgfscope}%
\begin{pgfscope}%
\definecolor{textcolor}{rgb}{1.000000,1.000000,1.000000}%
\pgfsetstrokecolor{textcolor}%
\pgfsetfillcolor{textcolor}%
\pgftext[x=0.240000in,y=0.520000in,,]{\color{textcolor}\rmfamily\fontsize{7.000000}{8.400000}\selectfont 3.59}%
\end{pgfscope}%
\begin{pgfscope}%
\definecolor{textcolor}{rgb}{1.000000,1.000000,1.000000}%
\pgfsetstrokecolor{textcolor}%
\pgfsetfillcolor{textcolor}%
\pgftext[x=0.520000in,y=0.520000in,,]{\color{textcolor}\rmfamily\fontsize{7.000000}{8.400000}\selectfont 2.43}%
\end{pgfscope}%
\begin{pgfscope}%
\definecolor{textcolor}{rgb}{1.000000,1.000000,1.000000}%
\pgfsetstrokecolor{textcolor}%
\pgfsetfillcolor{textcolor}%
\pgftext[x=0.800000in,y=0.520000in,,]{\color{textcolor}\rmfamily\fontsize{7.000000}{8.400000}\selectfont 1.22}%
\end{pgfscope}%
\begin{pgfscope}%
\definecolor{textcolor}{rgb}{1.000000,1.000000,1.000000}%
\pgfsetstrokecolor{textcolor}%
\pgfsetfillcolor{textcolor}%
\pgftext[x=1.080000in,y=0.520000in,,]{\color{textcolor}\rmfamily\fontsize{7.000000}{8.400000}\selectfont 1.29}%
\end{pgfscope}%
\begin{pgfscope}%
\definecolor{textcolor}{rgb}{1.000000,1.000000,1.000000}%
\pgfsetstrokecolor{textcolor}%
\pgfsetfillcolor{textcolor}%
\pgftext[x=1.360000in,y=0.520000in,,]{\color{textcolor}\rmfamily\fontsize{7.000000}{8.400000}\selectfont 1.81}%
\end{pgfscope}%
\begin{pgfscope}%
\definecolor{textcolor}{rgb}{1.000000,1.000000,1.000000}%
\pgfsetstrokecolor{textcolor}%
\pgfsetfillcolor{textcolor}%
\pgftext[x=0.240000in,y=0.240000in,,]{\color{textcolor}\rmfamily\fontsize{7.000000}{8.400000}\selectfont 0.73}%
\end{pgfscope}%
\begin{pgfscope}%
\definecolor{textcolor}{rgb}{1.000000,1.000000,1.000000}%
\pgfsetstrokecolor{textcolor}%
\pgfsetfillcolor{textcolor}%
\pgftext[x=0.520000in,y=0.240000in,,]{\color{textcolor}\rmfamily\fontsize{7.000000}{8.400000}\selectfont 0.91}%
\end{pgfscope}%
\begin{pgfscope}%
\definecolor{textcolor}{rgb}{1.000000,1.000000,1.000000}%
\pgfsetstrokecolor{textcolor}%
\pgfsetfillcolor{textcolor}%
\pgftext[x=0.800000in,y=0.240000in,,]{\color{textcolor}\rmfamily\fontsize{7.000000}{8.400000}\selectfont 1.03}%
\end{pgfscope}%
\begin{pgfscope}%
\definecolor{textcolor}{rgb}{1.000000,1.000000,1.000000}%
\pgfsetstrokecolor{textcolor}%
\pgfsetfillcolor{textcolor}%
\pgftext[x=1.080000in,y=0.240000in,,]{\color{textcolor}\rmfamily\fontsize{7.000000}{8.400000}\selectfont 1.11}%
\end{pgfscope}%
\begin{pgfscope}%
\definecolor{textcolor}{rgb}{1.000000,1.000000,1.000000}%
\pgfsetstrokecolor{textcolor}%
\pgfsetfillcolor{textcolor}%
\pgftext[x=1.360000in,y=0.240000in,,]{\color{textcolor}\rmfamily\fontsize{7.000000}{8.400000}\selectfont 1.32}%
\end{pgfscope}%
\begin{pgfscope}%
\definecolor{textcolor}{rgb}{0.000000,0.000000,0.000000}%
\pgfsetstrokecolor{textcolor}%
\pgfsetfillcolor{textcolor}%
\pgftext[x=0.800000in,y=1.583333in,,base]{\color{textcolor}\rmfamily\fontsize{8.400000}{10.080000}\selectfont envelope misfit of \(\displaystyle u_f\)}%
\end{pgfscope}%
\begin{pgfscope}%
\pgfsetbuttcap%
\pgfsetmiterjoin%
\definecolor{currentfill}{rgb}{1.000000,1.000000,1.000000}%
\pgfsetfillcolor{currentfill}%
\pgfsetlinewidth{0.000000pt}%
\definecolor{currentstroke}{rgb}{0.000000,0.000000,0.000000}%
\pgfsetstrokecolor{currentstroke}%
\pgfsetstrokeopacity{0.000000}%
\pgfsetdash{}{0pt}%
\pgfpathmoveto{\pgfqpoint{2.145455in}{0.100000in}}%
\pgfpathlineto{\pgfqpoint{3.545455in}{0.100000in}}%
\pgfpathlineto{\pgfqpoint{3.545455in}{1.500000in}}%
\pgfpathlineto{\pgfqpoint{2.145455in}{1.500000in}}%
\pgfpathclose%
\pgfusepath{fill}%
\end{pgfscope}%
\begin{pgfscope}%
\pgfpathrectangle{\pgfqpoint{2.145455in}{0.100000in}}{\pgfqpoint{1.400000in}{1.400000in}}%
\pgfusepath{clip}%
\pgfsys@transformshift{2.145455in}{0.100000in}%
\pgftext[left,bottom]{\includegraphics[interpolate=true,width=1.401667in,height=1.400000in]{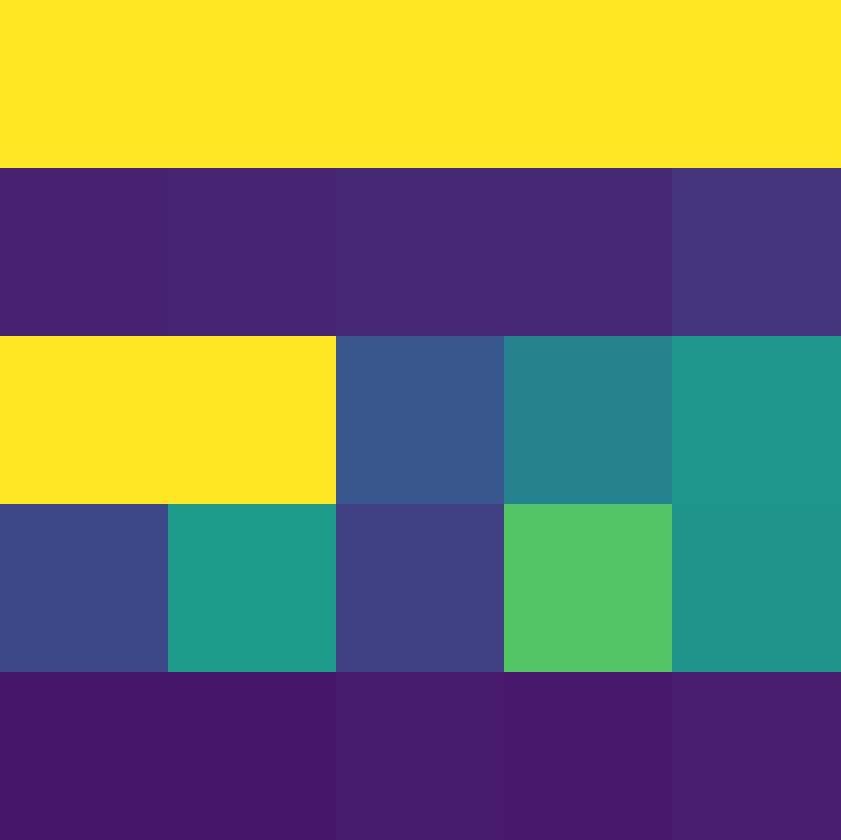}}%
\end{pgfscope}%
\begin{pgfscope}%
\pgfsetrectcap%
\pgfsetmiterjoin%
\pgfsetlinewidth{0.803000pt}%
\definecolor{currentstroke}{rgb}{0.000000,0.000000,0.000000}%
\pgfsetstrokecolor{currentstroke}%
\pgfsetdash{}{0pt}%
\pgfpathmoveto{\pgfqpoint{2.145455in}{0.100000in}}%
\pgfpathlineto{\pgfqpoint{2.145455in}{1.500000in}}%
\pgfusepath{stroke}%
\end{pgfscope}%
\begin{pgfscope}%
\pgfsetrectcap%
\pgfsetmiterjoin%
\pgfsetlinewidth{0.803000pt}%
\definecolor{currentstroke}{rgb}{0.000000,0.000000,0.000000}%
\pgfsetstrokecolor{currentstroke}%
\pgfsetdash{}{0pt}%
\pgfpathmoveto{\pgfqpoint{3.545455in}{0.100000in}}%
\pgfpathlineto{\pgfqpoint{3.545455in}{1.500000in}}%
\pgfusepath{stroke}%
\end{pgfscope}%
\begin{pgfscope}%
\pgfsetrectcap%
\pgfsetmiterjoin%
\pgfsetlinewidth{0.803000pt}%
\definecolor{currentstroke}{rgb}{0.000000,0.000000,0.000000}%
\pgfsetstrokecolor{currentstroke}%
\pgfsetdash{}{0pt}%
\pgfpathmoveto{\pgfqpoint{2.145455in}{0.100000in}}%
\pgfpathlineto{\pgfqpoint{3.545455in}{0.100000in}}%
\pgfusepath{stroke}%
\end{pgfscope}%
\begin{pgfscope}%
\pgfsetrectcap%
\pgfsetmiterjoin%
\pgfsetlinewidth{0.803000pt}%
\definecolor{currentstroke}{rgb}{0.000000,0.000000,0.000000}%
\pgfsetstrokecolor{currentstroke}%
\pgfsetdash{}{0pt}%
\pgfpathmoveto{\pgfqpoint{2.145455in}{1.500000in}}%
\pgfpathlineto{\pgfqpoint{3.545455in}{1.500000in}}%
\pgfusepath{stroke}%
\end{pgfscope}%
\begin{pgfscope}%
\definecolor{textcolor}{rgb}{0.000000,0.000000,0.000000}%
\pgfsetstrokecolor{textcolor}%
\pgfsetfillcolor{textcolor}%
\pgftext[x=2.285455in,y=1.360000in,,]{\color{textcolor}\rmfamily\fontsize{7.000000}{8.400000}\selectfont 190}%
\end{pgfscope}%
\begin{pgfscope}%
\definecolor{textcolor}{rgb}{0.000000,0.000000,0.000000}%
\pgfsetstrokecolor{textcolor}%
\pgfsetfillcolor{textcolor}%
\pgftext[x=2.565455in,y=1.360000in,,]{\color{textcolor}\rmfamily\fontsize{7.000000}{8.400000}\selectfont 199}%
\end{pgfscope}%
\begin{pgfscope}%
\definecolor{textcolor}{rgb}{0.000000,0.000000,0.000000}%
\pgfsetstrokecolor{textcolor}%
\pgfsetfillcolor{textcolor}%
\pgftext[x=2.845455in,y=1.360000in,,]{\color{textcolor}\rmfamily\fontsize{7.000000}{8.400000}\selectfont 161}%
\end{pgfscope}%
\begin{pgfscope}%
\definecolor{textcolor}{rgb}{0.000000,0.000000,0.000000}%
\pgfsetstrokecolor{textcolor}%
\pgfsetfillcolor{textcolor}%
\pgftext[x=3.125455in,y=1.360000in,,]{\color{textcolor}\rmfamily\fontsize{7.000000}{8.400000}\selectfont 224}%
\end{pgfscope}%
\begin{pgfscope}%
\definecolor{textcolor}{rgb}{0.000000,0.000000,0.000000}%
\pgfsetstrokecolor{textcolor}%
\pgfsetfillcolor{textcolor}%
\pgftext[x=3.405455in,y=1.360000in,,]{\color{textcolor}\rmfamily\fontsize{7.000000}{8.400000}\selectfont 138}%
\end{pgfscope}%
\begin{pgfscope}%
\definecolor{textcolor}{rgb}{1.000000,1.000000,1.000000}%
\pgfsetstrokecolor{textcolor}%
\pgfsetfillcolor{textcolor}%
\pgftext[x=2.285455in,y=1.080000in,,]{\color{textcolor}\rmfamily\fontsize{7.000000}{8.400000}\selectfont 0.97}%
\end{pgfscope}%
\begin{pgfscope}%
\definecolor{textcolor}{rgb}{1.000000,1.000000,1.000000}%
\pgfsetstrokecolor{textcolor}%
\pgfsetfillcolor{textcolor}%
\pgftext[x=2.565455in,y=1.080000in,,]{\color{textcolor}\rmfamily\fontsize{7.000000}{8.400000}\selectfont 1.09}%
\end{pgfscope}%
\begin{pgfscope}%
\definecolor{textcolor}{rgb}{1.000000,1.000000,1.000000}%
\pgfsetstrokecolor{textcolor}%
\pgfsetfillcolor{textcolor}%
\pgftext[x=2.845455in,y=1.080000in,,]{\color{textcolor}\rmfamily\fontsize{7.000000}{8.400000}\selectfont 1.15}%
\end{pgfscope}%
\begin{pgfscope}%
\definecolor{textcolor}{rgb}{1.000000,1.000000,1.000000}%
\pgfsetstrokecolor{textcolor}%
\pgfsetfillcolor{textcolor}%
\pgftext[x=3.125455in,y=1.080000in,,]{\color{textcolor}\rmfamily\fontsize{7.000000}{8.400000}\selectfont 1.13}%
\end{pgfscope}%
\begin{pgfscope}%
\definecolor{textcolor}{rgb}{1.000000,1.000000,1.000000}%
\pgfsetstrokecolor{textcolor}%
\pgfsetfillcolor{textcolor}%
\pgftext[x=3.405455in,y=1.080000in,,]{\color{textcolor}\rmfamily\fontsize{7.000000}{8.400000}\selectfont 1.55}%
\end{pgfscope}%
\begin{pgfscope}%
\definecolor{textcolor}{rgb}{0.000000,0.000000,0.000000}%
\pgfsetstrokecolor{textcolor}%
\pgfsetfillcolor{textcolor}%
\pgftext[x=2.285455in,y=0.800000in,,]{\color{textcolor}\rmfamily\fontsize{7.000000}{8.400000}\selectfont 15.7}%
\end{pgfscope}%
\begin{pgfscope}%
\definecolor{textcolor}{rgb}{0.000000,0.000000,0.000000}%
\pgfsetstrokecolor{textcolor}%
\pgfsetfillcolor{textcolor}%
\pgftext[x=2.565455in,y=0.800000in,,]{\color{textcolor}\rmfamily\fontsize{7.000000}{8.400000}\selectfont 20.5}%
\end{pgfscope}%
\begin{pgfscope}%
\definecolor{textcolor}{rgb}{1.000000,1.000000,1.000000}%
\pgfsetstrokecolor{textcolor}%
\pgfsetfillcolor{textcolor}%
\pgftext[x=2.845455in,y=0.800000in,,]{\color{textcolor}\rmfamily\fontsize{7.000000}{8.400000}\selectfont 2.73}%
\end{pgfscope}%
\begin{pgfscope}%
\definecolor{textcolor}{rgb}{1.000000,1.000000,1.000000}%
\pgfsetstrokecolor{textcolor}%
\pgfsetfillcolor{textcolor}%
\pgftext[x=3.125455in,y=0.800000in,,]{\color{textcolor}\rmfamily\fontsize{7.000000}{8.400000}\selectfont 4.44}%
\end{pgfscope}%
\begin{pgfscope}%
\definecolor{textcolor}{rgb}{0.000000,0.000000,0.000000}%
\pgfsetstrokecolor{textcolor}%
\pgfsetfillcolor{textcolor}%
\pgftext[x=3.405455in,y=0.800000in,,]{\color{textcolor}\rmfamily\fontsize{7.000000}{8.400000}\selectfont 5.33}%
\end{pgfscope}%
\begin{pgfscope}%
\definecolor{textcolor}{rgb}{1.000000,1.000000,1.000000}%
\pgfsetstrokecolor{textcolor}%
\pgfsetfillcolor{textcolor}%
\pgftext[x=2.285455in,y=0.520000in,,]{\color{textcolor}\rmfamily\fontsize{7.000000}{8.400000}\selectfont 2.18}%
\end{pgfscope}%
\begin{pgfscope}%
\definecolor{textcolor}{rgb}{0.000000,0.000000,0.000000}%
\pgfsetstrokecolor{textcolor}%
\pgfsetfillcolor{textcolor}%
\pgftext[x=2.565455in,y=0.520000in,,]{\color{textcolor}\rmfamily\fontsize{7.000000}{8.400000}\selectfont 5.51}%
\end{pgfscope}%
\begin{pgfscope}%
\definecolor{textcolor}{rgb}{1.000000,1.000000,1.000000}%
\pgfsetstrokecolor{textcolor}%
\pgfsetfillcolor{textcolor}%
\pgftext[x=2.845455in,y=0.520000in,,]{\color{textcolor}\rmfamily\fontsize{7.000000}{8.400000}\selectfont 1.96}%
\end{pgfscope}%
\begin{pgfscope}%
\definecolor{textcolor}{rgb}{0.000000,0.000000,0.000000}%
\pgfsetstrokecolor{textcolor}%
\pgfsetfillcolor{textcolor}%
\pgftext[x=3.125455in,y=0.520000in,,]{\color{textcolor}\rmfamily\fontsize{7.000000}{8.400000}\selectfont 7.31}%
\end{pgfscope}%
\begin{pgfscope}%
\definecolor{textcolor}{rgb}{0.000000,0.000000,0.000000}%
\pgfsetstrokecolor{textcolor}%
\pgfsetfillcolor{textcolor}%
\pgftext[x=3.405455in,y=0.520000in,,]{\color{textcolor}\rmfamily\fontsize{7.000000}{8.400000}\selectfont 5.23}%
\end{pgfscope}%
\begin{pgfscope}%
\definecolor{textcolor}{rgb}{1.000000,1.000000,1.000000}%
\pgfsetstrokecolor{textcolor}%
\pgfsetfillcolor{textcolor}%
\pgftext[x=2.285455in,y=0.240000in,,]{\color{textcolor}\rmfamily\fontsize{7.000000}{8.400000}\selectfont 0.59}%
\end{pgfscope}%
\begin{pgfscope}%
\definecolor{textcolor}{rgb}{1.000000,1.000000,1.000000}%
\pgfsetstrokecolor{textcolor}%
\pgfsetfillcolor{textcolor}%
\pgftext[x=2.565455in,y=0.240000in,,]{\color{textcolor}\rmfamily\fontsize{7.000000}{8.400000}\selectfont 0.65}%
\end{pgfscope}%
\begin{pgfscope}%
\definecolor{textcolor}{rgb}{1.000000,1.000000,1.000000}%
\pgfsetstrokecolor{textcolor}%
\pgfsetfillcolor{textcolor}%
\pgftext[x=2.845455in,y=0.240000in,,]{\color{textcolor}\rmfamily\fontsize{7.000000}{8.400000}\selectfont 0.74}%
\end{pgfscope}%
\begin{pgfscope}%
\definecolor{textcolor}{rgb}{1.000000,1.000000,1.000000}%
\pgfsetstrokecolor{textcolor}%
\pgfsetfillcolor{textcolor}%
\pgftext[x=3.125455in,y=0.240000in,,]{\color{textcolor}\rmfamily\fontsize{7.000000}{8.400000}\selectfont 0.67}%
\end{pgfscope}%
\begin{pgfscope}%
\definecolor{textcolor}{rgb}{1.000000,1.000000,1.000000}%
\pgfsetstrokecolor{textcolor}%
\pgfsetfillcolor{textcolor}%
\pgftext[x=3.405455in,y=0.240000in,,]{\color{textcolor}\rmfamily\fontsize{7.000000}{8.400000}\selectfont 0.83}%
\end{pgfscope}%
\begin{pgfscope}%
\definecolor{textcolor}{rgb}{0.000000,0.000000,0.000000}%
\pgfsetstrokecolor{textcolor}%
\pgfsetfillcolor{textcolor}%
\pgftext[x=2.845455in,y=1.583333in,,base]{\color{textcolor}\rmfamily\fontsize{8.400000}{10.080000}\selectfont envelope misfit of \(\displaystyle w_f\)}%
\end{pgfscope}%
\begin{pgfscope}%
\pgfsetbuttcap%
\pgfsetmiterjoin%
\definecolor{currentfill}{rgb}{1.000000,1.000000,1.000000}%
\pgfsetfillcolor{currentfill}%
\pgfsetlinewidth{0.000000pt}%
\definecolor{currentstroke}{rgb}{0.000000,0.000000,0.000000}%
\pgfsetstrokecolor{currentstroke}%
\pgfsetstrokeopacity{0.000000}%
\pgfsetdash{}{0pt}%
\pgfpathmoveto{\pgfqpoint{3.747955in}{0.870000in}}%
\pgfpathlineto{\pgfqpoint{3.824955in}{0.870000in}}%
\pgfpathlineto{\pgfqpoint{3.824955in}{2.410000in}}%
\pgfpathlineto{\pgfqpoint{3.747955in}{2.410000in}}%
\pgfpathclose%
\pgfusepath{fill}%
\end{pgfscope}%
\begin{pgfscope}%
\pgfpathrectangle{\pgfqpoint{3.747955in}{0.870000in}}{\pgfqpoint{0.077000in}{1.540000in}}%
\pgfusepath{clip}%
\pgfsetbuttcap%
\pgfsetmiterjoin%
\definecolor{currentfill}{rgb}{1.000000,1.000000,1.000000}%
\pgfsetfillcolor{currentfill}%
\pgfsetlinewidth{0.010037pt}%
\definecolor{currentstroke}{rgb}{1.000000,1.000000,1.000000}%
\pgfsetstrokecolor{currentstroke}%
\pgfsetdash{}{0pt}%
\pgfpathmoveto{\pgfqpoint{3.747955in}{0.870000in}}%
\pgfpathlineto{\pgfqpoint{3.747955in}{0.876016in}}%
\pgfpathlineto{\pgfqpoint{3.747955in}{2.403984in}}%
\pgfpathlineto{\pgfqpoint{3.747955in}{2.410000in}}%
\pgfpathlineto{\pgfqpoint{3.824955in}{2.410000in}}%
\pgfpathlineto{\pgfqpoint{3.824955in}{2.403984in}}%
\pgfpathlineto{\pgfqpoint{3.824955in}{0.876016in}}%
\pgfpathlineto{\pgfqpoint{3.824955in}{0.870000in}}%
\pgfpathlineto{\pgfqpoint{3.824955in}{0.870000in}}%
\pgfpathclose%
\pgfusepath{stroke,fill}%
\end{pgfscope}%
\begin{pgfscope}%
\pgfsys@transformshift{3.748333in}{0.870494in}%
\pgftext[left,bottom]{\includegraphics[interpolate=true,width=0.076667in,height=1.540000in]{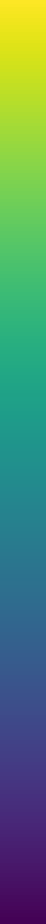}}%
\end{pgfscope}%
\begin{pgfscope}%
\pgfsetbuttcap%
\pgfsetroundjoin%
\definecolor{currentfill}{rgb}{0.000000,0.000000,0.000000}%
\pgfsetfillcolor{currentfill}%
\pgfsetlinewidth{0.803000pt}%
\definecolor{currentstroke}{rgb}{0.000000,0.000000,0.000000}%
\pgfsetstrokecolor{currentstroke}%
\pgfsetdash{}{0pt}%
\pgfsys@defobject{currentmarker}{\pgfqpoint{0.000000in}{0.000000in}}{\pgfqpoint{0.048611in}{0.000000in}}{%
\pgfpathmoveto{\pgfqpoint{0.000000in}{0.000000in}}%
\pgfpathlineto{\pgfqpoint{0.048611in}{0.000000in}}%
\pgfusepath{stroke,fill}%
}%
\begin{pgfscope}%
\pgfsys@transformshift{3.824955in}{0.870000in}%
\pgfsys@useobject{currentmarker}{}%
\end{pgfscope}%
\end{pgfscope}%
\begin{pgfscope}%
\definecolor{textcolor}{rgb}{0.000000,0.000000,0.000000}%
\pgfsetstrokecolor{textcolor}%
\pgfsetfillcolor{textcolor}%
\pgftext[x=3.922177in, y=0.836242in, left, base]{\color{textcolor}\rmfamily\fontsize{7.000000}{8.400000}\selectfont \(\displaystyle {0}\)}%
\end{pgfscope}%
\begin{pgfscope}%
\pgfsetbuttcap%
\pgfsetroundjoin%
\definecolor{currentfill}{rgb}{0.000000,0.000000,0.000000}%
\pgfsetfillcolor{currentfill}%
\pgfsetlinewidth{0.803000pt}%
\definecolor{currentstroke}{rgb}{0.000000,0.000000,0.000000}%
\pgfsetstrokecolor{currentstroke}%
\pgfsetdash{}{0pt}%
\pgfsys@defobject{currentmarker}{\pgfqpoint{0.000000in}{0.000000in}}{\pgfqpoint{0.048611in}{0.000000in}}{%
\pgfpathmoveto{\pgfqpoint{0.000000in}{0.000000in}}%
\pgfpathlineto{\pgfqpoint{0.048611in}{0.000000in}}%
\pgfusepath{stroke,fill}%
}%
\begin{pgfscope}%
\pgfsys@transformshift{3.824955in}{1.178000in}%
\pgfsys@useobject{currentmarker}{}%
\end{pgfscope}%
\end{pgfscope}%
\begin{pgfscope}%
\definecolor{textcolor}{rgb}{0.000000,0.000000,0.000000}%
\pgfsetstrokecolor{textcolor}%
\pgfsetfillcolor{textcolor}%
\pgftext[x=3.922177in, y=1.144242in, left, base]{\color{textcolor}\rmfamily\fontsize{7.000000}{8.400000}\selectfont \(\displaystyle {2}\)}%
\end{pgfscope}%
\begin{pgfscope}%
\pgfsetbuttcap%
\pgfsetroundjoin%
\definecolor{currentfill}{rgb}{0.000000,0.000000,0.000000}%
\pgfsetfillcolor{currentfill}%
\pgfsetlinewidth{0.803000pt}%
\definecolor{currentstroke}{rgb}{0.000000,0.000000,0.000000}%
\pgfsetstrokecolor{currentstroke}%
\pgfsetdash{}{0pt}%
\pgfsys@defobject{currentmarker}{\pgfqpoint{0.000000in}{0.000000in}}{\pgfqpoint{0.048611in}{0.000000in}}{%
\pgfpathmoveto{\pgfqpoint{0.000000in}{0.000000in}}%
\pgfpathlineto{\pgfqpoint{0.048611in}{0.000000in}}%
\pgfusepath{stroke,fill}%
}%
\begin{pgfscope}%
\pgfsys@transformshift{3.824955in}{1.486000in}%
\pgfsys@useobject{currentmarker}{}%
\end{pgfscope}%
\end{pgfscope}%
\begin{pgfscope}%
\definecolor{textcolor}{rgb}{0.000000,0.000000,0.000000}%
\pgfsetstrokecolor{textcolor}%
\pgfsetfillcolor{textcolor}%
\pgftext[x=3.922177in, y=1.452242in, left, base]{\color{textcolor}\rmfamily\fontsize{7.000000}{8.400000}\selectfont \(\displaystyle {4}\)}%
\end{pgfscope}%
\begin{pgfscope}%
\pgfsetbuttcap%
\pgfsetroundjoin%
\definecolor{currentfill}{rgb}{0.000000,0.000000,0.000000}%
\pgfsetfillcolor{currentfill}%
\pgfsetlinewidth{0.803000pt}%
\definecolor{currentstroke}{rgb}{0.000000,0.000000,0.000000}%
\pgfsetstrokecolor{currentstroke}%
\pgfsetdash{}{0pt}%
\pgfsys@defobject{currentmarker}{\pgfqpoint{0.000000in}{0.000000in}}{\pgfqpoint{0.048611in}{0.000000in}}{%
\pgfpathmoveto{\pgfqpoint{0.000000in}{0.000000in}}%
\pgfpathlineto{\pgfqpoint{0.048611in}{0.000000in}}%
\pgfusepath{stroke,fill}%
}%
\begin{pgfscope}%
\pgfsys@transformshift{3.824955in}{1.794000in}%
\pgfsys@useobject{currentmarker}{}%
\end{pgfscope}%
\end{pgfscope}%
\begin{pgfscope}%
\definecolor{textcolor}{rgb}{0.000000,0.000000,0.000000}%
\pgfsetstrokecolor{textcolor}%
\pgfsetfillcolor{textcolor}%
\pgftext[x=3.922177in, y=1.760242in, left, base]{\color{textcolor}\rmfamily\fontsize{7.000000}{8.400000}\selectfont \(\displaystyle {6}\)}%
\end{pgfscope}%
\begin{pgfscope}%
\pgfsetbuttcap%
\pgfsetroundjoin%
\definecolor{currentfill}{rgb}{0.000000,0.000000,0.000000}%
\pgfsetfillcolor{currentfill}%
\pgfsetlinewidth{0.803000pt}%
\definecolor{currentstroke}{rgb}{0.000000,0.000000,0.000000}%
\pgfsetstrokecolor{currentstroke}%
\pgfsetdash{}{0pt}%
\pgfsys@defobject{currentmarker}{\pgfqpoint{0.000000in}{0.000000in}}{\pgfqpoint{0.048611in}{0.000000in}}{%
\pgfpathmoveto{\pgfqpoint{0.000000in}{0.000000in}}%
\pgfpathlineto{\pgfqpoint{0.048611in}{0.000000in}}%
\pgfusepath{stroke,fill}%
}%
\begin{pgfscope}%
\pgfsys@transformshift{3.824955in}{2.102000in}%
\pgfsys@useobject{currentmarker}{}%
\end{pgfscope}%
\end{pgfscope}%
\begin{pgfscope}%
\definecolor{textcolor}{rgb}{0.000000,0.000000,0.000000}%
\pgfsetstrokecolor{textcolor}%
\pgfsetfillcolor{textcolor}%
\pgftext[x=3.922177in, y=2.068242in, left, base]{\color{textcolor}\rmfamily\fontsize{7.000000}{8.400000}\selectfont \(\displaystyle {8}\)}%
\end{pgfscope}%
\begin{pgfscope}%
\pgfsetbuttcap%
\pgfsetroundjoin%
\definecolor{currentfill}{rgb}{0.000000,0.000000,0.000000}%
\pgfsetfillcolor{currentfill}%
\pgfsetlinewidth{0.803000pt}%
\definecolor{currentstroke}{rgb}{0.000000,0.000000,0.000000}%
\pgfsetstrokecolor{currentstroke}%
\pgfsetdash{}{0pt}%
\pgfsys@defobject{currentmarker}{\pgfqpoint{0.000000in}{0.000000in}}{\pgfqpoint{0.048611in}{0.000000in}}{%
\pgfpathmoveto{\pgfqpoint{0.000000in}{0.000000in}}%
\pgfpathlineto{\pgfqpoint{0.048611in}{0.000000in}}%
\pgfusepath{stroke,fill}%
}%
\begin{pgfscope}%
\pgfsys@transformshift{3.824955in}{2.410000in}%
\pgfsys@useobject{currentmarker}{}%
\end{pgfscope}%
\end{pgfscope}%
\begin{pgfscope}%
\definecolor{textcolor}{rgb}{0.000000,0.000000,0.000000}%
\pgfsetstrokecolor{textcolor}%
\pgfsetfillcolor{textcolor}%
\pgftext[x=3.922177in, y=2.376242in, left, base]{\color{textcolor}\rmfamily\fontsize{7.000000}{8.400000}\selectfont \(\displaystyle {10}\)}%
\end{pgfscope}%
\begin{pgfscope}%
\definecolor{textcolor}{rgb}{0.000000,0.000000,0.000000}%
\pgfsetstrokecolor{textcolor}%
\pgfsetfillcolor{textcolor}%
\pgftext[x=4.088458in,y=1.640000in,,top,rotate=90.000000]{\color{textcolor}\rmfamily\fontsize{7.000000}{8.400000}\selectfont [\%]}%
\end{pgfscope}%
\begin{pgfscope}%
\pgfsetrectcap%
\pgfsetmiterjoin%
\pgfsetlinewidth{0.803000pt}%
\definecolor{currentstroke}{rgb}{0.000000,0.000000,0.000000}%
\pgfsetstrokecolor{currentstroke}%
\pgfsetdash{}{0pt}%
\pgfpathmoveto{\pgfqpoint{3.747955in}{0.870000in}}%
\pgfpathlineto{\pgfqpoint{3.747955in}{0.876016in}}%
\pgfpathlineto{\pgfqpoint{3.747955in}{2.403984in}}%
\pgfpathlineto{\pgfqpoint{3.747955in}{2.410000in}}%
\pgfpathlineto{\pgfqpoint{3.824955in}{2.410000in}}%
\pgfpathlineto{\pgfqpoint{3.824955in}{2.403984in}}%
\pgfpathlineto{\pgfqpoint{3.824955in}{0.876016in}}%
\pgfpathlineto{\pgfqpoint{3.824955in}{0.870000in}}%
\pgfpathclose%
\pgfusepath{stroke}%
\end{pgfscope}%
\end{pgfpicture}%
\makeatother%
\endgroup%

%% file: figures/loh_plots/misfit_depth.pgf
\begingroup%
\makeatletter%
\begin{pgfpicture}%
\pgfpathrectangle{\pgfpointorigin}{\pgfqpoint{3.815726in}{2.313913in}}%
\pgfusepath{use as bounding box, clip}%
\begin{pgfscope}%
\pgfsetbuttcap%
\pgfsetmiterjoin%
\definecolor{currentfill}{rgb}{1.000000,1.000000,1.000000}%
\pgfsetfillcolor{currentfill}%
\pgfsetlinewidth{0.000000pt}%
\definecolor{currentstroke}{rgb}{1.000000,1.000000,1.000000}%
\pgfsetstrokecolor{currentstroke}%
\pgfsetdash{}{0pt}%
\pgfpathmoveto{\pgfqpoint{0.000000in}{-0.000000in}}%
\pgfpathlineto{\pgfqpoint{3.815726in}{-0.000000in}}%
\pgfpathlineto{\pgfqpoint{3.815726in}{2.313913in}}%
\pgfpathlineto{\pgfqpoint{0.000000in}{2.313913in}}%
\pgfpathclose%
\pgfusepath{fill}%
\end{pgfscope}%
\begin{pgfscope}%
\pgfsetbuttcap%
\pgfsetmiterjoin%
\definecolor{currentfill}{rgb}{1.000000,1.000000,1.000000}%
\pgfsetfillcolor{currentfill}%
\pgfsetlinewidth{0.000000pt}%
\definecolor{currentstroke}{rgb}{0.000000,0.000000,0.000000}%
\pgfsetstrokecolor{currentstroke}%
\pgfsetstrokeopacity{0.000000}%
\pgfsetdash{}{0pt}%
\pgfpathmoveto{\pgfqpoint{0.460726in}{0.436419in}}%
\pgfpathlineto{\pgfqpoint{3.715726in}{0.436419in}}%
\pgfpathlineto{\pgfqpoint{3.715726in}{2.053419in}}%
\pgfpathlineto{\pgfqpoint{0.460726in}{2.053419in}}%
\pgfpathclose%
\pgfusepath{fill}%
\end{pgfscope}%
\begin{pgfscope}%
\pgfsetbuttcap%
\pgfsetroundjoin%
\definecolor{currentfill}{rgb}{0.000000,0.000000,0.000000}%
\pgfsetfillcolor{currentfill}%
\pgfsetlinewidth{0.803000pt}%
\definecolor{currentstroke}{rgb}{0.000000,0.000000,0.000000}%
\pgfsetstrokecolor{currentstroke}%
\pgfsetdash{}{0pt}%
\pgfsys@defobject{currentmarker}{\pgfqpoint{0.000000in}{-0.048611in}}{\pgfqpoint{0.000000in}{0.000000in}}{%
\pgfpathmoveto{\pgfqpoint{0.000000in}{0.000000in}}%
\pgfpathlineto{\pgfqpoint{0.000000in}{-0.048611in}}%
\pgfusepath{stroke,fill}%
}%
\begin{pgfscope}%
\pgfsys@transformshift{0.608680in}{0.436419in}%
\pgfsys@useobject{currentmarker}{}%
\end{pgfscope}%
\end{pgfscope}%
\begin{pgfscope}%
\definecolor{textcolor}{rgb}{0.000000,0.000000,0.000000}%
\pgfsetstrokecolor{textcolor}%
\pgfsetfillcolor{textcolor}%
\pgftext[x=0.608680in,y=0.339197in,,top]{\color{textcolor}\rmfamily\fontsize{7.000000}{8.400000}\selectfont   0}%
\end{pgfscope}%
\begin{pgfscope}%
\pgfsetbuttcap%
\pgfsetroundjoin%
\definecolor{currentfill}{rgb}{0.000000,0.000000,0.000000}%
\pgfsetfillcolor{currentfill}%
\pgfsetlinewidth{0.803000pt}%
\definecolor{currentstroke}{rgb}{0.000000,0.000000,0.000000}%
\pgfsetstrokecolor{currentstroke}%
\pgfsetdash{}{0pt}%
\pgfsys@defobject{currentmarker}{\pgfqpoint{0.000000in}{-0.048611in}}{\pgfqpoint{0.000000in}{0.000000in}}{%
\pgfpathmoveto{\pgfqpoint{0.000000in}{0.000000in}}%
\pgfpathlineto{\pgfqpoint{0.000000in}{-0.048611in}}%
\pgfusepath{stroke,fill}%
}%
\begin{pgfscope}%
\pgfsys@transformshift{1.200499in}{0.436419in}%
\pgfsys@useobject{currentmarker}{}%
\end{pgfscope}%
\end{pgfscope}%
\begin{pgfscope}%
\definecolor{textcolor}{rgb}{0.000000,0.000000,0.000000}%
\pgfsetstrokecolor{textcolor}%
\pgfsetfillcolor{textcolor}%
\pgftext[x=1.200499in,y=0.339197in,,top]{\color{textcolor}\rmfamily\fontsize{7.000000}{8.400000}\selectfont  50}%
\end{pgfscope}%
\begin{pgfscope}%
\pgfsetbuttcap%
\pgfsetroundjoin%
\definecolor{currentfill}{rgb}{0.000000,0.000000,0.000000}%
\pgfsetfillcolor{currentfill}%
\pgfsetlinewidth{0.803000pt}%
\definecolor{currentstroke}{rgb}{0.000000,0.000000,0.000000}%
\pgfsetstrokecolor{currentstroke}%
\pgfsetdash{}{0pt}%
\pgfsys@defobject{currentmarker}{\pgfqpoint{0.000000in}{-0.048611in}}{\pgfqpoint{0.000000in}{0.000000in}}{%
\pgfpathmoveto{\pgfqpoint{0.000000in}{0.000000in}}%
\pgfpathlineto{\pgfqpoint{0.000000in}{-0.048611in}}%
\pgfusepath{stroke,fill}%
}%
\begin{pgfscope}%
\pgfsys@transformshift{1.792317in}{0.436419in}%
\pgfsys@useobject{currentmarker}{}%
\end{pgfscope}%
\end{pgfscope}%
\begin{pgfscope}%
\definecolor{textcolor}{rgb}{0.000000,0.000000,0.000000}%
\pgfsetstrokecolor{textcolor}%
\pgfsetfillcolor{textcolor}%
\pgftext[x=1.792317in,y=0.339197in,,top]{\color{textcolor}\rmfamily\fontsize{7.000000}{8.400000}\selectfont 100}%
\end{pgfscope}%
\begin{pgfscope}%
\pgfsetbuttcap%
\pgfsetroundjoin%
\definecolor{currentfill}{rgb}{0.000000,0.000000,0.000000}%
\pgfsetfillcolor{currentfill}%
\pgfsetlinewidth{0.803000pt}%
\definecolor{currentstroke}{rgb}{0.000000,0.000000,0.000000}%
\pgfsetstrokecolor{currentstroke}%
\pgfsetdash{}{0pt}%
\pgfsys@defobject{currentmarker}{\pgfqpoint{0.000000in}{-0.048611in}}{\pgfqpoint{0.000000in}{0.000000in}}{%
\pgfpathmoveto{\pgfqpoint{0.000000in}{0.000000in}}%
\pgfpathlineto{\pgfqpoint{0.000000in}{-0.048611in}}%
\pgfusepath{stroke,fill}%
}%
\begin{pgfscope}%
\pgfsys@transformshift{2.384135in}{0.436419in}%
\pgfsys@useobject{currentmarker}{}%
\end{pgfscope}%
\end{pgfscope}%
\begin{pgfscope}%
\definecolor{textcolor}{rgb}{0.000000,0.000000,0.000000}%
\pgfsetstrokecolor{textcolor}%
\pgfsetfillcolor{textcolor}%
\pgftext[x=2.384135in,y=0.339197in,,top]{\color{textcolor}\rmfamily\fontsize{7.000000}{8.400000}\selectfont 150}%
\end{pgfscope}%
\begin{pgfscope}%
\pgfsetbuttcap%
\pgfsetroundjoin%
\definecolor{currentfill}{rgb}{0.000000,0.000000,0.000000}%
\pgfsetfillcolor{currentfill}%
\pgfsetlinewidth{0.803000pt}%
\definecolor{currentstroke}{rgb}{0.000000,0.000000,0.000000}%
\pgfsetstrokecolor{currentstroke}%
\pgfsetdash{}{0pt}%
\pgfsys@defobject{currentmarker}{\pgfqpoint{0.000000in}{-0.048611in}}{\pgfqpoint{0.000000in}{0.000000in}}{%
\pgfpathmoveto{\pgfqpoint{0.000000in}{0.000000in}}%
\pgfpathlineto{\pgfqpoint{0.000000in}{-0.048611in}}%
\pgfusepath{stroke,fill}%
}%
\begin{pgfscope}%
\pgfsys@transformshift{2.975953in}{0.436419in}%
\pgfsys@useobject{currentmarker}{}%
\end{pgfscope}%
\end{pgfscope}%
\begin{pgfscope}%
\definecolor{textcolor}{rgb}{0.000000,0.000000,0.000000}%
\pgfsetstrokecolor{textcolor}%
\pgfsetfillcolor{textcolor}%
\pgftext[x=2.975953in,y=0.339197in,,top]{\color{textcolor}\rmfamily\fontsize{7.000000}{8.400000}\selectfont 200}%
\end{pgfscope}%
\begin{pgfscope}%
\pgfsetbuttcap%
\pgfsetroundjoin%
\definecolor{currentfill}{rgb}{0.000000,0.000000,0.000000}%
\pgfsetfillcolor{currentfill}%
\pgfsetlinewidth{0.803000pt}%
\definecolor{currentstroke}{rgb}{0.000000,0.000000,0.000000}%
\pgfsetstrokecolor{currentstroke}%
\pgfsetdash{}{0pt}%
\pgfsys@defobject{currentmarker}{\pgfqpoint{0.000000in}{-0.048611in}}{\pgfqpoint{0.000000in}{0.000000in}}{%
\pgfpathmoveto{\pgfqpoint{0.000000in}{0.000000in}}%
\pgfpathlineto{\pgfqpoint{0.000000in}{-0.048611in}}%
\pgfusepath{stroke,fill}%
}%
\begin{pgfscope}%
\pgfsys@transformshift{3.567771in}{0.436419in}%
\pgfsys@useobject{currentmarker}{}%
\end{pgfscope}%
\end{pgfscope}%
\begin{pgfscope}%
\definecolor{textcolor}{rgb}{0.000000,0.000000,0.000000}%
\pgfsetstrokecolor{textcolor}%
\pgfsetfillcolor{textcolor}%
\pgftext[x=3.567771in,y=0.339197in,,top]{\color{textcolor}\rmfamily\fontsize{7.000000}{8.400000}\selectfont 250}%
\end{pgfscope}%
\begin{pgfscope}%
\definecolor{textcolor}{rgb}{0.000000,0.000000,0.000000}%
\pgfsetstrokecolor{textcolor}%
\pgfsetfillcolor{textcolor}%
\pgftext[x=2.088226in,y=0.197222in,,top]{\color{textcolor}\rmfamily\fontsize{7.000000}{8.400000}\selectfont depth [m]}%
\end{pgfscope}%
\begin{pgfscope}%
\pgfsetbuttcap%
\pgfsetroundjoin%
\definecolor{currentfill}{rgb}{0.000000,0.000000,0.000000}%
\pgfsetfillcolor{currentfill}%
\pgfsetlinewidth{0.803000pt}%
\definecolor{currentstroke}{rgb}{0.000000,0.000000,0.000000}%
\pgfsetstrokecolor{currentstroke}%
\pgfsetdash{}{0pt}%
\pgfsys@defobject{currentmarker}{\pgfqpoint{-0.048611in}{0.000000in}}{\pgfqpoint{-0.000000in}{0.000000in}}{%
\pgfpathmoveto{\pgfqpoint{-0.000000in}{0.000000in}}%
\pgfpathlineto{\pgfqpoint{-0.048611in}{0.000000in}}%
\pgfusepath{stroke,fill}%
}%
\begin{pgfscope}%
\pgfsys@transformshift{0.460726in}{0.501464in}%
\pgfsys@useobject{currentmarker}{}%
\end{pgfscope}%
\end{pgfscope}%
\begin{pgfscope}%
\definecolor{textcolor}{rgb}{0.000000,0.000000,0.000000}%
\pgfsetstrokecolor{textcolor}%
\pgfsetfillcolor{textcolor}%
\pgftext[x=0.308141in, y=0.467707in, left, base]{\color{textcolor}\rmfamily\fontsize{7.000000}{8.400000}\selectfont \(\displaystyle {0}\)}%
\end{pgfscope}%
\begin{pgfscope}%
\pgfsetbuttcap%
\pgfsetroundjoin%
\definecolor{currentfill}{rgb}{0.000000,0.000000,0.000000}%
\pgfsetfillcolor{currentfill}%
\pgfsetlinewidth{0.803000pt}%
\definecolor{currentstroke}{rgb}{0.000000,0.000000,0.000000}%
\pgfsetstrokecolor{currentstroke}%
\pgfsetdash{}{0pt}%
\pgfsys@defobject{currentmarker}{\pgfqpoint{-0.048611in}{0.000000in}}{\pgfqpoint{-0.000000in}{0.000000in}}{%
\pgfpathmoveto{\pgfqpoint{-0.000000in}{0.000000in}}%
\pgfpathlineto{\pgfqpoint{-0.048611in}{0.000000in}}%
\pgfusepath{stroke,fill}%
}%
\begin{pgfscope}%
\pgfsys@transformshift{0.460726in}{0.740953in}%
\pgfsys@useobject{currentmarker}{}%
\end{pgfscope}%
\end{pgfscope}%
\begin{pgfscope}%
\definecolor{textcolor}{rgb}{0.000000,0.000000,0.000000}%
\pgfsetstrokecolor{textcolor}%
\pgfsetfillcolor{textcolor}%
\pgftext[x=0.252778in, y=0.707196in, left, base]{\color{textcolor}\rmfamily\fontsize{7.000000}{8.400000}\selectfont \(\displaystyle {10}\)}%
\end{pgfscope}%
\begin{pgfscope}%
\pgfsetbuttcap%
\pgfsetroundjoin%
\definecolor{currentfill}{rgb}{0.000000,0.000000,0.000000}%
\pgfsetfillcolor{currentfill}%
\pgfsetlinewidth{0.803000pt}%
\definecolor{currentstroke}{rgb}{0.000000,0.000000,0.000000}%
\pgfsetstrokecolor{currentstroke}%
\pgfsetdash{}{0pt}%
\pgfsys@defobject{currentmarker}{\pgfqpoint{-0.048611in}{0.000000in}}{\pgfqpoint{-0.000000in}{0.000000in}}{%
\pgfpathmoveto{\pgfqpoint{-0.000000in}{0.000000in}}%
\pgfpathlineto{\pgfqpoint{-0.048611in}{0.000000in}}%
\pgfusepath{stroke,fill}%
}%
\begin{pgfscope}%
\pgfsys@transformshift{0.460726in}{0.980443in}%
\pgfsys@useobject{currentmarker}{}%
\end{pgfscope}%
\end{pgfscope}%
\begin{pgfscope}%
\definecolor{textcolor}{rgb}{0.000000,0.000000,0.000000}%
\pgfsetstrokecolor{textcolor}%
\pgfsetfillcolor{textcolor}%
\pgftext[x=0.252778in, y=0.946685in, left, base]{\color{textcolor}\rmfamily\fontsize{7.000000}{8.400000}\selectfont \(\displaystyle {20}\)}%
\end{pgfscope}%
\begin{pgfscope}%
\pgfsetbuttcap%
\pgfsetroundjoin%
\definecolor{currentfill}{rgb}{0.000000,0.000000,0.000000}%
\pgfsetfillcolor{currentfill}%
\pgfsetlinewidth{0.803000pt}%
\definecolor{currentstroke}{rgb}{0.000000,0.000000,0.000000}%
\pgfsetstrokecolor{currentstroke}%
\pgfsetdash{}{0pt}%
\pgfsys@defobject{currentmarker}{\pgfqpoint{-0.048611in}{0.000000in}}{\pgfqpoint{-0.000000in}{0.000000in}}{%
\pgfpathmoveto{\pgfqpoint{-0.000000in}{0.000000in}}%
\pgfpathlineto{\pgfqpoint{-0.048611in}{0.000000in}}%
\pgfusepath{stroke,fill}%
}%
\begin{pgfscope}%
\pgfsys@transformshift{0.460726in}{1.219932in}%
\pgfsys@useobject{currentmarker}{}%
\end{pgfscope}%
\end{pgfscope}%
\begin{pgfscope}%
\definecolor{textcolor}{rgb}{0.000000,0.000000,0.000000}%
\pgfsetstrokecolor{textcolor}%
\pgfsetfillcolor{textcolor}%
\pgftext[x=0.252778in, y=1.186174in, left, base]{\color{textcolor}\rmfamily\fontsize{7.000000}{8.400000}\selectfont \(\displaystyle {30}\)}%
\end{pgfscope}%
\begin{pgfscope}%
\pgfsetbuttcap%
\pgfsetroundjoin%
\definecolor{currentfill}{rgb}{0.000000,0.000000,0.000000}%
\pgfsetfillcolor{currentfill}%
\pgfsetlinewidth{0.803000pt}%
\definecolor{currentstroke}{rgb}{0.000000,0.000000,0.000000}%
\pgfsetstrokecolor{currentstroke}%
\pgfsetdash{}{0pt}%
\pgfsys@defobject{currentmarker}{\pgfqpoint{-0.048611in}{0.000000in}}{\pgfqpoint{-0.000000in}{0.000000in}}{%
\pgfpathmoveto{\pgfqpoint{-0.000000in}{0.000000in}}%
\pgfpathlineto{\pgfqpoint{-0.048611in}{0.000000in}}%
\pgfusepath{stroke,fill}%
}%
\begin{pgfscope}%
\pgfsys@transformshift{0.460726in}{1.459421in}%
\pgfsys@useobject{currentmarker}{}%
\end{pgfscope}%
\end{pgfscope}%
\begin{pgfscope}%
\definecolor{textcolor}{rgb}{0.000000,0.000000,0.000000}%
\pgfsetstrokecolor{textcolor}%
\pgfsetfillcolor{textcolor}%
\pgftext[x=0.252778in, y=1.425663in, left, base]{\color{textcolor}\rmfamily\fontsize{7.000000}{8.400000}\selectfont \(\displaystyle {40}\)}%
\end{pgfscope}%
\begin{pgfscope}%
\pgfsetbuttcap%
\pgfsetroundjoin%
\definecolor{currentfill}{rgb}{0.000000,0.000000,0.000000}%
\pgfsetfillcolor{currentfill}%
\pgfsetlinewidth{0.803000pt}%
\definecolor{currentstroke}{rgb}{0.000000,0.000000,0.000000}%
\pgfsetstrokecolor{currentstroke}%
\pgfsetdash{}{0pt}%
\pgfsys@defobject{currentmarker}{\pgfqpoint{-0.048611in}{0.000000in}}{\pgfqpoint{-0.000000in}{0.000000in}}{%
\pgfpathmoveto{\pgfqpoint{-0.000000in}{0.000000in}}%
\pgfpathlineto{\pgfqpoint{-0.048611in}{0.000000in}}%
\pgfusepath{stroke,fill}%
}%
\begin{pgfscope}%
\pgfsys@transformshift{0.460726in}{1.698910in}%
\pgfsys@useobject{currentmarker}{}%
\end{pgfscope}%
\end{pgfscope}%
\begin{pgfscope}%
\definecolor{textcolor}{rgb}{0.000000,0.000000,0.000000}%
\pgfsetstrokecolor{textcolor}%
\pgfsetfillcolor{textcolor}%
\pgftext[x=0.252778in, y=1.665153in, left, base]{\color{textcolor}\rmfamily\fontsize{7.000000}{8.400000}\selectfont \(\displaystyle {50}\)}%
\end{pgfscope}%
\begin{pgfscope}%
\pgfsetbuttcap%
\pgfsetroundjoin%
\definecolor{currentfill}{rgb}{0.000000,0.000000,0.000000}%
\pgfsetfillcolor{currentfill}%
\pgfsetlinewidth{0.803000pt}%
\definecolor{currentstroke}{rgb}{0.000000,0.000000,0.000000}%
\pgfsetstrokecolor{currentstroke}%
\pgfsetdash{}{0pt}%
\pgfsys@defobject{currentmarker}{\pgfqpoint{-0.048611in}{0.000000in}}{\pgfqpoint{-0.000000in}{0.000000in}}{%
\pgfpathmoveto{\pgfqpoint{-0.000000in}{0.000000in}}%
\pgfpathlineto{\pgfqpoint{-0.048611in}{0.000000in}}%
\pgfusepath{stroke,fill}%
}%
\begin{pgfscope}%
\pgfsys@transformshift{0.460726in}{1.938399in}%
\pgfsys@useobject{currentmarker}{}%
\end{pgfscope}%
\end{pgfscope}%
\begin{pgfscope}%
\definecolor{textcolor}{rgb}{0.000000,0.000000,0.000000}%
\pgfsetstrokecolor{textcolor}%
\pgfsetfillcolor{textcolor}%
\pgftext[x=0.252778in, y=1.904642in, left, base]{\color{textcolor}\rmfamily\fontsize{7.000000}{8.400000}\selectfont \(\displaystyle {60}\)}%
\end{pgfscope}%
\begin{pgfscope}%
\definecolor{textcolor}{rgb}{0.000000,0.000000,0.000000}%
\pgfsetstrokecolor{textcolor}%
\pgfsetfillcolor{textcolor}%
\pgftext[x=0.197222in,y=1.244919in,,bottom,rotate=90.000000]{\color{textcolor}\rmfamily\fontsize{7.000000}{8.400000}\selectfont [\%]}%
\end{pgfscope}%
\begin{pgfscope}%
\pgfpathrectangle{\pgfqpoint{0.460726in}{0.436419in}}{\pgfqpoint{3.255000in}{1.617000in}}%
\pgfusepath{clip}%
\pgfsetrectcap%
\pgfsetroundjoin%
\pgfsetlinewidth{1.505625pt}%
\definecolor{currentstroke}{rgb}{0.121569,0.466667,0.705882}%
\pgfsetstrokecolor{currentstroke}%
\pgfsetdash{}{0pt}%
\pgfpathmoveto{\pgfqpoint{0.608680in}{1.979919in}}%
\pgfpathlineto{\pgfqpoint{0.667862in}{1.260368in}}%
\pgfpathlineto{\pgfqpoint{0.727044in}{1.112045in}}%
\pgfpathlineto{\pgfqpoint{0.786226in}{0.936508in}}%
\pgfpathlineto{\pgfqpoint{0.845408in}{0.693924in}}%
\pgfpathlineto{\pgfqpoint{0.904589in}{0.589229in}}%
\pgfpathlineto{\pgfqpoint{0.963771in}{0.591868in}}%
\pgfpathlineto{\pgfqpoint{1.022953in}{0.587040in}}%
\pgfpathlineto{\pgfqpoint{1.082135in}{0.555907in}}%
\pgfpathlineto{\pgfqpoint{1.141317in}{0.580087in}}%
\pgfpathlineto{\pgfqpoint{1.200499in}{0.663047in}}%
\pgfpathlineto{\pgfqpoint{1.259680in}{0.604821in}}%
\pgfpathlineto{\pgfqpoint{1.318862in}{0.802512in}}%
\pgfpathlineto{\pgfqpoint{1.378044in}{0.591475in}}%
\pgfpathlineto{\pgfqpoint{1.437226in}{0.536523in}}%
\pgfpathlineto{\pgfqpoint{1.496408in}{0.521835in}}%
\pgfpathlineto{\pgfqpoint{1.555589in}{0.511541in}}%
\pgfpathlineto{\pgfqpoint{1.614771in}{0.518057in}}%
\pgfpathlineto{\pgfqpoint{1.673953in}{0.528760in}}%
\pgfpathlineto{\pgfqpoint{1.733135in}{0.517515in}}%
\pgfpathlineto{\pgfqpoint{1.792317in}{0.512609in}}%
\pgfpathlineto{\pgfqpoint{1.851499in}{0.511138in}}%
\pgfpathlineto{\pgfqpoint{1.910680in}{0.512213in}}%
\pgfpathlineto{\pgfqpoint{1.969862in}{0.512878in}}%
\pgfpathlineto{\pgfqpoint{2.029044in}{0.511795in}}%
\pgfpathlineto{\pgfqpoint{2.088226in}{0.514146in}}%
\pgfpathlineto{\pgfqpoint{2.147408in}{0.512139in}}%
\pgfpathlineto{\pgfqpoint{2.206589in}{0.512807in}}%
\pgfpathlineto{\pgfqpoint{2.265771in}{0.511796in}}%
\pgfpathlineto{\pgfqpoint{2.324953in}{0.515429in}}%
\pgfpathlineto{\pgfqpoint{2.384135in}{0.514445in}}%
\pgfpathlineto{\pgfqpoint{2.443317in}{0.512716in}}%
\pgfpathlineto{\pgfqpoint{2.502499in}{0.513049in}}%
\pgfpathlineto{\pgfqpoint{2.561680in}{0.513489in}}%
\pgfpathlineto{\pgfqpoint{2.620862in}{0.518715in}}%
\pgfpathlineto{\pgfqpoint{2.680044in}{0.512624in}}%
\pgfpathlineto{\pgfqpoint{2.739226in}{0.512234in}}%
\pgfpathlineto{\pgfqpoint{2.798408in}{0.511660in}}%
\pgfpathlineto{\pgfqpoint{2.857589in}{0.511377in}}%
\pgfpathlineto{\pgfqpoint{2.916771in}{0.511168in}}%
\pgfpathlineto{\pgfqpoint{2.975953in}{0.511048in}}%
\pgfpathlineto{\pgfqpoint{3.035135in}{0.511012in}}%
\pgfpathlineto{\pgfqpoint{3.094317in}{0.511017in}}%
\pgfpathlineto{\pgfqpoint{3.153499in}{0.511041in}}%
\pgfpathlineto{\pgfqpoint{3.212680in}{0.553341in}}%
\pgfpathlineto{\pgfqpoint{3.271862in}{0.530607in}}%
\pgfpathlineto{\pgfqpoint{3.331044in}{0.537378in}}%
\pgfpathlineto{\pgfqpoint{3.390226in}{0.510869in}}%
\pgfpathlineto{\pgfqpoint{3.449408in}{0.510657in}}%
\pgfpathlineto{\pgfqpoint{3.508589in}{0.511315in}}%
\pgfpathlineto{\pgfqpoint{3.567771in}{0.509919in}}%
\pgfusepath{stroke}%
\end{pgfscope}%
\begin{pgfscope}%
\pgfsetrectcap%
\pgfsetmiterjoin%
\pgfsetlinewidth{0.803000pt}%
\definecolor{currentstroke}{rgb}{0.000000,0.000000,0.000000}%
\pgfsetstrokecolor{currentstroke}%
\pgfsetdash{}{0pt}%
\pgfpathmoveto{\pgfqpoint{0.460726in}{0.436419in}}%
\pgfpathlineto{\pgfqpoint{0.460726in}{2.053419in}}%
\pgfusepath{stroke}%
\end{pgfscope}%
\begin{pgfscope}%
\pgfsetrectcap%
\pgfsetmiterjoin%
\pgfsetlinewidth{0.803000pt}%
\definecolor{currentstroke}{rgb}{0.000000,0.000000,0.000000}%
\pgfsetstrokecolor{currentstroke}%
\pgfsetdash{}{0pt}%
\pgfpathmoveto{\pgfqpoint{3.715726in}{0.436419in}}%
\pgfpathlineto{\pgfqpoint{3.715726in}{2.053419in}}%
\pgfusepath{stroke}%
\end{pgfscope}%
\begin{pgfscope}%
\pgfsetrectcap%
\pgfsetmiterjoin%
\pgfsetlinewidth{0.803000pt}%
\definecolor{currentstroke}{rgb}{0.000000,0.000000,0.000000}%
\pgfsetstrokecolor{currentstroke}%
\pgfsetdash{}{0pt}%
\pgfpathmoveto{\pgfqpoint{0.460726in}{0.436419in}}%
\pgfpathlineto{\pgfqpoint{3.715726in}{0.436419in}}%
\pgfusepath{stroke}%
\end{pgfscope}%
\begin{pgfscope}%
\pgfsetrectcap%
\pgfsetmiterjoin%
\pgfsetlinewidth{0.803000pt}%
\definecolor{currentstroke}{rgb}{0.000000,0.000000,0.000000}%
\pgfsetstrokecolor{currentstroke}%
\pgfsetdash{}{0pt}%
\pgfpathmoveto{\pgfqpoint{0.460726in}{2.053419in}}%
\pgfpathlineto{\pgfqpoint{3.715726in}{2.053419in}}%
\pgfusepath{stroke}%
\end{pgfscope}%
\begin{pgfscope}%
\definecolor{textcolor}{rgb}{0.000000,0.000000,0.000000}%
\pgfsetstrokecolor{textcolor}%
\pgfsetfillcolor{textcolor}%
\pgftext[x=2.088226in,y=2.136753in,,base]{\color{textcolor}\rmfamily\fontsize{8.400000}{10.080000}\selectfont envelope misfit of \(\displaystyle w_f\)}%
\end{pgfscope}%
\end{pgfpicture}%
\makeatother%
\endgroup%

%% file: Performance.tex
\section{Performance}
\label{sec:performance}
SeisSol is optimised for large-scale simulations on supercomputers.
Such simulations can require meshes consisting of several hundred millions of elements to resolve all phenomena accurately.
Hence the number of unknowns can reach $10^{11}$ or more (e.g.~\cite{heinecke_petascale_2014,uphoff_extreme_2017,krenz_3d_2021}).
Therefore, we present a performance and scalability analysis of our extension of SeisSol towards poroelastic wave propagation.
All experiments are carried out on SuperMUC-NG (
2 $\times$ Intel Xeon Platinum 8174 with \num{48} cores @ \SI{2.5}{\giga\Hertz}, \SI{96}{\giga\byte} RAM per node~\cite{lrz_hardware_2021}), which is installed at the Leibniz Supercomputing Centre, Garching, Germany.

\subsection{Implementation using \texttt{YATeTo}}
\label{sec:implementation}
The space-time predictor algorithm as presented in \cref{sec:stp} has proven to produce accurate simulation results, see \cref{sec:benchmarks}.
Its implementation relies heavily on the code generator \texttt{YATeTo}~\cite{uphoff_yet_2020}.
\cref{alg:algorithm-3} is already formulated as a sequence of tensor operations.
\texttt{YATeTo} provides a domain-specific language, embedded into Python, to express these tensor operations in Einstein sum convention.
\Cref{lst:yateto} shows an example of how parts of \cref{alg:algorithm-3} are implemented in \texttt{YATeTo}.
Here \texttt{selectModes(n)} is a matrix, which extracts the basis functions in $(B_{n-1}, B_n]$, similarly, \texttt{selectQuantity(o)} selects the $o^{th}$ quantity and \texttt{Zinv(o)} is the matrix $(Z - E^*_{oo}I)^{-1}$.
\begin{listing}[H]
  \caption{Example of a tensor contraction, which appears in the space-time predictor solver, implemented using \texttt{YATeTo}. This example shows lines $2$ to $6$ of \cref{alg:algorithm-3}.}
\label{lst:yateto}
\begin{minted}[frame=lines]{python}
for n in range(N,-1,-1):
      for o in range(numberOfQuantities-1,-1,-1):
        kernels.append(stp['kpt'] <= stp['kpt'] + selectModes(n)['kl'] 
            * selectQuantity(o)['pq'] * stpRhs['lqu'] * Zinv(o)['ut'] )
\end{minted}
\end{listing}
\texttt{YATeTo} then builds an abstract syntax tree for these tensor operations and maps the tensor contractions to matrix-matrix multiplications.
Specialised code-generators are available for these matrix-matrix multiplications.
For processors of the Intel Skylake generation, we use a combination of \texttt{libxsmm}~\cite{heinecke_libxsmm_2016} for dense-dense multiplications and \texttt{PSpaMM}\footnote{\url{https://github.com/peterwauligmann/PSpaMM}} for multiplications with sparse matrices.
These backends are used to generate the operational code, which is then used as compute kernel during the  simulation phase.
\subsection{Roofline model}
\label{sec:roofline}
\begin{figure}
  \center{
    \input{figures/performance/roofline.pgf} 
  }
  \caption{Roofline model for the SeisSol performance proxy for different polynomial degrees with maximal attainable performance and \SI{40}{\percent} roofline.}
  \label{fig:roofline}
\end{figure}
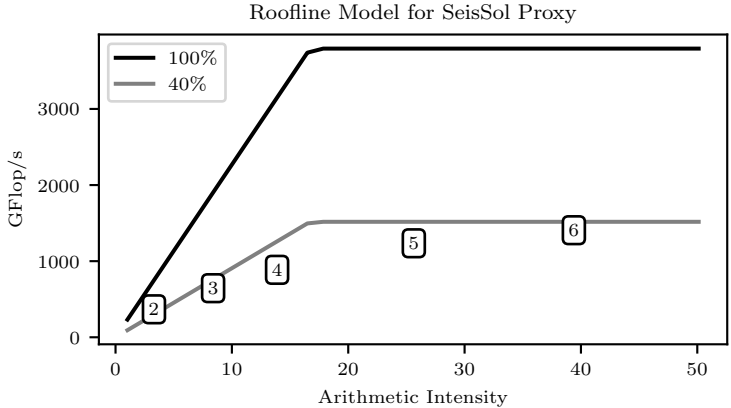

First, we examine the single-node performance of the new back-substitution algorithm for the space-time predictor (i.e., \cref{alg:algorithm-3}).
The roofline model in \cref{fig:roofline} provides an overview of how well our implementation of the space-time predictor utilises the available computer resources~\cite{williams_roofline_2009}.
It provides insight, whether a computation is memory or compute-bound and thus also gives hints where to further optimise.

We utilise a SeisSol performance proxy application, which executes the compute kernels on random data and omits other aspects such as I/O or communication~\cite{uphoff_yet_2020,uphoff_extreme_2017}.
Node-level performance is subject to variation across nodes.
Therefore, we first use the \texttt{likwid} suite~\cite{treibig_likwid_2010} to measure the theoretical node performance.
Running the test on 10 nodes individually, we obtain a mean floating-point performance of \SI[round-mode=places,round-precision=1]{3792.904793475193}{\giga\flop\per\second} and a memory bandwidth of \SI[round-mode=places,round-precision=1]{226.81458981397873}{\giga\byte\per\second}.
Then we use the SeisSol performance proxy on 10 nodes individually with \num{1e6} cells for \num{10} time steps.
The results in \cref{fig:roofline} show that the performance of the SeisSol proxy roughly follows the \SI{40}{\percent} roofline.
The maximum performance of \SI[round-mode=places,round-precision=1]{1406.7173197112272}{\giga\flop\per\second} is achieved for polynomial degree $6$.
With a high arithmetic intensity, our implementation is compute-bound for polynomial degrees $5$ and $6$.
Although our approach attains a reasonable performance on SuperMUC-NG, it does not achieve performance similar to the kernels for elastic wave propagation (\SI{2241}{\giga \flop \per \second} for polynomial degree $6$~\cite{uphoff_yet_2020}).
Architecture-specific backends for general matrix-matrix multiplications ($C = \alpha A \cdot B$) are the key part of our compute kernels.
The GEMM generator from \texttt{libxsmm}, does only support $\alpha=1$, whereas in our case we need $\alpha = \Delta t \neq 1$.
Hence, a fallback to a standard \texttt{for} loop is needed for the scalar multiplication.

Still, for polynomial degree \num{6}, our approach is about a factor of 10 faster compared to an ideally performing LU solver.
From \cref{tab:flop-count}, we observe that our approach requires only \SI{4}{\percent} of the floating-point operations needed for a back-substitution with precomputed LU decomposition.
If that solution was perfectly implemented to achieve full performance, time to solution with our approach ($T_{STP}$) would still be only \SI{11}{\percent} of time to solution with an LU decomposition ($T_{LU}$):
\begin{equation*}
    T_{STP} = 
    \frac{0.04 \cdot \# OP_{LU}}{0.37\cdot \text{performance}_{LU}} \approx
    0.108 \cdot \frac{\# OP_{LU}}{\text{performance}_{LU}} = 
    0.108 \cdot T_{LU}.
\end{equation*}

\subsection{Scaling}
SeisSol adopts a hybrid MPI+OpenMP parallelisation strategy.
Among MPI ranks, we parallelise using graph-based mesh partitioning.
In the initialisation phase, the elements of the mesh are distributed to the available MPI ranks, such that the load per rank is equally distributed.
We distinguish between local (LTS) and global (GTS) time stepping.
For GTS, the workload per element is homogeneous. 
For LTS, elements that need a smaller time step are updated more often than others, consequently, these elements generate a higher workload.
Hence, respective element weights are provided for mesh partitioning. 
Within each rank, we use OpenMP to assign the available elements to compute cores.
A dedicated thread is reserved for asynchronous I/O and communication between ranks~\cite{uphoff_extreme_2017, krenz_3d_2021}.

\begin{figure}
  \center{
    \input{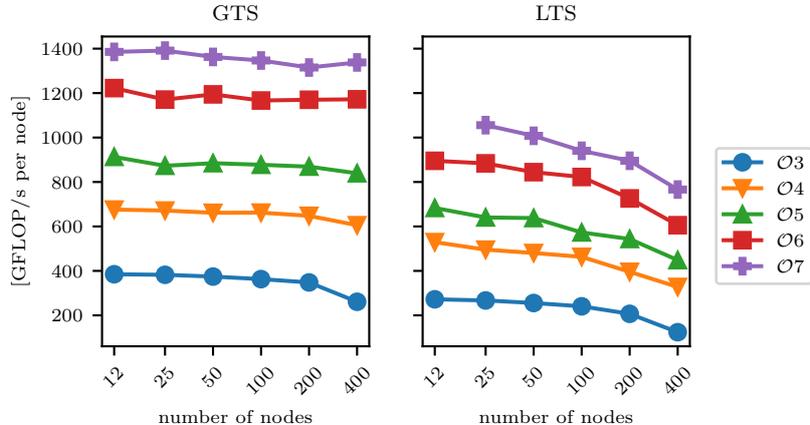}
  }
  \caption{Parallel efficiency for global and local time stepping using the LOHp model with \num{7.33e6} elements on SuperMUC-NG.}
  \label{fig:scaling}
\end{figure}

We use the LOHp benchmark (c.f \cref{sec:lohp}) for a strong scaling test.
Since we do not need to compare with a reference 2D solution, we consider only a single point source.
We set the final time to \SI{0.1}{\second} to test LTS and to \SI{0.01}{\second} to test GTS.
We scale from \num{12} to \num{400} nodes of SuperMUC-NG using a mesh with \num{7334942} elements.
The results are plotted in \cref{fig:scaling}.

In the GTS results, we observe nearly constant node performance, which implies we efficiently use the available parallel resources.
With \num{400} nodes, each node computes less than \num{20000} elements, which is remarkably little compared to non-poroelastic SeisSol applications.
For example, Krenz et al.\ use more than \num{150000} elements per node~\cite{krenz_3d_2021} for their largest mesh in a SeisSol simulation with an elastic-acoustic material model.
We attribute this to the higher workload per element, which is due to the more complex space-time pedictor.
With a peak performance of \SI{1385}{\giga\flop\per\second}, we achieve roughly the same performance as measured with the proxy in \cref{sec:roofline}.

For the LTS results, we observe that the absolute speed is slower than for GTS, which is expected due to the more complicated LTS scheme.
We also observe that the scaling is not as good as with GTS and decays with increasing order.
Still, for polynomial degree \num{6}, we obtain \SI{1056}{\giga\flop\per\second} on \num{25} nodes and reach \SI{766}{\giga\flop\per\second} on \num{400} nodes, which resembles a parallel efficiency of $\approx \SI{72.5}{\percent}$.
More importantly, by using LTS, time to solution is reduced by a factor of \num{6} to \num{10} compared to GTS, depending on the number of nodes and convergence order.

%% file: figures/performance/roofline.pgf
\begingroup%
\makeatletter%
\begin{pgfpicture}%
\pgfpathrectangle{\pgfpointorigin}{\pgfqpoint{3.926452in}{2.303111in}}%
\pgfusepath{use as bounding box, clip}%
\begin{pgfscope}%
\pgfsetbuttcap%
\pgfsetmiterjoin%
\definecolor{currentfill}{rgb}{1.000000,1.000000,1.000000}%
\pgfsetfillcolor{currentfill}%
\pgfsetlinewidth{0.000000pt}%
\definecolor{currentstroke}{rgb}{1.000000,1.000000,1.000000}%
\pgfsetstrokecolor{currentstroke}%
\pgfsetdash{}{0pt}%
\pgfpathmoveto{\pgfqpoint{0.000000in}{0.000000in}}%
\pgfpathlineto{\pgfqpoint{3.926452in}{0.000000in}}%
\pgfpathlineto{\pgfqpoint{3.926452in}{2.303111in}}%
\pgfpathlineto{\pgfqpoint{0.000000in}{2.303111in}}%
\pgfpathclose%
\pgfusepath{fill}%
\end{pgfscope}%
\begin{pgfscope}%
\pgfsetbuttcap%
\pgfsetmiterjoin%
\definecolor{currentfill}{rgb}{1.000000,1.000000,1.000000}%
\pgfsetfillcolor{currentfill}%
\pgfsetlinewidth{0.000000pt}%
\definecolor{currentstroke}{rgb}{0.000000,0.000000,0.000000}%
\pgfsetstrokecolor{currentstroke}%
\pgfsetstrokeopacity{0.000000}%
\pgfsetdash{}{0pt}%
\pgfpathmoveto{\pgfqpoint{0.571452in}{0.425617in}}%
\pgfpathlineto{\pgfqpoint{3.826452in}{0.425617in}}%
\pgfpathlineto{\pgfqpoint{3.826452in}{2.042617in}}%
\pgfpathlineto{\pgfqpoint{0.571452in}{2.042617in}}%
\pgfpathclose%
\pgfusepath{fill}%
\end{pgfscope}%
\begin{pgfscope}%
\pgfsetbuttcap%
\pgfsetroundjoin%
\definecolor{currentfill}{rgb}{0.000000,0.000000,0.000000}%
\pgfsetfillcolor{currentfill}%
\pgfsetlinewidth{0.803000pt}%
\definecolor{currentstroke}{rgb}{0.000000,0.000000,0.000000}%
\pgfsetstrokecolor{currentstroke}%
\pgfsetdash{}{0pt}%
\pgfsys@defobject{currentmarker}{\pgfqpoint{0.000000in}{-0.048611in}}{\pgfqpoint{0.000000in}{0.000000in}}{%
\pgfpathmoveto{\pgfqpoint{0.000000in}{0.000000in}}%
\pgfpathlineto{\pgfqpoint{0.000000in}{-0.048611in}}%
\pgfusepath{stroke,fill}%
}%
\begin{pgfscope}%
\pgfsys@transformshift{0.657730in}{0.425617in}%
\pgfsys@useobject{currentmarker}{}%
\end{pgfscope}%
\end{pgfscope}%
\begin{pgfscope}%
\definecolor{textcolor}{rgb}{0.000000,0.000000,0.000000}%
\pgfsetstrokecolor{textcolor}%
\pgfsetfillcolor{textcolor}%
\pgftext[x=0.657730in,y=0.328394in,,top]{\color{textcolor}\rmfamily\fontsize{7.000000}{8.400000}\selectfont \(\displaystyle {0}\)}%
\end{pgfscope}%
\begin{pgfscope}%
\pgfsetbuttcap%
\pgfsetroundjoin%
\definecolor{currentfill}{rgb}{0.000000,0.000000,0.000000}%
\pgfsetfillcolor{currentfill}%
\pgfsetlinewidth{0.803000pt}%
\definecolor{currentstroke}{rgb}{0.000000,0.000000,0.000000}%
\pgfsetstrokecolor{currentstroke}%
\pgfsetdash{}{0pt}%
\pgfsys@defobject{currentmarker}{\pgfqpoint{0.000000in}{-0.048611in}}{\pgfqpoint{0.000000in}{0.000000in}}{%
\pgfpathmoveto{\pgfqpoint{0.000000in}{0.000000in}}%
\pgfpathlineto{\pgfqpoint{0.000000in}{-0.048611in}}%
\pgfusepath{stroke,fill}%
}%
\begin{pgfscope}%
\pgfsys@transformshift{1.260452in}{0.425617in}%
\pgfsys@useobject{currentmarker}{}%
\end{pgfscope}%
\end{pgfscope}%
\begin{pgfscope}%
\definecolor{textcolor}{rgb}{0.000000,0.000000,0.000000}%
\pgfsetstrokecolor{textcolor}%
\pgfsetfillcolor{textcolor}%
\pgftext[x=1.260452in,y=0.328394in,,top]{\color{textcolor}\rmfamily\fontsize{7.000000}{8.400000}\selectfont \(\displaystyle {10}\)}%
\end{pgfscope}%
\begin{pgfscope}%
\pgfsetbuttcap%
\pgfsetroundjoin%
\definecolor{currentfill}{rgb}{0.000000,0.000000,0.000000}%
\pgfsetfillcolor{currentfill}%
\pgfsetlinewidth{0.803000pt}%
\definecolor{currentstroke}{rgb}{0.000000,0.000000,0.000000}%
\pgfsetstrokecolor{currentstroke}%
\pgfsetdash{}{0pt}%
\pgfsys@defobject{currentmarker}{\pgfqpoint{0.000000in}{-0.048611in}}{\pgfqpoint{0.000000in}{0.000000in}}{%
\pgfpathmoveto{\pgfqpoint{0.000000in}{0.000000in}}%
\pgfpathlineto{\pgfqpoint{0.000000in}{-0.048611in}}%
\pgfusepath{stroke,fill}%
}%
\begin{pgfscope}%
\pgfsys@transformshift{1.863175in}{0.425617in}%
\pgfsys@useobject{currentmarker}{}%
\end{pgfscope}%
\end{pgfscope}%
\begin{pgfscope}%
\definecolor{textcolor}{rgb}{0.000000,0.000000,0.000000}%
\pgfsetstrokecolor{textcolor}%
\pgfsetfillcolor{textcolor}%
\pgftext[x=1.863175in,y=0.328394in,,top]{\color{textcolor}\rmfamily\fontsize{7.000000}{8.400000}\selectfont \(\displaystyle {20}\)}%
\end{pgfscope}%
\begin{pgfscope}%
\pgfsetbuttcap%
\pgfsetroundjoin%
\definecolor{currentfill}{rgb}{0.000000,0.000000,0.000000}%
\pgfsetfillcolor{currentfill}%
\pgfsetlinewidth{0.803000pt}%
\definecolor{currentstroke}{rgb}{0.000000,0.000000,0.000000}%
\pgfsetstrokecolor{currentstroke}%
\pgfsetdash{}{0pt}%
\pgfsys@defobject{currentmarker}{\pgfqpoint{0.000000in}{-0.048611in}}{\pgfqpoint{0.000000in}{0.000000in}}{%
\pgfpathmoveto{\pgfqpoint{0.000000in}{0.000000in}}%
\pgfpathlineto{\pgfqpoint{0.000000in}{-0.048611in}}%
\pgfusepath{stroke,fill}%
}%
\begin{pgfscope}%
\pgfsys@transformshift{2.465897in}{0.425617in}%
\pgfsys@useobject{currentmarker}{}%
\end{pgfscope}%
\end{pgfscope}%
\begin{pgfscope}%
\definecolor{textcolor}{rgb}{0.000000,0.000000,0.000000}%
\pgfsetstrokecolor{textcolor}%
\pgfsetfillcolor{textcolor}%
\pgftext[x=2.465897in,y=0.328394in,,top]{\color{textcolor}\rmfamily\fontsize{7.000000}{8.400000}\selectfont \(\displaystyle {30}\)}%
\end{pgfscope}%
\begin{pgfscope}%
\pgfsetbuttcap%
\pgfsetroundjoin%
\definecolor{currentfill}{rgb}{0.000000,0.000000,0.000000}%
\pgfsetfillcolor{currentfill}%
\pgfsetlinewidth{0.803000pt}%
\definecolor{currentstroke}{rgb}{0.000000,0.000000,0.000000}%
\pgfsetstrokecolor{currentstroke}%
\pgfsetdash{}{0pt}%
\pgfsys@defobject{currentmarker}{\pgfqpoint{0.000000in}{-0.048611in}}{\pgfqpoint{0.000000in}{0.000000in}}{%
\pgfpathmoveto{\pgfqpoint{0.000000in}{0.000000in}}%
\pgfpathlineto{\pgfqpoint{0.000000in}{-0.048611in}}%
\pgfusepath{stroke,fill}%
}%
\begin{pgfscope}%
\pgfsys@transformshift{3.068619in}{0.425617in}%
\pgfsys@useobject{currentmarker}{}%
\end{pgfscope}%
\end{pgfscope}%
\begin{pgfscope}%
\definecolor{textcolor}{rgb}{0.000000,0.000000,0.000000}%
\pgfsetstrokecolor{textcolor}%
\pgfsetfillcolor{textcolor}%
\pgftext[x=3.068619in,y=0.328394in,,top]{\color{textcolor}\rmfamily\fontsize{7.000000}{8.400000}\selectfont \(\displaystyle {40}\)}%
\end{pgfscope}%
\begin{pgfscope}%
\pgfsetbuttcap%
\pgfsetroundjoin%
\definecolor{currentfill}{rgb}{0.000000,0.000000,0.000000}%
\pgfsetfillcolor{currentfill}%
\pgfsetlinewidth{0.803000pt}%
\definecolor{currentstroke}{rgb}{0.000000,0.000000,0.000000}%
\pgfsetstrokecolor{currentstroke}%
\pgfsetdash{}{0pt}%
\pgfsys@defobject{currentmarker}{\pgfqpoint{0.000000in}{-0.048611in}}{\pgfqpoint{0.000000in}{0.000000in}}{%
\pgfpathmoveto{\pgfqpoint{0.000000in}{0.000000in}}%
\pgfpathlineto{\pgfqpoint{0.000000in}{-0.048611in}}%
\pgfusepath{stroke,fill}%
}%
\begin{pgfscope}%
\pgfsys@transformshift{3.671341in}{0.425617in}%
\pgfsys@useobject{currentmarker}{}%
\end{pgfscope}%
\end{pgfscope}%
\begin{pgfscope}%
\definecolor{textcolor}{rgb}{0.000000,0.000000,0.000000}%
\pgfsetstrokecolor{textcolor}%
\pgfsetfillcolor{textcolor}%
\pgftext[x=3.671341in,y=0.328394in,,top]{\color{textcolor}\rmfamily\fontsize{7.000000}{8.400000}\selectfont \(\displaystyle {50}\)}%
\end{pgfscope}%
\begin{pgfscope}%
\definecolor{textcolor}{rgb}{0.000000,0.000000,0.000000}%
\pgfsetstrokecolor{textcolor}%
\pgfsetfillcolor{textcolor}%
\pgftext[x=2.198952in,y=0.186419in,,top]{\color{textcolor}\rmfamily\fontsize{7.000000}{8.400000}\selectfont Arithmetic Intensity}%
\end{pgfscope}%
\begin{pgfscope}%
\pgfsetbuttcap%
\pgfsetroundjoin%
\definecolor{currentfill}{rgb}{0.000000,0.000000,0.000000}%
\pgfsetfillcolor{currentfill}%
\pgfsetlinewidth{0.803000pt}%
\definecolor{currentstroke}{rgb}{0.000000,0.000000,0.000000}%
\pgfsetstrokecolor{currentstroke}%
\pgfsetdash{}{0pt}%
\pgfsys@defobject{currentmarker}{\pgfqpoint{-0.048611in}{0.000000in}}{\pgfqpoint{-0.000000in}{0.000000in}}{%
\pgfpathmoveto{\pgfqpoint{-0.000000in}{0.000000in}}%
\pgfpathlineto{\pgfqpoint{-0.048611in}{0.000000in}}%
\pgfusepath{stroke,fill}%
}%
\begin{pgfscope}%
\pgfsys@transformshift{0.571452in}{0.462233in}%
\pgfsys@useobject{currentmarker}{}%
\end{pgfscope}%
\end{pgfscope}%
\begin{pgfscope}%
\definecolor{textcolor}{rgb}{0.000000,0.000000,0.000000}%
\pgfsetstrokecolor{textcolor}%
\pgfsetfillcolor{textcolor}%
\pgftext[x=0.418867in, y=0.428475in, left, base]{\color{textcolor}\rmfamily\fontsize{7.000000}{8.400000}\selectfont \(\displaystyle {0}\)}%
\end{pgfscope}%
\begin{pgfscope}%
\pgfsetbuttcap%
\pgfsetroundjoin%
\definecolor{currentfill}{rgb}{0.000000,0.000000,0.000000}%
\pgfsetfillcolor{currentfill}%
\pgfsetlinewidth{0.803000pt}%
\definecolor{currentstroke}{rgb}{0.000000,0.000000,0.000000}%
\pgfsetstrokecolor{currentstroke}%
\pgfsetdash{}{0pt}%
\pgfsys@defobject{currentmarker}{\pgfqpoint{-0.048611in}{0.000000in}}{\pgfqpoint{-0.000000in}{0.000000in}}{%
\pgfpathmoveto{\pgfqpoint{-0.000000in}{0.000000in}}%
\pgfpathlineto{\pgfqpoint{-0.048611in}{0.000000in}}%
\pgfusepath{stroke,fill}%
}%
\begin{pgfscope}%
\pgfsys@transformshift{0.571452in}{0.859523in}%
\pgfsys@useobject{currentmarker}{}%
\end{pgfscope}%
\end{pgfscope}%
\begin{pgfscope}%
\definecolor{textcolor}{rgb}{0.000000,0.000000,0.000000}%
\pgfsetstrokecolor{textcolor}%
\pgfsetfillcolor{textcolor}%
\pgftext[x=0.252778in, y=0.825765in, left, base]{\color{textcolor}\rmfamily\fontsize{7.000000}{8.400000}\selectfont \(\displaystyle {1000}\)}%
\end{pgfscope}%
\begin{pgfscope}%
\pgfsetbuttcap%
\pgfsetroundjoin%
\definecolor{currentfill}{rgb}{0.000000,0.000000,0.000000}%
\pgfsetfillcolor{currentfill}%
\pgfsetlinewidth{0.803000pt}%
\definecolor{currentstroke}{rgb}{0.000000,0.000000,0.000000}%
\pgfsetstrokecolor{currentstroke}%
\pgfsetdash{}{0pt}%
\pgfsys@defobject{currentmarker}{\pgfqpoint{-0.048611in}{0.000000in}}{\pgfqpoint{-0.000000in}{0.000000in}}{%
\pgfpathmoveto{\pgfqpoint{-0.000000in}{0.000000in}}%
\pgfpathlineto{\pgfqpoint{-0.048611in}{0.000000in}}%
\pgfusepath{stroke,fill}%
}%
\begin{pgfscope}%
\pgfsys@transformshift{0.571452in}{1.256813in}%
\pgfsys@useobject{currentmarker}{}%
\end{pgfscope}%
\end{pgfscope}%
\begin{pgfscope}%
\definecolor{textcolor}{rgb}{0.000000,0.000000,0.000000}%
\pgfsetstrokecolor{textcolor}%
\pgfsetfillcolor{textcolor}%
\pgftext[x=0.252778in, y=1.223055in, left, base]{\color{textcolor}\rmfamily\fontsize{7.000000}{8.400000}\selectfont \(\displaystyle {2000}\)}%
\end{pgfscope}%
\begin{pgfscope}%
\pgfsetbuttcap%
\pgfsetroundjoin%
\definecolor{currentfill}{rgb}{0.000000,0.000000,0.000000}%
\pgfsetfillcolor{currentfill}%
\pgfsetlinewidth{0.803000pt}%
\definecolor{currentstroke}{rgb}{0.000000,0.000000,0.000000}%
\pgfsetstrokecolor{currentstroke}%
\pgfsetdash{}{0pt}%
\pgfsys@defobject{currentmarker}{\pgfqpoint{-0.048611in}{0.000000in}}{\pgfqpoint{-0.000000in}{0.000000in}}{%
\pgfpathmoveto{\pgfqpoint{-0.000000in}{0.000000in}}%
\pgfpathlineto{\pgfqpoint{-0.048611in}{0.000000in}}%
\pgfusepath{stroke,fill}%
}%
\begin{pgfscope}%
\pgfsys@transformshift{0.571452in}{1.654103in}%
\pgfsys@useobject{currentmarker}{}%
\end{pgfscope}%
\end{pgfscope}%
\begin{pgfscope}%
\definecolor{textcolor}{rgb}{0.000000,0.000000,0.000000}%
\pgfsetstrokecolor{textcolor}%
\pgfsetfillcolor{textcolor}%
\pgftext[x=0.252778in, y=1.620346in, left, base]{\color{textcolor}\rmfamily\fontsize{7.000000}{8.400000}\selectfont \(\displaystyle {3000}\)}%
\end{pgfscope}%
\begin{pgfscope}%
\definecolor{textcolor}{rgb}{0.000000,0.000000,0.000000}%
\pgfsetstrokecolor{textcolor}%
\pgfsetfillcolor{textcolor}%
\pgftext[x=0.197222in,y=1.234117in,,bottom,rotate=90.000000]{\color{textcolor}\rmfamily\fontsize{7.000000}{8.400000}\selectfont GFlop/s}%
\end{pgfscope}%
\begin{pgfscope}%
\pgfpathrectangle{\pgfqpoint{0.571452in}{0.425617in}}{\pgfqpoint{3.255000in}{1.617000in}}%
\pgfusepath{clip}%
\pgfsetrectcap%
\pgfsetroundjoin%
\pgfsetlinewidth{1.505625pt}%
\definecolor{currentstroke}{rgb}{0.000000,0.000000,0.000000}%
\pgfsetstrokecolor{currentstroke}%
\pgfsetdash{}{0pt}%
\pgfpathmoveto{\pgfqpoint{0.719406in}{0.554443in}}%
\pgfpathlineto{\pgfqpoint{0.724504in}{0.562064in}}%
\pgfpathlineto{\pgfqpoint{0.730023in}{0.570316in}}%
\pgfpathlineto{\pgfqpoint{0.735998in}{0.579249in}}%
\pgfpathlineto{\pgfqpoint{0.742468in}{0.588921in}}%
\pgfpathlineto{\pgfqpoint{0.749472in}{0.599393in}}%
\pgfpathlineto{\pgfqpoint{0.757054in}{0.610730in}}%
\pgfpathlineto{\pgfqpoint{0.765264in}{0.623003in}}%
\pgfpathlineto{\pgfqpoint{0.774152in}{0.636292in}}%
\pgfpathlineto{\pgfqpoint{0.783775in}{0.650679in}}%
\pgfpathlineto{\pgfqpoint{0.794193in}{0.666255in}}%
\pgfpathlineto{\pgfqpoint{0.805472in}{0.683118in}}%
\pgfpathlineto{\pgfqpoint{0.817684in}{0.701375in}}%
\pgfpathlineto{\pgfqpoint{0.830905in}{0.721141in}}%
\pgfpathlineto{\pgfqpoint{0.845218in}{0.742541in}}%
\pgfpathlineto{\pgfqpoint{0.860715in}{0.765710in}}%
\pgfpathlineto{\pgfqpoint{0.877493in}{0.790793in}}%
\pgfpathlineto{\pgfqpoint{0.895657in}{0.817950in}}%
\pgfpathlineto{\pgfqpoint{0.915323in}{0.847352in}}%
\pgfpathlineto{\pgfqpoint{0.936614in}{0.879184in}}%
\pgfpathlineto{\pgfqpoint{0.959665in}{0.913646in}}%
\pgfpathlineto{\pgfqpoint{0.984621in}{0.950958in}}%
\pgfpathlineto{\pgfqpoint{1.011640in}{0.991353in}}%
\pgfpathlineto{\pgfqpoint{1.040892in}{1.035087in}}%
\pgfpathlineto{\pgfqpoint{1.072562in}{1.082436in}}%
\pgfpathlineto{\pgfqpoint{1.106850in}{1.133698in}}%
\pgfpathlineto{\pgfqpoint{1.143971in}{1.189198in}}%
\pgfpathlineto{\pgfqpoint{1.184161in}{1.249285in}}%
\pgfpathlineto{\pgfqpoint{1.227673in}{1.314338in}}%
\pgfpathlineto{\pgfqpoint{1.274781in}{1.384768in}}%
\pgfpathlineto{\pgfqpoint{1.325783in}{1.461019in}}%
\pgfpathlineto{\pgfqpoint{1.381001in}{1.543573in}}%
\pgfpathlineto{\pgfqpoint{1.440782in}{1.632951in}}%
\pgfpathlineto{\pgfqpoint{1.505505in}{1.729716in}}%
\pgfpathlineto{\pgfqpoint{1.575577in}{1.834479in}}%
\pgfpathlineto{\pgfqpoint{1.651441in}{1.947901in}}%
\pgfpathlineto{\pgfqpoint{1.733576in}{1.969117in}}%
\pgfpathlineto{\pgfqpoint{1.822499in}{1.969117in}}%
\pgfpathlineto{\pgfqpoint{1.918772in}{1.969117in}}%
\pgfpathlineto{\pgfqpoint{2.023002in}{1.969117in}}%
\pgfpathlineto{\pgfqpoint{2.135848in}{1.969117in}}%
\pgfpathlineto{\pgfqpoint{2.258021in}{1.969117in}}%
\pgfpathlineto{\pgfqpoint{2.390292in}{1.969117in}}%
\pgfpathlineto{\pgfqpoint{2.533495in}{1.969117in}}%
\pgfpathlineto{\pgfqpoint{2.688535in}{1.969117in}}%
\pgfpathlineto{\pgfqpoint{2.856390in}{1.969117in}}%
\pgfpathlineto{\pgfqpoint{3.038119in}{1.969117in}}%
\pgfpathlineto{\pgfqpoint{3.234868in}{1.969117in}}%
\pgfpathlineto{\pgfqpoint{3.447879in}{1.969117in}}%
\pgfpathlineto{\pgfqpoint{3.678497in}{1.969117in}}%
\pgfusepath{stroke}%
\end{pgfscope}%
\begin{pgfscope}%
\pgfpathrectangle{\pgfqpoint{0.571452in}{0.425617in}}{\pgfqpoint{3.255000in}{1.617000in}}%
\pgfusepath{clip}%
\pgfsetrectcap%
\pgfsetroundjoin%
\pgfsetlinewidth{1.505625pt}%
\definecolor{currentstroke}{rgb}{0.501961,0.501961,0.501961}%
\pgfsetstrokecolor{currentstroke}%
\pgfsetdash{}{0pt}%
\pgfpathmoveto{\pgfqpoint{0.719406in}{0.499117in}}%
\pgfpathlineto{\pgfqpoint{0.724504in}{0.502165in}}%
\pgfpathlineto{\pgfqpoint{0.730023in}{0.505466in}}%
\pgfpathlineto{\pgfqpoint{0.735998in}{0.509039in}}%
\pgfpathlineto{\pgfqpoint{0.742468in}{0.512908in}}%
\pgfpathlineto{\pgfqpoint{0.749472in}{0.517097in}}%
\pgfpathlineto{\pgfqpoint{0.757054in}{0.521631in}}%
\pgfpathlineto{\pgfqpoint{0.765264in}{0.526541in}}%
\pgfpathlineto{\pgfqpoint{0.774152in}{0.531856in}}%
\pgfpathlineto{\pgfqpoint{0.783775in}{0.537611in}}%
\pgfpathlineto{\pgfqpoint{0.794193in}{0.543841in}}%
\pgfpathlineto{\pgfqpoint{0.805472in}{0.550587in}}%
\pgfpathlineto{\pgfqpoint{0.817684in}{0.557890in}}%
\pgfpathlineto{\pgfqpoint{0.830905in}{0.565796in}}%
\pgfpathlineto{\pgfqpoint{0.845218in}{0.574356in}}%
\pgfpathlineto{\pgfqpoint{0.860715in}{0.583623in}}%
\pgfpathlineto{\pgfqpoint{0.877493in}{0.593657in}}%
\pgfpathlineto{\pgfqpoint{0.895657in}{0.604520in}}%
\pgfpathlineto{\pgfqpoint{0.915323in}{0.616280in}}%
\pgfpathlineto{\pgfqpoint{0.936614in}{0.629013in}}%
\pgfpathlineto{\pgfqpoint{0.959665in}{0.642798in}}%
\pgfpathlineto{\pgfqpoint{0.984621in}{0.657723in}}%
\pgfpathlineto{\pgfqpoint{1.011640in}{0.673881in}}%
\pgfpathlineto{\pgfqpoint{1.040892in}{0.691374in}}%
\pgfpathlineto{\pgfqpoint{1.072562in}{0.710314in}}%
\pgfpathlineto{\pgfqpoint{1.106850in}{0.730819in}}%
\pgfpathlineto{\pgfqpoint{1.143971in}{0.753019in}}%
\pgfpathlineto{\pgfqpoint{1.184161in}{0.777053in}}%
\pgfpathlineto{\pgfqpoint{1.227673in}{0.803075in}}%
\pgfpathlineto{\pgfqpoint{1.274781in}{0.831247in}}%
\pgfpathlineto{\pgfqpoint{1.325783in}{0.861747in}}%
\pgfpathlineto{\pgfqpoint{1.381001in}{0.894769in}}%
\pgfpathlineto{\pgfqpoint{1.440782in}{0.930520in}}%
\pgfpathlineto{\pgfqpoint{1.505505in}{0.969226in}}%
\pgfpathlineto{\pgfqpoint{1.575577in}{1.011131in}}%
\pgfpathlineto{\pgfqpoint{1.651441in}{1.056500in}}%
\pgfpathlineto{\pgfqpoint{1.733576in}{1.064986in}}%
\pgfpathlineto{\pgfqpoint{1.822499in}{1.064986in}}%
\pgfpathlineto{\pgfqpoint{1.918772in}{1.064986in}}%
\pgfpathlineto{\pgfqpoint{2.023002in}{1.064986in}}%
\pgfpathlineto{\pgfqpoint{2.135848in}{1.064986in}}%
\pgfpathlineto{\pgfqpoint{2.258021in}{1.064986in}}%
\pgfpathlineto{\pgfqpoint{2.390292in}{1.064986in}}%
\pgfpathlineto{\pgfqpoint{2.533495in}{1.064986in}}%
\pgfpathlineto{\pgfqpoint{2.688535in}{1.064986in}}%
\pgfpathlineto{\pgfqpoint{2.856390in}{1.064986in}}%
\pgfpathlineto{\pgfqpoint{3.038119in}{1.064986in}}%
\pgfpathlineto{\pgfqpoint{3.234868in}{1.064986in}}%
\pgfpathlineto{\pgfqpoint{3.447879in}{1.064986in}}%
\pgfpathlineto{\pgfqpoint{3.678497in}{1.064986in}}%
\pgfusepath{stroke}%
\end{pgfscope}%
\begin{pgfscope}%
\pgfsetrectcap%
\pgfsetmiterjoin%
\pgfsetlinewidth{0.803000pt}%
\definecolor{currentstroke}{rgb}{0.000000,0.000000,0.000000}%
\pgfsetstrokecolor{currentstroke}%
\pgfsetdash{}{0pt}%
\pgfpathmoveto{\pgfqpoint{0.571452in}{0.425617in}}%
\pgfpathlineto{\pgfqpoint{0.571452in}{2.042617in}}%
\pgfusepath{stroke}%
\end{pgfscope}%
\begin{pgfscope}%
\pgfsetrectcap%
\pgfsetmiterjoin%
\pgfsetlinewidth{0.803000pt}%
\definecolor{currentstroke}{rgb}{0.000000,0.000000,0.000000}%
\pgfsetstrokecolor{currentstroke}%
\pgfsetdash{}{0pt}%
\pgfpathmoveto{\pgfqpoint{3.826452in}{0.425617in}}%
\pgfpathlineto{\pgfqpoint{3.826452in}{2.042617in}}%
\pgfusepath{stroke}%
\end{pgfscope}%
\begin{pgfscope}%
\pgfsetrectcap%
\pgfsetmiterjoin%
\pgfsetlinewidth{0.803000pt}%
\definecolor{currentstroke}{rgb}{0.000000,0.000000,0.000000}%
\pgfsetstrokecolor{currentstroke}%
\pgfsetdash{}{0pt}%
\pgfpathmoveto{\pgfqpoint{0.571452in}{0.425617in}}%
\pgfpathlineto{\pgfqpoint{3.826452in}{0.425617in}}%
\pgfusepath{stroke}%
\end{pgfscope}%
\begin{pgfscope}%
\pgfsetrectcap%
\pgfsetmiterjoin%
\pgfsetlinewidth{0.803000pt}%
\definecolor{currentstroke}{rgb}{0.000000,0.000000,0.000000}%
\pgfsetstrokecolor{currentstroke}%
\pgfsetdash{}{0pt}%
\pgfpathmoveto{\pgfqpoint{0.571452in}{2.042617in}}%
\pgfpathlineto{\pgfqpoint{3.826452in}{2.042617in}}%
\pgfusepath{stroke}%
\end{pgfscope}%
\begin{pgfscope}%
\pgfsetbuttcap%
\pgfsetmiterjoin%
\definecolor{currentfill}{rgb}{1.000000,1.000000,1.000000}%
\pgfsetfillcolor{currentfill}%
\pgfsetlinewidth{1.003750pt}%
\definecolor{currentstroke}{rgb}{0.000000,0.000000,0.000000}%
\pgfsetstrokecolor{currentstroke}%
\pgfsetdash{}{0pt}%
\pgfpathmoveto{\pgfqpoint{0.829119in}{0.535895in}}%
\pgfpathlineto{\pgfqpoint{0.884482in}{0.535895in}}%
\pgfpathquadraticcurveto{\pgfqpoint{0.913648in}{0.535895in}}{\pgfqpoint{0.913648in}{0.565062in}}%
\pgfpathlineto{\pgfqpoint{0.913648in}{0.651482in}}%
\pgfpathquadraticcurveto{\pgfqpoint{0.913648in}{0.680648in}}{\pgfqpoint{0.884482in}{0.680648in}}%
\pgfpathlineto{\pgfqpoint{0.829119in}{0.680648in}}%
\pgfpathquadraticcurveto{\pgfqpoint{0.799952in}{0.680648in}}{\pgfqpoint{0.799952in}{0.651482in}}%
\pgfpathlineto{\pgfqpoint{0.799952in}{0.565062in}}%
\pgfpathquadraticcurveto{\pgfqpoint{0.799952in}{0.535895in}}{\pgfqpoint{0.829119in}{0.535895in}}%
\pgfpathclose%
\pgfusepath{stroke,fill}%
\end{pgfscope}%
\begin{pgfscope}%
\definecolor{textcolor}{rgb}{0.000000,0.000000,0.000000}%
\pgfsetstrokecolor{textcolor}%
\pgfsetfillcolor{textcolor}%
\pgftext[x=0.856800in,y=0.608272in,,]{\color{textcolor}\rmfamily\fontsize{7.000000}{8.400000}\selectfont 2}%
\end{pgfscope}%
\begin{pgfscope}%
\pgfsetbuttcap%
\pgfsetmiterjoin%
\definecolor{currentfill}{rgb}{1.000000,1.000000,1.000000}%
\pgfsetfillcolor{currentfill}%
\pgfsetlinewidth{1.003750pt}%
\definecolor{currentstroke}{rgb}{0.000000,0.000000,0.000000}%
\pgfsetstrokecolor{currentstroke}%
\pgfsetdash{}{0pt}%
\pgfpathmoveto{\pgfqpoint{1.135456in}{0.646069in}}%
\pgfpathlineto{\pgfqpoint{1.190819in}{0.646069in}}%
\pgfpathquadraticcurveto{\pgfqpoint{1.219986in}{0.646069in}}{\pgfqpoint{1.219986in}{0.675236in}}%
\pgfpathlineto{\pgfqpoint{1.219986in}{0.761655in}}%
\pgfpathquadraticcurveto{\pgfqpoint{1.219986in}{0.790822in}}{\pgfqpoint{1.190819in}{0.790822in}}%
\pgfpathlineto{\pgfqpoint{1.135456in}{0.790822in}}%
\pgfpathquadraticcurveto{\pgfqpoint{1.106289in}{0.790822in}}{\pgfqpoint{1.106289in}{0.761655in}}%
\pgfpathlineto{\pgfqpoint{1.106289in}{0.675236in}}%
\pgfpathquadraticcurveto{\pgfqpoint{1.106289in}{0.646069in}}{\pgfqpoint{1.135456in}{0.646069in}}%
\pgfpathclose%
\pgfusepath{stroke,fill}%
\end{pgfscope}%
\begin{pgfscope}%
\definecolor{textcolor}{rgb}{0.000000,0.000000,0.000000}%
\pgfsetstrokecolor{textcolor}%
\pgfsetfillcolor{textcolor}%
\pgftext[x=1.163137in,y=0.718445in,,]{\color{textcolor}\rmfamily\fontsize{7.000000}{8.400000}\selectfont 3}%
\end{pgfscope}%
\begin{pgfscope}%
\pgfsetbuttcap%
\pgfsetmiterjoin%
\definecolor{currentfill}{rgb}{1.000000,1.000000,1.000000}%
\pgfsetfillcolor{currentfill}%
\pgfsetlinewidth{1.003750pt}%
\definecolor{currentstroke}{rgb}{0.000000,0.000000,0.000000}%
\pgfsetstrokecolor{currentstroke}%
\pgfsetdash{}{0pt}%
\pgfpathmoveto{\pgfqpoint{1.466953in}{0.741960in}}%
\pgfpathlineto{\pgfqpoint{1.522316in}{0.741960in}}%
\pgfpathquadraticcurveto{\pgfqpoint{1.551483in}{0.741960in}}{\pgfqpoint{1.551483in}{0.771127in}}%
\pgfpathlineto{\pgfqpoint{1.551483in}{0.857546in}}%
\pgfpathquadraticcurveto{\pgfqpoint{1.551483in}{0.886713in}}{\pgfqpoint{1.522316in}{0.886713in}}%
\pgfpathlineto{\pgfqpoint{1.466953in}{0.886713in}}%
\pgfpathquadraticcurveto{\pgfqpoint{1.437787in}{0.886713in}}{\pgfqpoint{1.437787in}{0.857546in}}%
\pgfpathlineto{\pgfqpoint{1.437787in}{0.771127in}}%
\pgfpathquadraticcurveto{\pgfqpoint{1.437787in}{0.741960in}}{\pgfqpoint{1.466953in}{0.741960in}}%
\pgfpathclose%
\pgfusepath{stroke,fill}%
\end{pgfscope}%
\begin{pgfscope}%
\definecolor{textcolor}{rgb}{0.000000,0.000000,0.000000}%
\pgfsetstrokecolor{textcolor}%
\pgfsetfillcolor{textcolor}%
\pgftext[x=1.494635in,y=0.814336in,,]{\color{textcolor}\rmfamily\fontsize{7.000000}{8.400000}\selectfont 4}%
\end{pgfscope}%
\begin{pgfscope}%
\pgfsetbuttcap%
\pgfsetmiterjoin%
\definecolor{currentfill}{rgb}{1.000000,1.000000,1.000000}%
\pgfsetfillcolor{currentfill}%
\pgfsetlinewidth{1.003750pt}%
\definecolor{currentstroke}{rgb}{0.000000,0.000000,0.000000}%
\pgfsetstrokecolor{currentstroke}%
\pgfsetdash{}{0pt}%
\pgfpathmoveto{\pgfqpoint{2.175222in}{0.880516in}}%
\pgfpathlineto{\pgfqpoint{2.230585in}{0.880516in}}%
\pgfpathquadraticcurveto{\pgfqpoint{2.259751in}{0.880516in}}{\pgfqpoint{2.259751in}{0.909683in}}%
\pgfpathlineto{\pgfqpoint{2.259751in}{0.996102in}}%
\pgfpathquadraticcurveto{\pgfqpoint{2.259751in}{1.025269in}}{\pgfqpoint{2.230585in}{1.025269in}}%
\pgfpathlineto{\pgfqpoint{2.175222in}{1.025269in}}%
\pgfpathquadraticcurveto{\pgfqpoint{2.146055in}{1.025269in}}{\pgfqpoint{2.146055in}{0.996102in}}%
\pgfpathlineto{\pgfqpoint{2.146055in}{0.909683in}}%
\pgfpathquadraticcurveto{\pgfqpoint{2.146055in}{0.880516in}}{\pgfqpoint{2.175222in}{0.880516in}}%
\pgfpathclose%
\pgfusepath{stroke,fill}%
\end{pgfscope}%
\begin{pgfscope}%
\definecolor{textcolor}{rgb}{0.000000,0.000000,0.000000}%
\pgfsetstrokecolor{textcolor}%
\pgfsetfillcolor{textcolor}%
\pgftext[x=2.202903in,y=0.952892in,,]{\color{textcolor}\rmfamily\fontsize{7.000000}{8.400000}\selectfont 5}%
\end{pgfscope}%
\begin{pgfscope}%
\pgfsetbuttcap%
\pgfsetmiterjoin%
\definecolor{currentfill}{rgb}{1.000000,1.000000,1.000000}%
\pgfsetfillcolor{currentfill}%
\pgfsetlinewidth{1.003750pt}%
\definecolor{currentstroke}{rgb}{0.000000,0.000000,0.000000}%
\pgfsetstrokecolor{currentstroke}%
\pgfsetdash{}{0pt}%
\pgfpathmoveto{\pgfqpoint{3.004331in}{0.948731in}}%
\pgfpathlineto{\pgfqpoint{3.059694in}{0.948731in}}%
\pgfpathquadraticcurveto{\pgfqpoint{3.088860in}{0.948731in}}{\pgfqpoint{3.088860in}{0.977898in}}%
\pgfpathlineto{\pgfqpoint{3.088860in}{1.064317in}}%
\pgfpathquadraticcurveto{\pgfqpoint{3.088860in}{1.093484in}}{\pgfqpoint{3.059694in}{1.093484in}}%
\pgfpathlineto{\pgfqpoint{3.004331in}{1.093484in}}%
\pgfpathquadraticcurveto{\pgfqpoint{2.975164in}{1.093484in}}{\pgfqpoint{2.975164in}{1.064317in}}%
\pgfpathlineto{\pgfqpoint{2.975164in}{0.977898in}}%
\pgfpathquadraticcurveto{\pgfqpoint{2.975164in}{0.948731in}}{\pgfqpoint{3.004331in}{0.948731in}}%
\pgfpathclose%
\pgfusepath{stroke,fill}%
\end{pgfscope}%
\begin{pgfscope}%
\definecolor{textcolor}{rgb}{0.000000,0.000000,0.000000}%
\pgfsetstrokecolor{textcolor}%
\pgfsetfillcolor{textcolor}%
\pgftext[x=3.032012in,y=1.021108in,,]{\color{textcolor}\rmfamily\fontsize{7.000000}{8.400000}\selectfont 6}%
\end{pgfscope}%
\begin{pgfscope}%
\definecolor{textcolor}{rgb}{0.000000,0.000000,0.000000}%
\pgfsetstrokecolor{textcolor}%
\pgfsetfillcolor{textcolor}%
\pgftext[x=2.198952in,y=2.125950in,,base]{\color{textcolor}\rmfamily\fontsize{8.400000}{10.080000}\selectfont Roofline Model for SeisSol Proxy}%
\end{pgfscope}%
\begin{pgfscope}%
\pgfsetbuttcap%
\pgfsetmiterjoin%
\definecolor{currentfill}{rgb}{1.000000,1.000000,1.000000}%
\pgfsetfillcolor{currentfill}%
\pgfsetfillopacity{0.800000}%
\pgfsetlinewidth{1.003750pt}%
\definecolor{currentstroke}{rgb}{0.800000,0.800000,0.800000}%
\pgfsetstrokecolor{currentstroke}%
\pgfsetstrokeopacity{0.800000}%
\pgfsetdash{}{0pt}%
\pgfpathmoveto{\pgfqpoint{0.639507in}{1.693697in}}%
\pgfpathlineto{\pgfqpoint{1.207950in}{1.693697in}}%
\pgfpathquadraticcurveto{\pgfqpoint{1.227394in}{1.693697in}}{\pgfqpoint{1.227394in}{1.713142in}}%
\pgfpathlineto{\pgfqpoint{1.227394in}{1.974561in}}%
\pgfpathquadraticcurveto{\pgfqpoint{1.227394in}{1.994006in}}{\pgfqpoint{1.207950in}{1.994006in}}%
\pgfpathlineto{\pgfqpoint{0.639507in}{1.994006in}}%
\pgfpathquadraticcurveto{\pgfqpoint{0.620063in}{1.994006in}}{\pgfqpoint{0.620063in}{1.974561in}}%
\pgfpathlineto{\pgfqpoint{0.620063in}{1.713142in}}%
\pgfpathquadraticcurveto{\pgfqpoint{0.620063in}{1.693697in}}{\pgfqpoint{0.639507in}{1.693697in}}%
\pgfpathclose%
\pgfusepath{stroke,fill}%
\end{pgfscope}%
\begin{pgfscope}%
\pgfsetrectcap%
\pgfsetroundjoin%
\pgfsetlinewidth{1.505625pt}%
\definecolor{currentstroke}{rgb}{0.000000,0.000000,0.000000}%
\pgfsetstrokecolor{currentstroke}%
\pgfsetdash{}{0pt}%
\pgfpathmoveto{\pgfqpoint{0.658952in}{1.921089in}}%
\pgfpathlineto{\pgfqpoint{0.853396in}{1.921089in}}%
\pgfusepath{stroke}%
\end{pgfscope}%
\begin{pgfscope}%
\definecolor{textcolor}{rgb}{0.000000,0.000000,0.000000}%
\pgfsetstrokecolor{textcolor}%
\pgfsetfillcolor{textcolor}%
\pgftext[x=0.931174in,y=1.887061in,left,base]{\color{textcolor}\rmfamily\fontsize{7.000000}{8.400000}\selectfont 100\%}%
\end{pgfscope}%
\begin{pgfscope}%
\pgfsetrectcap%
\pgfsetroundjoin%
\pgfsetlinewidth{1.505625pt}%
\definecolor{currentstroke}{rgb}{0.501961,0.501961,0.501961}%
\pgfsetstrokecolor{currentstroke}%
\pgfsetdash{}{0pt}%
\pgfpathmoveto{\pgfqpoint{0.658952in}{1.785518in}}%
\pgfpathlineto{\pgfqpoint{0.853396in}{1.785518in}}%
\pgfusepath{stroke}%
\end{pgfscope}%
\begin{pgfscope}%
\definecolor{textcolor}{rgb}{0.000000,0.000000,0.000000}%
\pgfsetstrokecolor{textcolor}%
\pgfsetfillcolor{textcolor}%
\pgftext[x=0.931174in,y=1.751490in,left,base]{\color{textcolor}\rmfamily\fontsize{7.000000}{8.400000}\selectfont 40\%}%
\end{pgfscope}%
\end{pgfpicture}%
\makeatother%
\endgroup%

%% file: Discussion.tex
\section{Discussion}
\label{sec:discussion}
\subsection{Comparison to elastic wave propagation kernels}
The poroelastic material model is inherently computationally more expensive than the elastic model.
First of all, we increase the number of quantities ($\mathcal{Q}$) from \num{9} to \num{13}, thus, the number of total DOFs in a simulation increases.
Secondly, the space-time predictor, to compute the predicted element-local solution with a stiff source term, is substantially more complicated than the \ck procedure used in the elastic case.
For the predictor step, the elastic kernel requires \num[round-mode=places,round-precision=3]{0.476712} million floating-point operations with polynomials of degree \num{6}.
For the poroelastic model, the predictor kernel requires \num[round-mode=places,round-precision=3]{1.086859} million floating-point operations.
This increase in computational workload can be attributed to the increased number of quantities, but also to the source term, which is absent in the elastic case.
Furthermore, the poroelasticity kernel does not achieve the same performance as the elasticity kernel (c.f. \cref{sec:roofline}).
In conclusion, we estimate that the per-element cost of a simulation using poroelastic materials is about \num{3.6} times higher than a simulation with elastic materials.
Naturally, this does not include differences in time step size (e.g., in case of different P-wave speeds for poroelastic and elastic materials), mesh refinement requirements (consider, e.g., the refinement necessary for resolving a slow P-wave, as in the LOHp scenario) or in parallel scalability (where the higher per-element costs may be beneficial). 

\subsection{Limitations}
With applications using seismic wavefield synthetics up to $\approx$\SI{10}{Hz} in mind, we focus on the low-frequency case.
If an application requires the high-frequency regime, i.e. simulation of waves with frequencies comparable or larger than Biot's frequency (tens of \si{\Hz} to hundreds of \si{\kilo\Hz} for geo-reservoirs), the frequency-dependent permeability and resistive friction ($b=\kappa/\nu$) have to be taken into account by Darcy's law in the equations of motion.
\citet{gregor_subcell-resolution_2021} show how to incorporate the high-frequency case in the 2D FD framework.

An additional limitation is, that, in the derivation of our scheme, we assumed constant material parameters per element.
We could achieve subcell resolution if we computed the second integral in \cref{eq:poroelastic-weak}
by quadrature on each element as detailed in~\cite{castro_seismic_2010}.
However, in this case, the stiffness matrices are not matrices anymore but 3D tensors.
The scheme would become more complex and would require careful performance evaluation and optimisation.

\subsection{Future GPU and multi-physics implementation}
SeisSol is currently available as a CPU and a GPU version, which share a large portion of the codebase. 
All compute kernels of SeisSol are expressed in a domain-specific language (c.f. \cref{sec:implementation}), which is then translated to machine code, for either CPUs or GPUs~\cite{dorozhinskii_seissol_2021}.
Therefore, it does not pose a major challenge to run also simulations with the poroelastic model on a GPU cluster, but a careful performance study has to be done again.

Up to now only point sources have been considered.
For physics--based earthquake simulations and to study fault-fluid interaction we have to consider more complex sources. 
These will include moment-tensor based double-couple point source implementations, kinematic finite earthquake source models~\cite{mai_srcmod_2014} and non-linear earthquake rupture dynamics taking the interaction of frictional shear fracture and propagating waves into account~\cite{de_la_puente_dynamic_2009,pelties_three-dimensional_2012, pelties_verification_2014}.
To do so, a fault will be embedded as an internal boundary in the mesh.
At this interface, we do not just exchange information by numerical fluxes, but instead, we solve a nonlinear friction problem.
Dynamic rupture simulations in (visco-)elastic media can already be simulated with SeisSol.
To combine this source mechanism with poroelastic materials, the coupling between poroelastic parameters and parameters of friction laws have to be investigated.
To fully capture the interaction of fluids, fault slip and seismic waves, additional multi-physics interactions can be accounted for that describe the thermal pressurisation of pore fluids~\cite{sibson_interactions_1973, noda_earthquake_2009, viesca_ubiquitous_2015} during earthquake rupture.
The thermal pressurisation model was recently implemented in SeisSol~\cite{gabriel_3D_2020}.
In the context of geo-reservoirs, pressure increase can drive fluid flow and in turn govern earthquake dynamics e.g.~\cite{galis_induced_2017}, however, a holistic method allowing to couple poroelastic effects on wave propagation and on rupture dynamics at the same time is currently not available.

\subsection{Other applications}
The solution approach presented in \cref{sec:stp} is general and can be applied to a broader class of problems described by linear hyperbolic PDEs with a stiff reactive source term. 
In our derivations, we made two assumptions on the sparsity pattern of the stiffness matrices $K^\alpha$ (c.f. \cref{sec:stp-structure}) and the source matrix $E$.
The matrix $K^\alpha$ is problem independent, but the matrix $E$ depends on the PDE which we consider.
For example, in the case of viscoelastic attenuation, it takes an upper triangular form~\cite{kaser_arbitrary_2007}, just as in our poroelastic case.

An additional applicable example are the damped Maxwell equations,
which model the interaction of electric ($\mathcal{E}$) and magnetic ($\mathcal{H}$) fields~\cite{grote_explicit_2010}:
\begin{equation*}
        \epsilon \derivative{\mathcal{E}}{t} = \nabla \times \mathcal{H} - \sigma \mathcal{E} +j \qquad
        \mu \derivative{\mathcal{H}}{t} = \nabla \times \mathcal{E}
\end{equation*}
Here, $\epsilon$ is the relative electric permeability, $\mu$ the relative magnetic permeability and $\sigma$ the conductivity.
The current density $j$ is a source term comparable to seismic sources in the context of poroelasticity.
With a non-zero conductivity, the Maxwell equations contain a possibly stiff source term.
In order to apply our proposed scheme, we expand the rotation operator and can write down the equation in a similar fashion as \cref{eq:poroelastic-wave}.
The source matrix is then upper triangular again, such that we can apply the space-time DG method and \cref{alg:algorithm-3} to solve the resulting linear system of equations.

%% file: appendix.tex
\section{Convergence results in the \texorpdfstring{$L^1$}{L1} and \texorpdfstring{$L^2$}{L2} norm}
\label{sec:convergence_other_norms}
\begin{figure}
  \center{
    \input{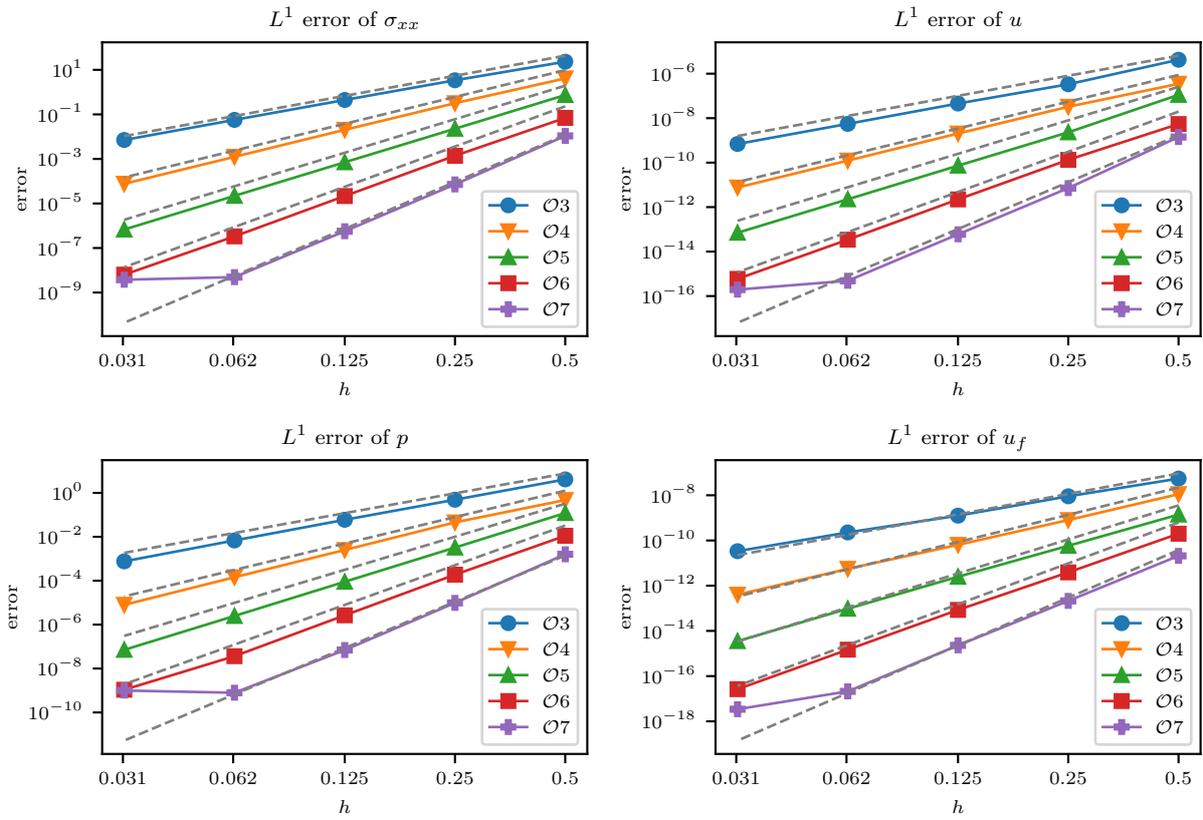}
  }
  \caption{Convergence plots for selected quantities of the planar wave convergence test in the $L^1$ norm. The expected convergence order is plotted in grey dashed lines. For $\mathcal{O}7$ we are close to machine precision on the finest mesh.}
  \label{fig:conv-l1}
\end{figure}

In \cref{sec:convergence}, we show the high-order convergence of our scheme with a planar wave scenario.
\cref{fig:conv-l1,fig:conv-l2} show the convergence results for the same setup in the $L^1$ and $L^2$ norm.
Also in these norms, we observe the same convergence behaviour as for the $L^\infty$ norm.
\begin{figure}
  \center{
    \input{figures/convergence_plots/dskx_poroelastic_s_xx-u-p-u_f_L2.pgf}
  }
  \caption{Convergence plots for selected quantities of the planar wave convergence test in the $L^2$ norm. The expected convergence order is plotted in grey dashed lines. For $\mathcal{O}7$ we are close to machine precision on the finest mesh.}
  \label{fig:conv-l2}
\end{figure}

\section{Convergence of the FD solutions}
\label{sec:convergence_fd}
\begin{figure}
    \center{
        \input{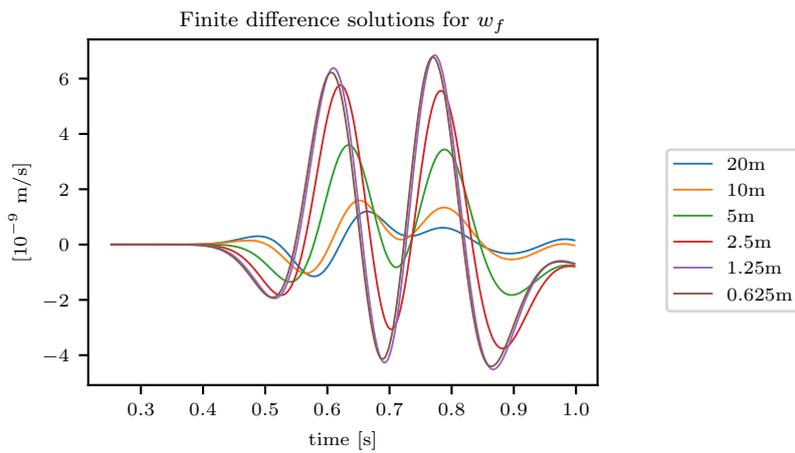}
    }
    \caption{Convergence of the vertical component of the relative fluid velocity ($w_f$) obtained with different grid spacings using the FD method.}
    \label{fig:convergence_fd}
\end{figure}

For the LOHp model (\cref{sec:lohp}), we use a 2D FD code as a reference.
Because the solutions obtained with a coarse resolution did not resolve the slow P-wave on the vertical component of relative fluid velocity ($w_f$) accurately enough, we conducted a convergence study for the FD solutions.
We used grid spacings of \SI{20}{\meter}, \SI{10}{\meter}, \SI{5}{\meter}, \SI{2.5}{\meter}, \SI{1.25}{\meter} and \SI{0.625}{\meter}.
\Cref{fig:convergence_fd} shows only a small difference between the solutions for grid spacings \SI{1.25}{\m} and \SI{0.625}{\m}, indicating that the solution converged.
We note that we choose the solution for grid spacing \SI{0.625}{\m} as the reference solution in \cref{sec:lohp}.
